\documentclass[twocolumn]{30onr_art}

\pdfoutput = 1

\usepackage[colorlinks=true,
            urlcolor=blue,
           linkcolor=docgreen,
            citecolor=docgreen,
            bookmarks=false,
            pdftitle={Direct Simulations of Wind-Driven Breaking Ocean Waves with Data Assimilation},
            pdfauthor={Douglas G. Dommermuth},
            pdfsubject={Proceedings of the 30th ONR Symposium on Naval Hydrodynamics, Hobart, Austrailia},
            pdfkeywords={Ocean Waves, Breaking Waves, Nonlinear Waves, HOS, NFA},
	   bookmarksopen=false,
            pdfpagemode=UseNone]{hyperref}

\usepackage{natbib}
\usepackage{url}
\usepackage{hyperref}
\usepackage{ulem}
\usepackage{graphicx}
\usepackage{fixltx2e}
\usepackage{amssymb,amsmath}
\usepackage{dblfloatfix}
\usepackage[compact]{titlesec}
\usepackage{color}
\definecolor{docgreen}{rgb}{0,.5,0}
\usepackage[top=1in,bottom=1in,left=1in,right=1in,paperheight=11in,paperwidth=8.5in]{geometry}
\usepackage{pdflscape}
\usepackage{textcomp}
\usepackage{setspace}

\titlespacing{\section}{0pt}{*0}{*0}
\titlespacing{\subsection}{0pt}{*0}{*0}
\titlespacing{\subsubsection}{0pt}{*0}{*0}

\newcommand\ie{i.e., }

\newcommand{\possessivecite}[1]{\citeauthor{#1}'s (\citeyear{#1})}
\newcommand{\comment}[1]{}

\def\curvered{\mbox{(~\textcolor{red}{-----------}~)}}

\def\curve{\mbox{(~-----------~)}}

\def\be{\begin{eqnarray}}
\def\ee{\end{eqnarray}}
\def\benl{\begin{eqnarray*}}
\def\eenl{\end{eqnarray*}}

\newcommand{\nwc}{\newcommand}
\nwc{\bm}{\boldmath}
\nwc{\m}{\mbox}
\nwc{\ubm}{\unboldmath}
\nwc{\bmU}{\m{\bm$U$\ubm}}
\nwc{\bmX}{\m{\bm$X$\ubm}}
\nwc{\bmu}{\m{\bm$u$\ubm}}
\nwc{\bmx}{\m{\bm$x$\ubm}}
\nwc{\bmz}{\m{\bm$z$\ubm}}
\nwc{\bmv}{\m{\bm$v$\ubm}}
\nwc{\bmw}{\m{\bm$w$\ubm}}
\nwc{\bmW}{\m{\bm$W$\ubm}}
\nwc{\bmn}{\m{\bm$n$\ubm}}
\nwc{\bmG}{\m{\bm$G$\ubm}}
\nwc{\bmF}{\m{\bm$F$\ubm}}
\nwc{\bmI}{\m{\bm$I$\ubm}}
\nwc{\bmN}{\m{\bm$N$\ubm}}
\nwc{\bmP}{\m{\bm$P$\ubm}}
\nwc{\bmcalP}{\m{\bm $\cal P$\ubm}}
\nwc{\bmV}{\m{\bm$V$\ubm}}
\nwc{\bmS}{\m{\bm$S$\ubm}}

\begin{document}
\addtocontents{toc}{\protect\setstretch{0.8}}

\title{Direct Simulations of Wind-Driven Breaking Ocean Waves with Data Assimilation}

\author{ Douglas G. Dommermuth \\
{\normalsize Naval Hydrodynamics Division, Leidos, Inc. \\ \vspace{-1.5mm}
10260 Campus Point Drive, MS C5, San Diego, CA  92121, USA} \\ \vspace{0.02in}
Christopher D. Lewis,  Vu H. Tran, and Miguel A. Valenciano \\ 
{\normalsize Data Analysis and Assessment Center, \\ \vspace{-1.5mm}
US Army Engineering Research and Development Center, MS 39180, USA}}

\maketitle

\begin{abstract}
A formulation is developed to assimilate ocean-wave data into the Numerical Flow Analysis (NFA) code. NFA is a Cartesian-based implicit Large-Eddy Simulation (LES) code with Volume of Fluid (VOF) interface capturing. The sequential assimilation of data into NFA permits detailed analysis of ocean-wave physics with higher bandwidths than is possible using either other formulations, such as High-Order Spectral (HOS) methods, or field measurements.  A framework is provided for assimilating the wavy and vortical portions of the flow.    Nudging is used to assimilate wave data at low wavenumbers, and the wave data at high wavenumbers form naturally through nonlinear interactions, wave breaking, and wind forcing.  Similarly, the vertical profiles of the mean vortical flow in the wind and the wind drift are nudged, and the turbulent fluctuations are allowed to form naturally.   Unlike subgrid-scale models of turbulence, nudging permits the direct enforcement of wave and turbulence statistics.  As a demonstration, the results of a HOS of a JONSWAP wave spectrum are assimilated to study short-crested seas in equilibrium with the wind.   Log profiles are assimilated for the mean wind and the mean wind drift.   The results of the data assimilations are (1) Windrows form under the action of breaking waves and the formation of swirling jets; (2) The crosswind and cross drift meander; (3) Swirling jets are organized into Langmuir cells in the upper oceanic boundary layer; (4) Swirling jets are organized into wind streaks in the lower atmospheric boundary layer; (5) The length and time scales of the Langmuir cells and the wind streaks increase away from the free surface; (6) Wave growth is very dynamic especially for breaking waves;  (7) The effects of the turbulent fluctuations in the upper ocean on wave growth need to be considered together with the turbulent fluctuations in the lower atmosphere; (8) Extreme events are most likely when waves are not in equilibrium; and (9) Vertical mixing is order 5 cm/s in the upper 15 meters of the ocean and 0.4 m/s in the lower 50 meters of the atmosphere.
\end{abstract}

\tableofcontents
\setcounter{tocdepth}{5}


\section{Introduction}

Data is sequentially assimilated into NFA every N time steps using nudging.   Here, we illustrate the assimilation of HOS simulations of the wavy portion of the flow and log profiles of the vortical portion of the flow into NFA.  The procedure can be generalized to assimilate radar or optical measurements of waves into NFA.   The wind profiles could be assimilated based on measurements using anemometers, SOnic Detection And Ranging (SODAR),  and lidar.  Wind-induced drift currents could be assimilated based on measurements using drifters.  

HOS, like measurements, is bandwidth limited.   In the case of HOS, approximations, including the Taylor series approximation, the perturbation expansion, and the single-valued free surface, limit the relative difference of the maximum wavenumber to the wavenumber at the peak of spectrum to about two decades.   Radar and optical measurements over a patch of the ocean surface have similar resolution limitations.  

HOS simulations or measured data are assimilated into NFA to drive the lowest wavenumbers of the wavy portion of the flow in the NFA simulation, and the highest wavenumbers in the NFA simulation are allowed to form naturally to enable modeling of wave breaking.   In the NFA data assimilation, energy cascades down from the lowest wavenumbers to the highest wavenumbers through the action of nonlinear wave interactions where it is dissipated due the effects of wave breaking, and forced by the wind and the wind drift.   Vertical profiles of the mean wind and the mean wind-drift are assimilated into NFA to drive the vortical portion of the flow.    The turbulent fluctuations form naturally through the energy cascade, the generation of free-surface vorticity, and the breaking of waves.   The wavy and vortical portions of the flow interact to form windrows, Langmuir cells, and wind streaks.  A separation of the flow into wavy and vortical components is similar to the Helmholtz decomposition that is used by \mbox{\cite{dommermuth93}} in his studies of the interaction of a vortex pair with a free surface.   \mbox{\cite{oceans13}} provide a formulation for assimilating just the wavy portion of the flow with no constraints on the vortical portion of the flow.  

Data assimilation has been used with HOS \citep{wu04,hassanaliaragh09,blondel10,yoon12}.   However, HOS data assimilations of phase-resolved ocean waves are limited to single-valued free surfaces with wave slopes no greater than one.   HOS is a potential-flow method with no capability to model turbulence or wave overturning.   Unlike HOS, data assimilations with NFA as the core solver provide the capability to simulate wave breaking directly including the effects of turbulence.

As  \mbox{\cite{perlin2013}} note in their annual review paper on breaking waves in deep and intermediate waters, as investigators  ``\ldots chip away at the difficult problem of quantifying wave breaking through laboratory investigations, field measurements, and numerical simulations, it is likely that progress will continue at a very slow pace.''  \mbox{\possessivecite{perlin2013}} assessment does not account for the impact that modern VOF methods in combination with data assimilation and supercomputers will have on understanding the effects of wave breaking. The authors of \mbox{\cite{perlin2013}} state that ``At this juncture, direct numerical simulations are not a viable option for ocean wave calculations over a large domain; therefore, a simpler approach should be adopted to model the breaking effects.''   As we show in this paper, the authors are mistaken.  Modern VOF methods with data assimilation implemented on today's supercomputers can resolve wave breaking over large patches of the ocean surface.

Data assimilations of breaking waves in equilibrium with the wind are used here to investigate the structures of the upper oceanic boundary layer (OBL) and the lower marine atmospheric boundary layer (ABL).    The short-crested seas are based on a JONSWAP wave spectrum.   The wavelength at the peak of the spectrum is 100 meters.   The significant wave height is 3.66 meters.    The wind speed at 10 meters height is 11.1 m/s.   The friction velocities in the atmosphere and the ocean are 0.814 m/s and 2.83 cm/s, respectively.   The ratio of the density of the air to the density of the water is 0.001207.  The wind drift on the ocean surface is 31.3 cm/s. The data is assimilated over a patch of the ocean surface that is 500 meters by 125 meters.   Coarse and medium-sized data assimilations are performed with respectively 12.2 and 6.10 cm resolution.   The total number of grid points for the coarse and medium assimilations are respectively 2.15 and 17.2 billion. The low wavenumber cutoff for the wavy portion of the flow is  0.70 rad/m, which corresponds to a wavelength of 8.98 m.   The durations of the coarse and medium assimilations are respectively  249.5 and 50.89 seconds.    Data are assimilated every 0.07983 and 0.03992 seconds for respectively the coarse and medium assimilations.  The high data rate is required to prevent ringing.   There is no stratification, and Coriolis effects are not considered.

\begin{figure*}[ht]
\begin{center}
\begin{tabular}{cc}
\includegraphics[width=0.45\linewidth]{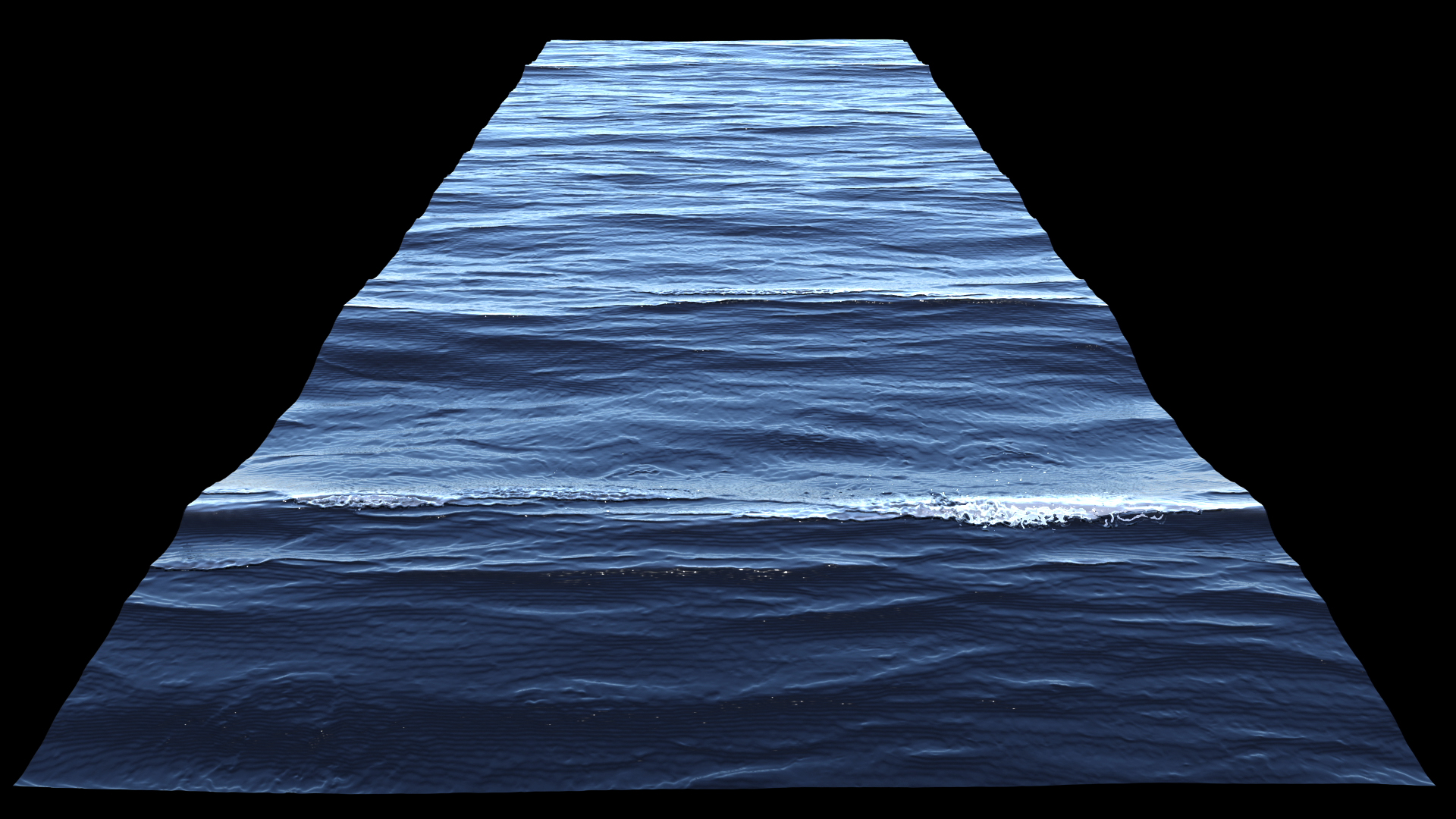} & 
\includegraphics[width=0.45\linewidth]{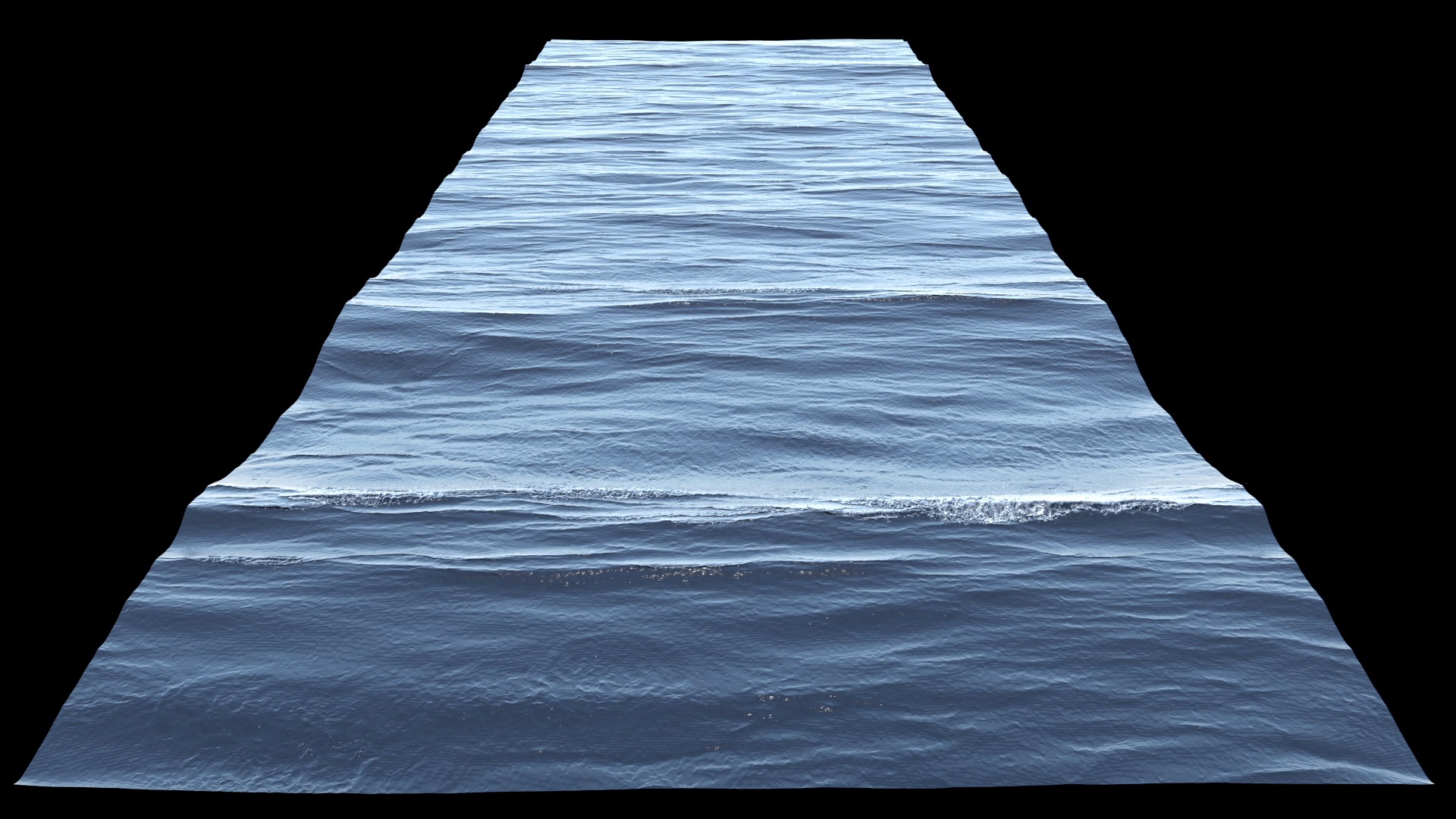}  \\
(a) & (b)
\end{tabular}
\end{center}
\caption{\label{fig:nfa_raytrace1} Perspective view of wave breaking.  (a) Coarse-sized data assimilation.  (b) Medium-sized data assimilation.  Animations of the preceding results are available at \mbox{\cite{daac4}} \href{http://youtu.be/ID3aAnlQlng}{coarse assimilation} and \mbox{\cite{daac10}} \href{http://youtu.be/GsjIZ36ytxw}{medium assimilation}.}
\end{figure*}

\begin{figure*}[ht]
\begin{center}
\begin{tabular}{cc}
\includegraphics[width=0.45\linewidth]{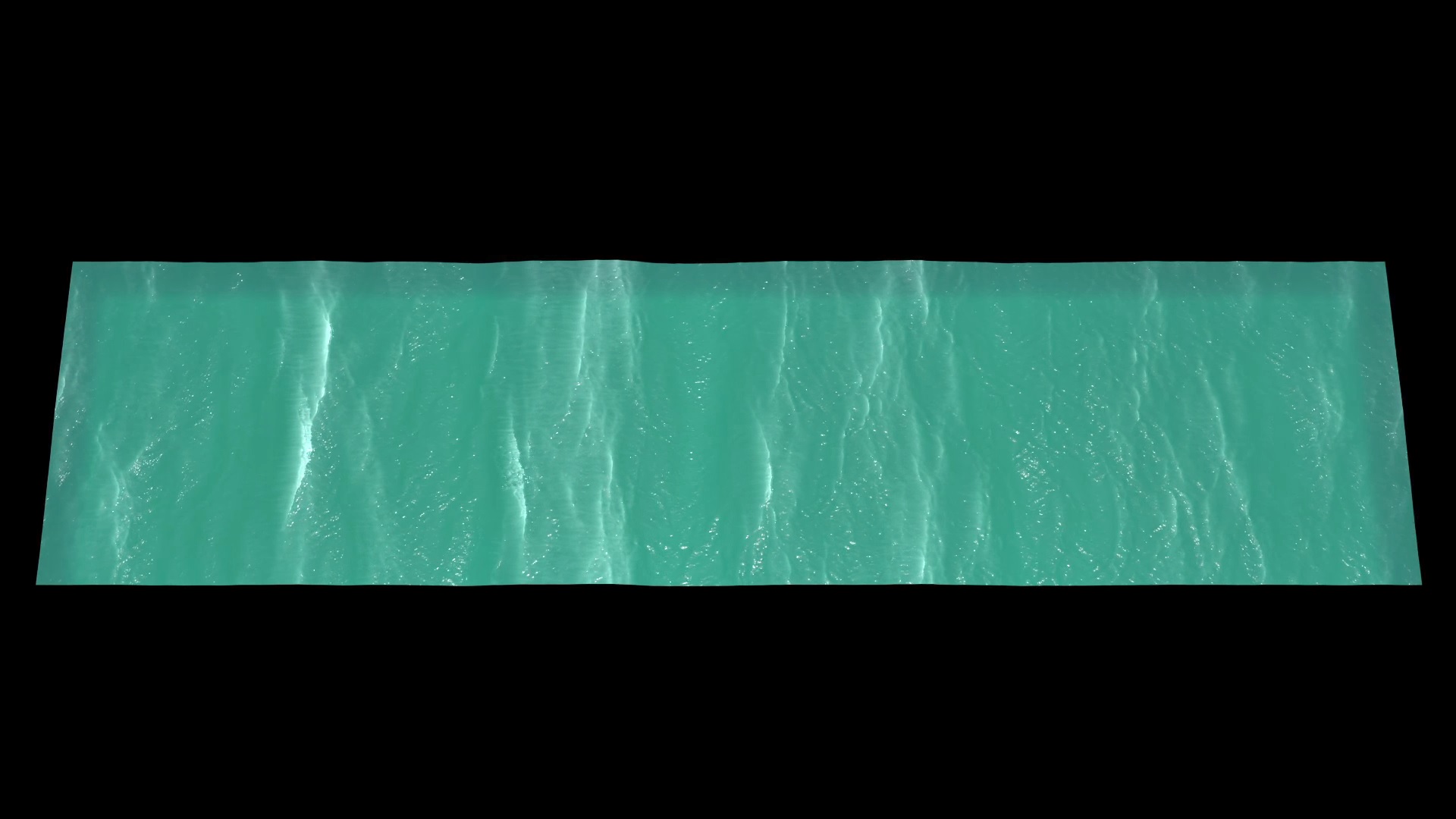} & 
\includegraphics[width=0.45\linewidth]{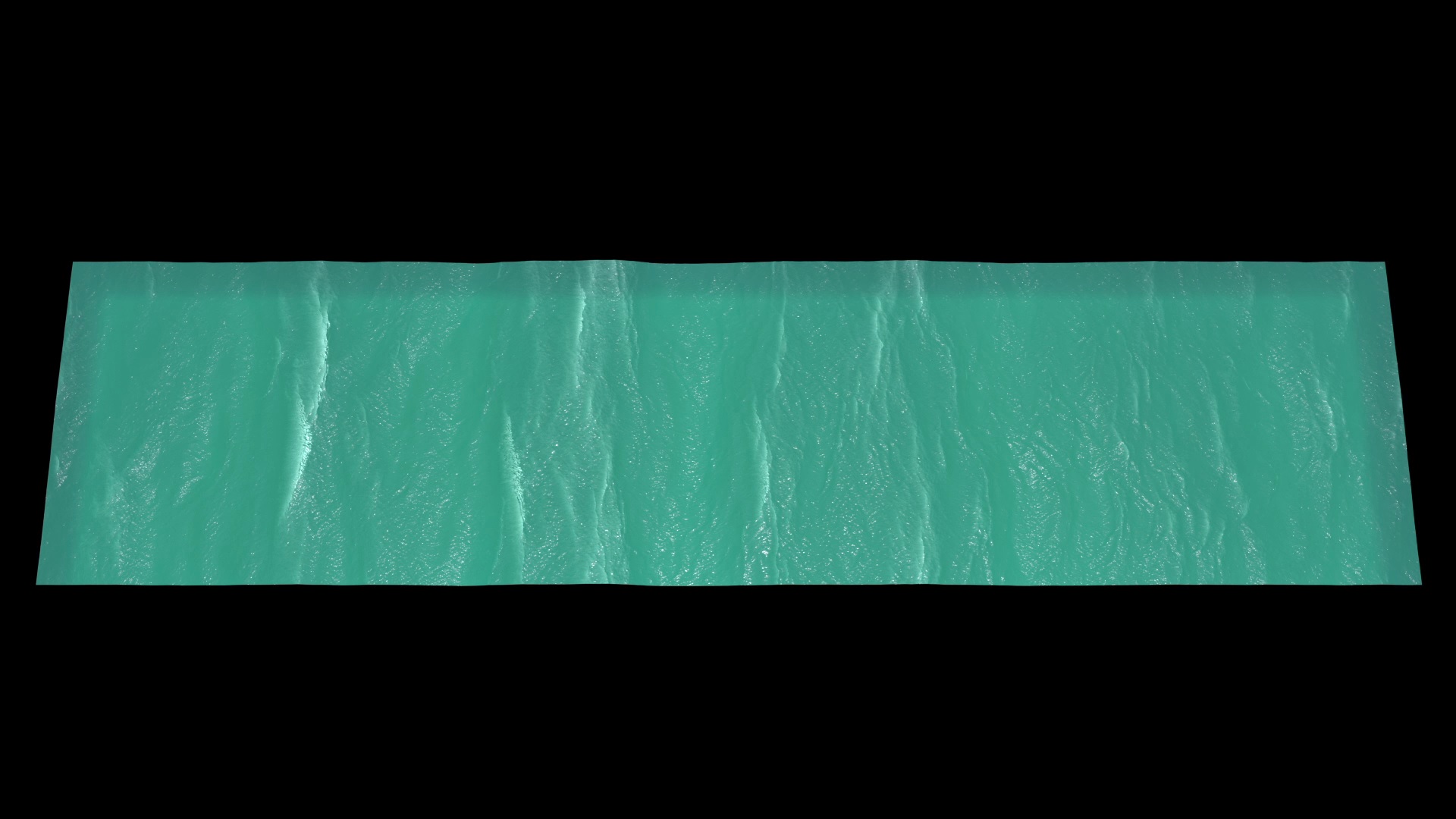}  \\
(a) & (b)
\end{tabular}
\end{center}
\caption{\label{fig:nfa_raytrace2} View looking down on wave breaking.  (a) Coarse-sized data assimilation.  (b) Medium-sized data assimilation.  Animations of the preceding results are available at  \mbox{\cite{daac7}} \href{http://youtu.be/n-0aFa0GAMw}{coarse assimilation} and \mbox{\cite{daac14}} \href{http://youtu.be/0NoBu9G-l5g}{medium assimilation}.}
\end{figure*}

As examples,  Figures \ref{fig:nfa_raytrace1} and \ref{fig:nfa_raytrace2} respectively show perspective and plane views of wave breaking for the coarse and medium-sized data assimilations.   The two assimilations are shown at the same instant of time.  Links to animations of the wave breaking are provided in the figure captions.   The figures in the electronic versions of this paper can be magnified to inspect details of the wave breaking that are occurring.   A plunging breaking event is occurring in the foregrounds of Figure \ref{fig:nfa_raytrace1}.  The animations show that white capping often occurs slightly behind the highest point of the underlying wave as the white water gets shed out the back.    The plane views in  Figure \ref{fig:nfa_raytrace2} show significant white capping.  The medium-sized data assimilation resolves the wave breaking better than the coarse assimilation.

Sections \S \ref{sec:langmuirintro},  \S \ref{sec:windstreaksintro}, and \S \ref{sec:growthintro} review respectively Langmuir circulations, wind streaks, and wave growth.   Sections  \S \ref{sec:bvp},  \S \ref{sec:vof}, and \S \ref{sec:assimilate} provide formulations for respectively HOS, NFA, and the data assimilation of HOS into NFA. Section \S \ref{sec:windrows} shows results for the formation of windrows.   Section \S \ref{sec:langmuirstreaks} shows results for the formation of Langmuir cells and wind streaks.   Section \S \ref{sec:profiles} discusses vertical profiles for crosswind meandering, cross drift meandering, and vertical streaming.  Section \S \ref{sec:growth} show results for wave growth.   Sections \S \ref{sec:statistics} and \S \ref{sec:mixing} provide respectively statistics of free-surface quantities and mixing probability distributions. 

\subsection{\label{sec:langmuirintro}Langmuir circulations}

On August 7, 1927, while crossing the Sargasso Sea, Irving Langmuir observed  bands of seaweed on the ocean surface \mbox{\citep{langmuir1938}}.   The bands of seaweed were aligned with the wind.   Langmuir called them streaks.   The streaks were upwards of 500 meters long with irregular spacings ranging from 100 to 200 meters for the larger streaks with smaller streaks interspersed among them.    Langmuir conjectured the following: ``At that time it seemed reasonable to me that the only reasonable hypothesis was that the seaweed accumulated in streaks because of transverse surface currents converging toward the streaks. The water in these converging currents descends under these streaks. Between the streaks rising currents, upon reaching the surface, flow out laterally toward the streaks."  Through a series of experiments, Langmuir showed that large-scale circulations did indeed exist beneath the ocean surface and those currents now bear his name.   However, numerical experiments that are reported in this paper show that the streaks form under the direct action of wave breaking as opposed to the large-scale circulations that are beneath the breaking waves.

Based on the results of our data assimilations, foam, biological material, flotsam, and jetsam surf the fronts of breaking waves.   The surfing action of spilling breaking waves scrubs the free surface.   The floating material is ejected out the sides and spills over the fronts of breaking waves.   Swirling jets are located beneath the floating material in the regions where shedding is occurring. The swirling jets on the ocean surface are aligned with the wind.   The width of the swirling jets is much less than the fronts of the breaking waves, and there are multiple swirling jets with regular spacings behind the breaking fronts.  (\mbox{\cite{gupta1984swirl}} and \mbox{\cite{shtern1999}} provide a review of swirling flows and jets.)  The floating material is shed by breaking waves in the same regions where swirling jets form because the velocities are lower there than at neighboring points.    As a result of this action, the floating material is aligned with the swirling jets.   Contrary to \mbox{\possessivecite{langmuir1938}} original hypothesis, windrows do not form due to flow converging transverse to the wind on the free surface due to the effects of Langmuir cells. For fully-developed seas our data assimilations show that windrows form under the action of breaking waves and the formation of swirling jets.

The jet portion of the flow within the swirling jets is downwind.   Windrows riding on top of jets that are in the direction  of the wind agrees qualitatively with measurements of Langmuir circulations that are reported in \mbox{\cite{smith2001}} and \mbox{\cite{thorpe2004}}.   Neutrally buoyant material that is placed along a windrow will be drawn down into the water by the swirl portion of the jet in agreement with Langmuir's original experimental observations.

In a Eulerian frame of reference, the swirling jets are very slender, but due to the meandering of the swirling jets, there are streaks of surface currents transverse to the wind in a Lagrangian frame of reference.   The correlation with the windrows of the convergence zones that are formed by these transverse surface currents is much less than the correlation of the windrows with the swirling jets themselves.    The meandering of the swirling jets also affects the spreading and enhancement of the wind drift  beneath the windrows in qualitative agreement with observations as show schematically in Figures 1 of \mbox{\cite{smith2001}},  \mbox{\cite{thorpe99a}},  and \mbox{\cite{thorpe2004}}.  In section \S \ref{sec:windrows}, we discuss the formation of windrows in greater detail.

In regard to the large-scale circulations beneath the surface, Langmuir's observations are as follows:  ``The effect of the wind is thus to produce a series of alternating right and left helical vortices in the water having horizontal axes parallel to the wind.   If we face in the direction toward which the wind blows we should observe that the water between two adjacent streaks forms a pair of vortices:  The water on the right-hand side of the vertical plane halfway between the streaks has a clockwise rotation (right helix), that on the left a counter clockwise rotation (left helix)."   

Indeed, the present numerical simulations show that pairs of helical vortices exist beneath the windrows.     The swirling jets are tilted by the vertical gradients in the horizontal component of the water-particle velocity that is in line with the wind.  The tilting that occurs in the crests of breaking waves is much stronger than that due to the Stokes drift.   Due to the effects of meandering on Lagrangian motions,  wide streaks form in the vertical component of the water-particle velocity and in the horizontal component of the water-particle velocity that is in the direction of the wind.  The swirling jets are also organized into streaks through merging.   Vortex breakdown of the swirling jets also contribute to the formation of the streaks.   Due to the interaction of the stream-wise vorticity with the orbital motion of the waves, the streaks in the crests and troughs are ninety degrees out of phase with each other giving rise to a helical orbit in the mixing that occurs.   Due to the changes in phase, the streaks in the horizontal component of velocity in the direction of the wind and the vertical component of velocity form an interweaving pattern.  Section \S \ref{sec:langmuirstreaks} shows the structure of Langmuir cells based on time-averaged velocity fields.

\mbox{\cite{dommermuth1992}} and  \mbox{\cite{dommermuth93}} observed similar mechanisms in numerical studies of vortex pairs interacting with walls and free surfaces.   As the vortex pair rises toward the free surface, striations  and whirls form on the free surface.   The striations are formed by transverse vorticity that is stretched across the vortex pair \citep{sarpkaya84,sarpkaya85,sarpkaya86,sarpkaya91}.   \mbox{\cite{dommermuth2009}} (\href{http://youtu.be/bBMQB8cRM9I}{Snail video}) provides a flow visualization of the primary and transverse vorticity for half of the vortex pair.   \mbox{\cite{walker2009}} (\href{http://news.bbc.co.uk/earth/hi/earth\_news/newsid\_8035000/8035593.stm}{BBC video}) shows the swirling jets that form on a plunging breaking wave (Source BBC News/bbc.co.uk - \textcopyright 1999 BBC).  The swirling jets are so strong that they entrain air.  The video shows the swirling jets breaking up and swirling jets pairing.   \mbox{\cite{brucker2009}} (\href{http://youtu.be/VSUitm-r8y0}{Breaking wave video}) shows transverse vorticity shed from an asymmetrical plunging breaking wave (see time 3:22).    \mbox{\cite{nfa1}} provides details of the breaking-wave analysis.   \mbox{\cite{watanabe2005}} show that transverse vortex structures also form between splash-up events for breaking waves in the surf zone.   The rib-vortex structures are composed of counter-rotating vortex tubes that connect from the top of one splash-up event to the bottom of the next splash-up event, which differs from the present mechanism for forming swirling jets.    The numerical simulations of \mbox{\cite{lubin2013}} show transverse jets that entrain air for plunging breaking waves but there are artifacts in their interface capturing scheme that impair their results.   

\mbox{\cite{handler2012}} observe transverse streaks  in their infrared imagery of small-scale spilling breaking waves.  The streaks are visible in \mbox{\cite{handler2010}} (\href{http://youtu.be/rGFoaD8kfdw}{Infrared video}) at 12 seconds.   The transverse spacings of the streaks are very regular, suggesting an instability.   Upwellings that are parallel to the breaking front are also visible in the video.   As shown in Figure 1 of \mbox{\cite{sullivan2010}}, windrows are often observed in the field, but there is no way of knowing the underlying vortical structure.   Slides 2 and 6 of \mbox{\cite{sullivan11}} show windrows for high-wind conditions.  Figure 2a and 2b of \mbox{\cite{thorpe99b}} respectively show streaks of foam transverse to the fronts of waves breaking on a beach and at a ship's bow.   The streaks of foam could be related to swirling jets.  Figures 3a and 3b of \mbox{\cite{thorpe99b}} show bands of foam that are parallel to the fronts of advancing breaking waves.    The bands of foam may be associated with the wake vorticity that is shed parallel to the front of breaking waves that is evident in our data assimilations.    Figure 6 shows of  \mbox{\cite{thorpe99b}} shows streaks of foam behind a wave breaking in deep water.    

The formation of windrows and Langmuir circulations are reviewed in \mbox{\cite{leibovich1983}}, \mbox{\cite{smith2001}}, \mbox{\cite{thorpe2004}}, and \mbox{\cite{sullivan2010}}.    The leading theory is based on a wave-current instability as discussed in \mbox{\cite{craik1976}} and \mbox{\cite{leibovich1983}}.    The key mechanism is a ``vortex force" that is included in the governing equations to model the interaction of the stokes drift and the vorticity.  The CL 1   mechanism as proposed by \mbox{\cite{craik1976}} relies on cross-wave interactions to induce stream-wise rolls.   The CL 2 mechanism as originally proposed by \mbox{\cite{craik1977}} is an inviscid instability that does not require a coherent surface-wave structure.   Span-wise perturbations to the component of velocity in the direction of the wind produces a vertical vorticity.    The cross-product of the stokes drift with the vertical component of vorticity produces a vortex force that is directed toward the position of the maximum velocity in the windward direction where a convergence zone forms.    Vertical currents are generated beneath the convergence zones by continuity.  As \mbox{\cite{thorpe2004}} discusses, the vertical gradients of the Stokes drift tilt the vertical components of vorticity to create stream-wise vorticity in the direction of the wind.  Of the two different mechanisms, CL 2 is considered to be the most robust \mbox{\citep{leibovich1983}}.   Windrows are formed according to this mechanism due to the small currents transverse to the wind that are induced by the circulations. 

The Stokes drift is key to generating Langmuir circulations according to Craik-Leibovich theory, but based on wave steepness, the Stokes drift is two orders smaller than the water-particle velocity in the crest of a spilling breaking wave where the fluid velocity is equal to the phase speed of the underlying wave.   If the wave steepness is 0.05, then the water-particle velocity in the crest of a spilling breaker is 400 times greater than the Stokes drift, and yet this effect is not included in Craik-Leibovich theory.    The averaging process that is used to derive Craik-Leibovich theory eliminates the very strong Eulerian velocity in the crest of a spilling breaking wave in favor of the very weak Lagrangian Stokes drift.   Notwithstanding the differences in magnitude, the vertical gradients of the water-particle velocities in the crests of breaking waves tilt vertical components of vorticity to form steam wise vorticity in line with the wind like the Stokes drift.    

Craik-Leibovich theory does not account for the surfing effect of breaking waves.    The water-particle velocity in the direction of the wind is equal to the phase speed in the crests of spilling breaking waves, and the vorticity in the crests moves with the phase speed of spilling breakers.   The effects of surfing scrub the free surface clean behind a spilling breaker.   Surfing is the first stage of the formation of windrows.    During the second stage, floating objects spill over the sides and fronts of spilling breaking waves at the same points where swirling jets form because the water-particle velocities at those points are slightly less than neighboring points.   Floating matter and swirling jets are collinear due to this effect of surfing.

Craik-Leibovich theory does not account for the periodic shedding of vorticity into the wakes of steep waves and spilling breakers.    The shed vorticity is parallel to the fronts of steep and breaking waves and transverse to the wind.   As wake vorticity moves away from the crests of steep waves or the fronts of breaking waves, swirling jets are stretched across the wake vorticity and the vorticity at the front.   The stretching forms long streaks of vorticity with a helical structure due to the jet and swirl portions of the flow interacting with each other.    The swirling jets are key to the formation of windrows and Langmuir circulations.

The free surface is modeled as a flat wall with an applied tangential stress in Craik-Leibovich theory.   There is no accounting for the effects of free-surface vorticity as discussed by \mbox{\cite{lundgren1989}}.  As Lundgren shows, vorticity is generated on the free surface wherever there is flow past regions of surface curvature.    \mbox{\cite{mui1994}} confirm the effect of free-surface curvature on free-surface vorticity in their direct simulations of parasitic capillary waves forming on gravity waves.  The flow separation that occurs in regions where there is high surface curvature such is in the crests of steep non-breaking waves and the uneven fronts of spilling breaking waves is not modeled in Craik-Leibovich theory.  The  vorticity that is generated in these regions is very strong because both the water-particle velocities and the free-surface curvature are high.

Craik-Leibovich theory does include the interactions of the wavy and vortical portions of flow over length and time scales comparable to the length and period of the wave.   The interaction between the wavy and vortical portions of the flow is limited to the Stokes drift, which has very long length and time scales.   The Stokes drift through the vortex force accounts for one type of effect of the wavy portion of the flow on the vortical portion of the flow.    As will be shown, the helical structure of  Langmuir circulations is in large part due to interactions of the orbital velocities of the waves with the vortical portion of the flow.   As formulated, Craik-Leibovich theory does not account for the effect of the vortical flow on the waves, but current results show indications that vortical portion of the flow affects wave growth. 

 \mbox{\cite{mcwilliams1997}} study Langmuir turbulence in the ocean using the phased-averaged Craik-Leibovich equations and large-eddy simulation (LES).   \mbox{\cite{mcwilliams1997}}  also include the effects of Coriolis forces and stratification that are not included in the present analysis.   As defined by \mbox{\cite{mcwilliams1997}}, the turbulent Langmuir number ${\rm La}_{tur}=\sqrt(w_*/U_s)$ measures the relative importance of the wind drift and the Stokes drift, where $w_*$ is the friction velocity in the water and $U_s$  is the magnitude of the Stokes drift at the ocean surface.  They discuss results for  ${\rm La}_{tur}=0.3$, which is within empirically observed conditions for the formation of Langmuir circulations, where $0.2 \lesssim {\rm La}_{tur} \lesssim 0.5$.     In comparison, ${\rm La}_{tur} \approx 0.53$ for the numerical simulations that are discussed in this paper.   The Stokes drift that is used in \mbox{\cite{mcwilliams1997}}  is based on a wavelength of 60 meters and the wave steepness is 0.08, which would typically be breaking.    The wavelength at the peak of the JONSWAP spectrum that is used in this paper is 100 meters.   The significant wave height is 3.66 meters.
 
Figures 20a-h in \mbox{\cite{mcwilliams1997}}  show particle traces as a function of time that are strongly correlated with the convergence zones that form between streaks of positive and negative phase-averaged stream-wise vorticity.    Compared to observations, the particle traces in \mbox{\cite{mcwilliams1997}} are not very straight with large excursions transverse to the wind.   Unlike the vermiculation of  \mbox{\cite{mcwilliams1997}}, the results of the present numerical simulations show that the positions of the windrows are most closely correlated with the swirling jets rather than the convergence zones that are formed by the stream-wise vorticity.   There is no surfing in \mbox{\cite{mcwilliams1997}}  because there are no breaking waves.  In section \S \ref{sec:windrows},  we show the correlation of the windrows with the swirling jets that are beneath them.

Figures 2 and 3 of \mbox{\cite{melville1998}} show particle traces of laboratory experiments that are similar to the particle traces in \mbox{\cite{mcwilliams1997}}.   The experiments of \mbox{\cite{melville1998}} are based on a shear layer and wind waves forming from rest.   The breakup of the shear layer in  \mbox{\cite{melville1998}} gives particle traces that are different from the windrows that are formed by swirling jets when the waves are fully developed.

After the initial stages, the spacings transverse to the wind of the particle traces in Figures 20a-h of \mbox{\cite{mcwilliams1997}} do not show any tendency to get wider.    The spacings transverse to the wind of the  particle traces in the present results do get wider under the action of wave breaking.   The widening of transverse spacings between the streaks is accompanied by an increase in density of the particles in each streak.    The widening of the distances between the streaks and the increase in density of the particles agrees with simple model that is described in \mbox{\cite{thorpe09}}.   He distinguishes between streaks that are formed continuously, such as those due to foam, and streaks that are formed by an initial injection of floating material, such as algae.   Also, \mbox{\cite{thorpe09}} notes that spacings between the streaks do not necessarily conform to the distances between the convergence zones that are formed by Langmuir circulations, which agrees with the present results and the observations of bands of macro algae by \mbox{\cite{Qiao2009}}.

Figures 12a-d in \mbox{\cite{mcwilliams1997}} show the results 3 meters below the free surface for the stream-wise vorticity, the vertical velocity, the fluctuating horizontal velocity transverse to the wind, and the fluctuating horizontal velocity in the stream-wise direction, respectively.   To emulate the results in Figures 12a-d \mbox{\cite{mcwilliams1997}}, which are instantaneous values based on the phased-averaged Craik-Lebovich equations with a flat free surface, the results in this paper are phased averaged over two wave periods in a surface-following coordinate system that 3.2 meters below the free surface.   

The magnitudes of the velocities in the vertical and stream-wise directions in the present study are over 10 times greater than those reported in \mbox{\cite{mcwilliams1997}}.    The magnitudes of the horizontal  velocity component that is transverse to the wind is over five times greater.   The lowest velocities in \mbox{\cite{mcwilliams1997}} are in the stream-wise direction, whereas in the present case, those are the highest velocities due to the effects of surfing and wave breaking.  The velocity measurements of \mbox{\cite{langmuir1938}} are as follows: ``In the streaks it was thus found that two meters below the surface there were descending currents of 2 to 3 cm/sec and rising currents of from 1 to 1.5 cm/sec midway between adjacent streaks."   Based on the results of \mbox{\cite{langmuir1938}}, the observed mean rise and fall of objects that are suspended in Langmuir circulations are greater than the peak values that are reported in \mbox{\cite{mcwilliams1997}}.   In section \S \ref{sec:langmuirstreaks}, we show the structure of Langmuir cells based on time-averaged velocity fields at various water depths.

The horizontal stream-wise and vertical velocities in Figures 12b and d of \mbox{\cite{mcwilliams1997}}  show very little organization in comparison to the present case.    For the current study,  the time-averaged horizontal stream-wise and vertical velocities have a distinct helical  structure that is not evident in the results of \mbox{\cite{mcwilliams1997}}.    The interweaving streaks are a result of interactions between the orbital velocity of the waves, surfing, and the vortical portion of the flow.    The helical orbits agree qualitatively with observations that are reported in \mbox{\cite{langmuir1938}}. 

\mbox{\cite{sullivan2007}} add a stochastic model of wave breaking to the vortex-force model that is used in  \mbox{\cite{mcwilliams1997}}.  \mbox{\cite{sullivan2004}} discuss the development of the breaking model.  As before, the effects of Coriolis forces and stratification are included, and the free surface is modeled as a flat wall.  The oscillatory motions of the waves and the effects of the vortical portion of the flow on the wavy portion are not modeled.   Their stochastic model of wave breaking is designed to match laboratory and field observations of breaking waves including mean fluxes of momentum and energy.   The laboratory experiments are based on measurements of breaking waves by  \mbox{\cite{melville2002}}.  The Stokes drift is input based on a spectrum of waves in equilibrium with the wind.   Their wind speed at 10 meters height is 15 m/sec.   In comparison, for the current study, the wind speed at 10 meters height is 11.1 m/sec.   For \mbox{\cite{sullivan2007}}, the turbulent Langmuir number is ${\rm La}_{tur}=0.3$. 

As \mbox{\cite{sullivan2004}} show,  simulations of a single breaking event compare well to digital particle image velocimetry measurements of \mbox{\cite{melville2002}} for velocities and mean kinetic energy, thus confirming their basic approach.    The velocity measurements of  \mbox{\cite{melville2002}}  are based on ensemble averages of plunging breaking waves.   A coherent vortex is generated that propagates downwind and descends vertically.   The sign of the coherent vortex is positive corresponding to the vorticity at the front of the breaking wave.   There is also a thin layer of negative vorticity that may be associated with wake vorticity, but  \mbox{\cite{melville2002}} note that it could be an artifact of their near-surface processing.  Figure 3  of \mbox{\cite{sullivan2004}} shows the velocity field for a single two-dimensional breaking event.   The velocities are two to four percent of the phase speed of the wave corresponding to post breaking.   There is no surfing mechanism, nor is there intermittent formation of wake vorticity and swirling jets.  

\mbox{\cite{sullivan2007}} observe downwelling jets in their numerical simulations that are depth filling.   As \mbox{\cite{sullivan2007}} discuss, the formation of the downwelling jets appears to be similar to a mechanism that is proposed by \mbox{\cite{csanady94}}.   As  \mbox{\cite{csanady94}} notes, the lateral spacings of windrows have a log normal distribution suggesting that they form at random times and locations on the sea surface.   \mbox{\cite{csanady94}} proceeds to speculate that breaking waves are capable of generating surface convergences in their wakes that could be responsible for the formation of windrows.    According to \mbox{\cite{csanady94}}, the Stokes drift could tilt vertical lines of vorticity at the edges of stress anomalies such as those formed by breaking waves.  The generation mechanism of \mbox{\cite{csanady94}} is a ``forced" version of Craik-Leibovich's  CL 2 theory that does not require feedback to grow infinitesimal span-wise disturbances.   The mechanism of \mbox{\cite{csanady94}} is also applicable to the tilting of the swirling jets that are observed in this study.

Figures 14 and 19 of \mbox{\cite{sullivan2007}} show the vertical component of the velocity 13.38 meters beneath the free surface.    Round downwelling jets that are  12 to 15 meters in diameter are visible.    In the very early stages of the formation of Langmuir cells, our research shows similar structures that are more intense for the w-component of velocity than those in \mbox{\cite{sullivan2007}}.   The difference in intensity may be attributable to the different turbulent Langmuir numbers.  

Our numerical results show that the other two components of velocity also have interesting features that are not discussed by  \mbox{\cite{sullivan2007}}.  In the present study, swirling jets are tilted by the wind drift and the Stokes drift.   Deep below the free surface and before the Langmuir cells are fully formed,  there are circular features in the u-component of velocity that are in the direction of the wind due to the jet portion of the swirling jets.   Also due to the jet portion of swirling jets, there are circular features in the w-component of velocity that are negative, which corresponds to the downwellings that are observed in \mbox{\cite{sullivan2007}} .   The v-component of velocity has quadrupole structures due to pairing of swirling jets.    The u-component of velocity also shows a weak swirling effect, but the jet portion of the flow tends to dominate the swirling portion for the u-component of velocity.   The swirling jets at depth are organized over time into Langmuir circulations in the present study.  Figure \ref{fig:langmuir}i shows the effects of the swirling jets 12.8 meters below the free surface before the Langmuir cells are fully formed.

Figure 5 of \mbox{\cite{sullivan2007}}  show the mean profiles of the horizontal velocity as a function of depth beneath the flat free surface.    \mbox{\cite{sullivan2007}}  do not discuss any variation of the profiles as a function of time.   In the present study, mean and fluctuating quantities are calculated using spatial averaging in a wave-following coordinate system as a function of time and distance to the free surface.   The profiles of the mean crosswind in the lower atmosphere and the mean cross drift in the upper ocean meander as a function of time and distance to the free surface.   The meandering of the cross drift  develops over time as the Langmuir circulations form with depth starting from rest.   Similarly, the meandering of the crosswind extends up into the lower atmosphere as large coherent structures propagate upward.   The frequency of oscillations for both the crosswind and cross drift is highest close to the free surface, and the frequency of the meandering decreases away from the free surface.     A weak streaming flow is also observed in the mean vertical velocities in the wind and the wind drift.

\mbox{\cite{thorpe95}} provides a simple analytical model for the meandering and dispersion of a plume of floating material under the action of a current and a steadily advecting array of Langmuir cells.    Figures 3 and 4 of  \mbox{\cite{thorpe95}} show the meandering and dispersion of floating diesel oil and crude oil, respectively.   The analysis of \mbox{\cite{thorpe95}} could conceivably be extended to three dimensions including the effects of waves and unsteady Langmuir circulations.

\mbox{\cite{huckle2011a}} (\href{http://youtu.be/enWfIGn9jBE}{u velocity video})  and \mbox{\cite{huckle2011b}} (\href{http://youtu.be/TrOsM0EtLtc}{w velocity video}) show animations of the u and w-components of velocity with Coriolis effects, vortex forcing, and a stochastic model of wave breaking.  The water depths are $z=-0.42$ and $z=-3.0$ meters for the u and w velocities, respectively.   The wind speed is 15 m/s.  The animations represent almost nine minutes of simulation.   No surfing due to wave breaking is evident in videos.   No swirling jets are shed from fronts of breaking waves.   Unnatural breaking events occur on top of other breaking events because of the stochastic nature of the breaking model.   The sudden appearance of breaking events in the animations is not consistent with the derivation of the phase-averaged Craik-Leibovich equations.  The breaking model of  \mbox{\cite{sullivan2007}} preserves energy and momentum due to breaking, but important dynamics are still missing that affect mixing in the ocean and the atmosphere.  

As \mbox{\cite{thorpe92}} discusses, bubbles are used as tracers in measurements of acoustic backscatter to study mixing in the upper ocean.   The acoustic scattering is used in range versus time plots to determine the depth of bubble penetration, the length of the bubble clouds, the near-surface structure of the bubble clouds, and the sources of bubbles.   

The depth of bubble plumes and the length of bubble clouds increases with the ratio of the wind speed ($U_{10}$) to the phase speed ($c_b$) of the dominant breaking wave.   According to  \mbox{\cite{thorpe92}}, the mean cloud depth ($d$) reaches about 10 meters for wind speeds of about 12 m/sec.   Figure 1a-c  of \mbox{\cite{thorpe92}} show the depths of the bubble clouds as a function of the wind speed.   The ratio of the depth to the wavelength ($\lambda_b$) of the dominant breaking waves increases from $d/\lambda_b=.04$ for $U_{10}/c_b=0.6$ to $d/\lambda_b=0.2$ for $U_{10}/c_b=2$.   The ratio of the length ($L_b$) of the bubble clouds to the wave length increases from $L_b/\lambda_b=.1$ for $U_{10}/c_b=0.6$ to $L_b/\lambda_b=1.2$ for $U_{10}/c_b=2$.  For this study, the ratio of the wind speed to the phase speed ($c_o=12.5 {\rm m/sec}$) at the peak of the spectrum is $U_{10}/c_o=0.89$.    Based on the current simulations, time-averaged velocity fields at various depths from \mbox{40 cm} to \mbox{6.4 m}  show large structures with lengths that are about 60\% of the wave length ($L_o$) at the peak of the spectrum, which agrees qualitatively with  \mbox{\cite{thorpe92}}.

In the present case, Probability Density Functions (PDFs) of mixing show that passive particles diffuse from the free surface down to depths exceeding 12 meters in less than 4 minutes of simulation time with no initial turbulence.  The depth of penetration agrees  qualitatively with the measurements of \mbox{\cite{thorpe92}} (see their Figures 1a-c).  We conjecture that the rate of turbulent diffusion would have been even greater if the flow had been fully developed.

Figures 2 and 3 of \mbox{\cite{thorpe92}} show range versus time plots for side-scan sonars looking up at the free surface at an acute angle.  Note that the frequency of side-scan sonar that is used in Figure 3 is tuned to illuminate wave breaking events.     The wind speeds are $U_{10}=15 \pm 2 {\rm m/sec}$ and $U_{10}=6 {\rm m/sec}$  for Figures 2 and 3, respectively.  The bubble clouds are spaced about 15 meters apart in Figure 2.   The time-averaged downwind and vertical velocities of our results have streaks corresponding to Langmuir cells that are about 11 meters apart for water depth that is 3.2 m.   The lateral spacing between the streaks gets smaller closer to the free surface.    

Figure 2 of \mbox{\cite{thorpe92}} shows small bands of bubble clouds merging into larger bands.   As some of the larger bands amalgamate they form characteristic downwind-pointing Y-junctions.   The moving time-averages of the downwind and vertical velocities in our results show the formation of numerous Y-junctions forming over time as streaks interweave in a helical pattern.   The interweaving of streaks and the formation of Y-junctions leading to mixing in the lateral direction is likely related to the meandering in the cross drift and crosswind that had been discussed earlier.  

Figure 3 of \mbox{\cite{thorpe92}} show trails of bubble clouds being shed from the fronts of breaking waves.   Some of the breaking waves appear to shed more than one trail of bubbles corresponding to the multiple swirling jets that are shed from breaking waves in our numerical simulations.

\comment{If possible, I need to reference the other paper that Steve Thorpe gave to me.  See thorpe03.}

\subsection{\label{sec:windstreaksintro}Wind streaks}

A Stokes drift, surfing effects, and swirling jets are also present in the air above the waves.   In the atmosphere, the wind shear dominates the Stokes drift and surfing effects except for a region that is very close to the free surface.     The results of the present numerical simulations show that wind streaks form above the waves.   The general structure of the wind streaks is similar to the coherent structures that are beneath the waves albeit on a much larger scale.  

\mbox{\cite{foster2006}} distinguish  two types of large-scale structures in the atmospheric boundary layer: roll vortices and wind streaks.    Roll vortices are persistent large-scale structures that are upwards of 100 km long with lateral spacing that is 1 to 2 km.   Roll vortices span the entire depth of the ABL.   As \mbox{\cite{foster2006}}  discuss, wind streaks are transient coherent structures that are 100 to 300 m long that are limited to the surface layer.   The formation of roll vortices relies on buoyancy effects, whereas streaks can form in neutrally-stratified boundary layers.    Roll vortices are so long that geostrophic effects are important.   

As noted by \mbox{\cite{mourand2000}}, the length scales of roll vortices are based on satellite images of clouds that obscure effects that can occur over shorter length and time scales.  \mbox{\cite{mourand2000}} caution that long cloud streaks are not necessarily indicative of equally long roll vortices - just as long windrows are not indicative of equally long Langmuir cells.    \mbox{\cite{mourand2000}} show a correlation between Synthetic Aperture Radar (SAR) streaks and roll vortices.   They attribute the SAR streaks to ocean-surface roughness.   The SAR streaks are 2 to 10 km long with lateral spacing that is 1.5 to 2.5 km.   The widths of the SAR streaks are 0.5 to 2.5 km.     

The LES results of  \mbox{\cite{foster2006}} show wind streaks with lateral spacing varying from 100 to 200 m.  The streaks exist up to 100 m  above the surface in the LES  of \mbox{\cite{foster2006}} .    \mbox{\cite{moeng1994}} also observe wind streaks in the results of their LES.   The streaks in the LES of \mbox{\cite{moeng1994}} are 0.5 to 2 km long and 250 m wide with a lateral spacing of 0.5 to 1 km.   The LES of \mbox{\cite{foster2006}} and \mbox{\cite{moeng1994}} do not have sufficient resolution to resolve the swirling jets that form on the ocean surface.

\cite{lemone1973,brown1980,foster1996} provide details of the instability mechanisms that lead to the formation of roll vortices. The roll vortices form through a combination of dynamic and convective instabilities.  The general structure of atmospheric roll vortices is similar to Langmuir cells.     As \mbox{\cite{foster2013}} discusses, the axes of the roll vortices are roughly aligned with the wind in the lower ABL.   The roll vortices form counter-rotating pairs of vortices with convergence and divergence zones, just like Langmuir cells.     We note that the updrafts that are observed due to the formation of roll vortices in the ABL are similar to the downwellings that are observed in Langmuir circulations in the OBL.  Figure 1 of \mbox{\cite{foster2013}} shows schematically that the structures of atmospheric roll vortices and Langmuir circulations are similar.

 Our numerical results show streaks in all three components of velocity.   The wind streaks are very narrow close to the free surface, especially for the z-component of velocity, which is indicative of swirling jets.   At higher altitudes, the wind streaks span the entire length of our computational domain and are fed by the swirling jets that are forming on the ocean surface.    As before, the swirling portion of the flow is associated with stream-wise vorticity in the direction of the wind, and the jet portion of the flow is due to components of vorticity that are transverse to the wind.   The stream-wise portion of the vorticity seeds the formation of the roll vortices.    The vertical portion of the vorticity that is transverse to the wind at the ocean surface also contributes to the seeding of roll vortices through tilting of vorticity in a manner that is similar to the formation of Langmuir cells.  Although there are differences due to the strength of the background shear, the formation of wind streaks and the seeding of  roll vortices in the ABL is similar to the formation of Langmuir cells in the OBL.  The wind streaks are long and narrow at higher altitudes, and  the streaks are aligned with the wind.   
 
At 12.8 m above the free surface, the wind streaks in our data assimilations are between 50 and 500 m long and 10 to 30 m wide with lateral spacings of about 80 m between streaks with the same sign.   The velocity in the streamwise direction is about 10 to 20\% of $U_{10}$, and the velocities in the plane that is transverse to the wind are about 5 to 10\% of $U_{10}$.    In comparison, according to \mbox{\cite{foster2013}}, the velocity in the stream wise direction for roll vortices is about 10 to 25\% of the wind speed above the ABL, and the velocity transverse to the wind in the horizontal plane is 5 to 10\% of the same wind speed.   

The relative velocities of the wind streaks in our data assimilations are very similar to observations of roll vortices.   The helical structure of the wind steaks is also similar to roll vortices.  The similarities suggest that wind streaks could support the formation of roll vortices.  In section \S \ref{sec:langmuirstreaks}, we show the structure of wind streaks based on time-averaged velocity fields at various altitudes above the free surface. 
 
\mbox{\cite{marusic2008}} show the formation of large-scale superstructures in the log layers of neutrally buoyant Atmospheric Surface Layers (ASLs).   The superstructures sinusoidally meander in a manner that is similar to the wind streaks in \mbox{\S \ref{sec:langmuirstreaks}}, which are also in the log layer of a neutrally buoyant fluid.   Figure 14 of \mbox{\cite{marusic2008}} shows a superstructure in the streamwise velocity that is 500 m long and about 20 meters wide at a height that is 2.14 m above a desert floor, which is comparable to the wind streaks that are observed in our data assimilations for heights greater than 3.2 m above the free surface.

The superstructures that are observed by  \mbox{\cite{marusic2008}}  are indeed very large, but we provide evidence that even larger  structures exist as seen in the vertical profiles of the cross wind in \mbox{\S \ref{sec:profiles}}.   The meandering of the cross wind is occurring at length scales that are much larger than the superstructures that are observed by \mbox{\cite{marusic2008}}.   Moreover, we show that length and temporal scales of the wind streaks and the cross wind get larger as the height above the free surface increases.    Since the turbulence in our data assimilations is organized into large-coherent structures as the turbulence diffuses up into the atmosphere, our results support a `bottom-up' model for the formation of wind streaks, meandering cross winds, and vertical streaming flows.   As noted by \mbox{\cite{marusic2012}}, there is some  ambiguity to whether a `bottom-up' or `top-down' model of turbulence is most appropriate in the log layer.  \mbox{\cite{marusic2012}} suggest that a `top-down' model dominates at high Reynolds numbers, whereby large coherent structures drive small-scale turbulence on the surface.

Once the wind streaks are formed, they can interact with the free surface to generate patches of ripples.   In fact, the wind streaks that we observe in our numerical simulations are also very similar to the wind streaks that are observed in radar images that are processed using the Surface Feature Monitoring System (SuFMoS).  The images had been provided to us by Dr.\ Jochen Horstmann \mbox{\citep{horstmann2014}}.   \mbox{\cite{dankert2005}}, \mbox{\cite{dankert2007}}, and \mbox{\cite{vincenbueno2013}} provide details of SuFMoS. SuFMoS processing is based on radar backscatter of the ocean surface, which is primarily driven by capillary waves that are particularly sensitive to wind gusts.   Dr.\ Jochen Horstmann notes that they observe streaks between \mbox{50 m} using SuFMoS and \mbox{10 km} long using SAR \mbox{\citep{horstmann2014}} .   The streaks that are observed using SuFMoS are 50 to 500 m long and 10 to 50 m wide with lateral spacings of about 200 to 400 m between streaks moving in the same direction.     Figure \ref{fig:windstreaks2} of this paper shows the SuFMoS image of wind streaks.
  
\subsection{\label{sec:growthintro}Wind-wave growth}

\mbox{\cite{phillips1957}} describes the initial growth of waves due to resonant interactions with turbulent fluctuations in the wind that have the same speed and wavenumber.   There is good agreement with measurements for initial growth.    \mbox{\cite{miles1957}} develops a  theory for the growth of waves under the action of wind over long periods of time.   According to \mbox{\cite{miles1957}}, the growth is proportional to the curvature of wind shear at the elevation where the wind speed is equal to the phase speed of the wave.   The flow in the air is based on inviscid flow with a prescribed logarithmic profile.   The waves are monochromatic and linear.    The waves include a correction for laminar viscosity.  \mbox{\cite{miles1957}} only considers the portion of the atmospheric pressure that is in phase with the wave slope. \mbox{\cite{miles1993}} extends his original quasi-laminar theory to include wave-induced perturbations to the Reynolds stresses.   For certain input parameters, his new critical-layer theory agrees well with the observations that \mbox{\cite{plant1982}} collects for inverse wave ages between $.05 < u_*/c < 1$, where $u_*$ is the friction velocity in the air and c is phase speed.   Neither \mbox{\cite{miles1957}}  nor \mbox{\cite{miles1993}} consider the work that the wind drift may do on the waves.   As \mbox{\cite{miles1993}} notes, the validity of critical-layer theory is questionable for $u_*/c > 0.2$ when flow separation occurs over the lee side of the waves. 

\mbox{\cite{zhou93}} studies the effect of the wind drift on the growth of the waves.  The analysis is based on a control-volume approach and linear wave theory.   \mbox{\cite{zhou93}} generally find that the effect of the wind drift is to increase wave growth as long as the drift is below a critical value.   In the extreme limit, when the surface drift is equal to the phase speed, there is no wave growth.   As the authors note, due to approximations in the approach, the results are qualitative. 

\mbox{\cite{belcher1993}}, \mbox{\cite{belcher1998}}, and \mbox{\cite{cohen1999}} consider the interaction of wind with waves in terms of three parameter regimes corresponding to slow, intermediate, and fast waves.   For slow waves with $u_*/c > 1$, the wind speed is high relative to the waves and the wind separates in the lee of the wave.  The critical layer is so close to the free surface for slow waves that it does not have a strong dynamical effect on the growth \mbox{\citep{belcher1993}}.  For intermediates waves, there is complex interplay between sheltering effects and critical layers that has not received much study \mbox{\citep{belcher1998}}.   For very fast waves, the critical layer is so high or even nonexistent for waves moving faster than the wind that the effect of the critical layer is small in comparison to the turbulent stresses \mbox{\citep{cohen1999}}.   

As  \mbox{\citep{cohen1999}} note, theories generally agree better with the results of wind over paddle-generated wave in the laboratory, but for pure wind-generated waves, theoretical predictions are about a factor of two less than measurements.   We conjecture that for wind-generated waves neglecting the work done by the wind drift, \ie the turbulence in the water, may explain the factor of two.  In our two-phase approximation, the pressure is continuous across the free-surface interface such that the effects of turbulent pressure fluctuations in the water and the air are considered simultaneously.     Our numerical results for short-crested seas suggest that  wind-wave growth is very unsteady especially for inverse wave ages  $u_*/c > 0.2$ where there is a lot of wave breaking.   Like \mbox{\citep{cohen1999}},  there are inverse wave ages where waves decay under the action of wind.  The wave growth is positive for $u_*/c > 0.2$ in our numerical studies but only in a time-averaged sense.   Our unsteady predictions of wave growth are within the upper and lower bounds of \mbox{\possessivecite{plant1982}} data.

\mbox{\cite{miles1957}}, \mbox{\cite{miles1993}}, \mbox{\cite{belcher1993}},  \mbox{\cite{belcher1998}}, and \mbox{\cite{cohen1999}} are all steady-state theories for wind-wave growth that do not consider unsteady effects.   For turbulent flow over the short-crested seas,  it seems very unlikely that the turbulent pressure fluctuations in the wind would act in such a manner that the wave growth would always be steady, let alone positive, for all waves throughout wavenumber space.  The theories do not account for the effect of organized structures such as wind streaks on the growth of waves.    \mbox{\cite{miles1957}}, \mbox{\cite{miles1993}}, \mbox{\cite{belcher1993}},  \mbox{\cite{belcher1998}}, and \mbox{\cite{cohen1999}} also do not consider the effect of the turbulent fluctuations in the water on the growth of waves. Yet, for example,  \mbox{\cite{longuet1992}} shows that vorticity in the crest of steep gravity waves forms capillary rollers and bores, and  \mbox{\cite{duncan1983}} shows that the wakes of breaking waves reduce the amplitudes of the waves that follow.  \mbox{\cite{melville1998}} show that waves initially form when the subsurface shear layer goes unstable.   For waves in equilibrium with the wind, the turbulent stresses in the air and in the water are equal across the free-surface interface, so considering the work done on the waves by the turbulence in the wind-drift layer is reasonable to consider for growth.    

\mbox{\cite{snyder1981}} use an array of wave sensors, an array of air-pressure sensors, and one wave-following pressure sensor to study wave growth under the action of wind. \mbox{\cite{plant1982}} gathers data from wind-wave tanks and ocean experiments to show the dependence of wave growth on wave age.    \mbox{\cite{plant1982}} discusses  pros and cons for various approaches for calculating wave growth-rates.   There is a gap in the data for inverse wave ages between $0.2 \leq c/u_* \leq 0.8$ where the current results show that the  growth rate is very unsteady.    \mbox{\cite{donelan2006}} use a wave-following pressure sensor and free-surface measurements to show that wave growth depends on wave steepness and  flow separation.  \mbox{\cite{savelyev2011}} use a wave-following pressure sensor to quantify the effects of wave slope on wave growth in high wind conditions over paddle-generated waves.   \mbox{\cite{grare2013}} measure tangential stresses and form drag in the laboratory.  \mbox{\cite{grare2013}} measurements suggest that the turbulent fluctuations in the water are important.  Generally, some form of ensemble averaging is used in the preceding experiments to predict mean growth rates, so unsteady effects are not considered.

\mbox{\cite{sullivan2000}} simulate turbulent flow over idealized waves.   \mbox{\cite{kihara2007}} use Direct Numerical Simulations (DNS) to compare critical-layer theory and non-separated sheltering.   \mbox{\cite{sullivan2000}} and  \mbox{\cite{kihara2007}} only consider the effect of monochromatic waves on the flow in the air.   The flow in the water is not modeled.  The results of \mbox{\cite{sullivan2000}} and  \mbox{\cite{kihara2007}} are generally lower than the experiments reported in \mbox{\cite{plant1982}}.   \mbox{\cite{lin2008}} use DNS of a coupled air-water model to study growth of short waves.  The flow is laminar and the waves are linear.  Their growth rates are slightly higher than the initial growth rates of \mbox{\cite{phillips1957}} and the collection of observations by \mbox{\cite{plant1982}}.  Turbulence in the water had minimal effect on their results.   

\mbox{\cite{yang2010}} simulate Couette flow over waves using DNS for different wave steepnesses and wave ages.   The waves are monochromatic.  \mbox{\cite{yang2010}}  do not model the flow in the water.  \mbox{\cite{yang2010}} identify coherent vortices in the turbulent boundary layer, including stream-wise vortices and horseshoe vortices.   \mbox{\cite{yang2010}}  predict negative form drags for faster waves.

\mbox{\cite{yang2013}} consider the growth of short-crested waves in their LES studies.    \mbox{\cite{yang2013}} calculate wave growth in terms of the wave drag and the linear phase speed.     \mbox{\cite{yang2013}}  use ensemble averaging across the wind to calculate steady stream-wise drag coefficients.   They do not model the effects of turbulence in the water on the growth of waves.  \mbox{\possessivecite{yang2013}} predictions of growth rates are lower than the data of \mbox{\cite{plant1982}}.   

We calculate unsteady growth rates directly with no linear approximations in terms of the atmospheric pressure and the water-particle velocity normal to the free surface.  We note that the pressure is continuous across the air-water interface in our two-phase formulation.  In addition, by integrating the cross-spectral density of the pressure with the normal velocity in the angular direction with no ensemble averaging across the wind, we account for the effects of directionality in our numerical simulations.  Section \S \ref{sec:growth} of this paper shows predictions for wave growth that are calculated based on the preceding approach.

\section{Formulation}

The HOS formulation that is assimilated into NFA to drive the wavy portion of the flow is described in \S\ref{sec:bvp}.  The HOS algorithm that is used in this paper is a slightly modified version of the algorithm that is described in \mbox{\cite{dommermuth87}} and \mbox{\cite{dommermuth87corr}}.  

The NFA formulation is described in \S\ref{sec:vof}.   Additional details of the NFA formulation and validation studies are provided in  \mbox{\cite{sussman2000}}, \mbox{\cite{dommermuth04}}, \mbox{\cite{dommermuth06}}, \mbox{\cite{dommermuth07}}, \mbox{\cite{wyatt08}}, \mbox{\cite{oshea08}}, \mbox{\cite{fu08}}, \mbox{\cite{nfa6}}, \mbox{\cite{nfa1}}, \mbox{\cite{nfa4}}, \mbox{\cite{brandt12}}, and \mbox{\cite{ikeda12}}.  

 \mbox{\cite{nfa1}} use NFA to analyze the balance of energy in their studies of plunging breaking waves.  \mbox{\cite{nfareview1}} (\href{http://youtu.be/F1OAy8hz5gw}{Breaking-wave video I}) shows a perspective view of the breaking waves, and \mbox{\cite{nfareview2}} (\href{http://youtu.be/mq3Pl9UH7Go}{Breaking-wave video II}) shows the analysis of various terms in the energy balance.  \mbox{\possessivecite{nfa4}} NFA simulations of the fissioning of an envelope soliton agree very well with experimental measurements, which illustrates the capability of simulating complex wave interactions that take place over very long periods of time with minimal numerical dissipation.    The ability of NFA to model spray is evident in \mbox{\possessivecite{Fu12}} studies of high-speed planing boats that show excellent agreement between predictions and experimental measurements of the structure of the spray root and spray sheet that is generated during planing.  NFA predictions of the flow behind a transom stern agree well with experimental measurements of the turbulent roughening of the free surface and the entrainment of air  \mbox{\citep{nfa5}}. \mbox{\cite{nfareview3}} (\href{http://youtu.be/cenIBqda8zA}{Transom-stern video}) shows the transom-stern flow with a comparison to experiments.

The Pierson-Moskowitz and JONSWAP spectra that are used to model the wavy portion of the flow are provided in \S\ref{sec:seaway}.  The profiles of the wind and the wind-drift currents that are used to characterize the vortical portion of the flow are provided in \S\ref{sec:profile}.   Finally, the assimilation of the wavy and vortical portions of the flow for wind-driven breaking ocean waves is described in \S\ref{sec:assimilate}.  Note that \mbox{\cite{oceans13}} provide a formulation for assimilating just the wavy portion of the flow.

\subsection{\label{sec:bvp}HOS formulation}

Laplace's equation is satisfied within the fluid:
\begin{eqnarray}
\label{laplace}
\phi_{xx}+\phi_{yy}+\phi_{zz}=0 \;\; {\rm for} \;\; -d \leq z \leq \eta \;\; ,
\end{eqnarray}
\noindent where $\phi({\bf x},z,t)$ is the velocity potential, $\eta({\bf x},t)$ is the free-surface elevation, and $d$ is the depth. ${\bf x}=(x,y)$ is a vector in the horizontal plane, and $t$ denotes time.  We assume that $\eta$ is continuous and single valued.

Following \mbox{\cite{Zakaharov}}, $\phi^s$ is the potential evaluated on the free surface:
\begin{eqnarray}
\phi^s({\bf x},t)=\left. \phi({\bf x},z,t) \right|_{z=\eta} \;\; .
\end{eqnarray}
The temporal and spatial derivatives of $\phi^s$ in terms of $\phi$ are 
\begin{eqnarray}
\frac{\partial \phi^s}{\partial t} & = & \left. \frac{\partial \phi}{\partial t} \right|_{z=\eta}+ \frac{\partial \eta}{\partial t} \left. \frac{\partial \phi}{\partial z}  \right|_{z=\eta}  \nonumber \\
\nabla_x \phi^s & = & \left. \nabla_x \phi \right|_{z=\eta} + \nabla_x \eta \; \left. \frac{\partial \phi}{\partial z} \right|_{z=\eta}  \;\; ,
\end{eqnarray}
where $\nabla_x$ denotes the horizontal gradient:
\begin{eqnarray}
\nabla_x \equiv \left( \frac{\partial}{\partial x}, \frac{\partial}{\partial y} \right) \;\; .
\end{eqnarray}

Length and velocity scales are respectively normalized by $L_o$ and $U_o$.   Based on this normalization, the dynamic and kinematic free-surface boundary conditions with weak viscous effects are
\begin{eqnarray}
\label{dfsbc}
\frac{\partial \phi^s}{\partial t} & = & - \frac{1}{F_r^2} \eta -\frac{1}{2} \nabla_x \phi^s \cdot \nabla_x \phi^s \nonumber \\
& + & \frac{1}{2} \left(1+\nabla_x \eta \cdot \nabla_x \eta \right) \frac{\partial \phi}{\partial z}^2 \nonumber \\
& + & \frac{2}{Re} (\frac{\partial^2 \phi}{\partial x^2} + \frac{\partial^2 \phi}{\partial y^2} )  \;\; {\rm on \; z=\eta}  \\
\label{kfsbc}
\frac{\partial \eta}{\partial t} & = & -\nabla_x \phi^s \cdot \nabla_x \eta \nonumber \\
& + &  ( 1+ \nabla_x \eta \cdot \nabla_x \eta )  \frac{\partial \phi}{\partial z} \nonumber \\ 
& + & \frac{2}{Re} (\frac{\partial^2 \eta}{\partial x^2} + \frac{\partial^2 \eta}{\partial y^2} ) \;\; {\rm on \; z=\eta} \;\; ,
\end{eqnarray}
\noindent where $F_r=U_o/\sqrt{g L_o}$ and $R_e=U_o L_o / \nu$ are respectively the Froude and Reynolds numbers.  $g$ is the acceleration of gravity, 
$\nu$ is the kinematic viscosity, and $\rho$ is the density of water. 

Following \mbox{\cite{dommermuth87}}, we assume that $\phi$ and $\eta$ are O($\epsilon$) quantities, where $\epsilon$, a small parameter, is the wave steepness.   The potential is expanded in a perturbation series up to order $M$ in $\epsilon$:
\begin{eqnarray}
\phi({\bf x},z,t) = \sum_{m=1}^{M} \phi^{(m)}({\bf x},z,t)  \;\; .
\end{eqnarray}
Each perturbation potential is further expanded in a Taylor series about $z=0$ to evaluate the surface potential:
\begin{eqnarray}
\label{perturb}
\lefteqn{\phi^s({\bf x},t) = \phi({\bf x},\eta,t)} & & \nonumber \\
 & & = \sum_{m=1}^{M} \sum_{k=0}^{M-m} \frac{\eta^k} {k!} \frac{\partial^k}{\partial z^k} \phi^{(m)}({\bf x},0,t) \;\; .
\end{eqnarray}
At any instant of time, $\phi^s$ and $\eta$ are known, so that (\ref{perturb}) provides a Dirichlet condition for the unknown $\phi^{(m)}$.   By collecting terms at each order, a sequence of boundary-value problems follows:
\begin{eqnarray}
\phi^{(m)}({\bf x},0,t) & = & R^{(m)} \; , \;\;  {\rm m=1,2,3, \ldots, M} \nonumber \\
R^{(1)} & = & \phi^s({\bf x},t) \nonumber \\
R^{(m)} & = & -\sum_{k=1}^{m-1} \frac{\eta^k} {k!} \frac{\partial^k}{\partial z^k} \phi^{(m-k)}({\bf x},0,t) \; , \nonumber \\
& & {\rm m=2,3, \ldots, M}
\end{eqnarray}
To solve the boundary-value problems, each $\phi^{(m)}$ is expanded in terms of a finite number ($N$) of  eigenfunctions ($\Psi_n$):
\begin{eqnarray}
\phi^{(m)}({\bf x},z,t) = \sum_{n=1} ^N \phi_n^{(m)}(t) \Psi_n({\bf x},z) \; , \; z \leq 0 \;\; .
\end{eqnarray}
The vertical velocity evaluated on $z=\eta$ is
\begin{eqnarray}
\label{zvelo}
\lefteqn{\phi_z({\bf x},\eta,t)} & & \nonumber \\
 & & = \sum_{m=1}^{M} \sum_{k=0}^{M-m} \frac{\eta^k} {k!}  \sum_{n=1} ^N \phi_n^{(m)}(t)  \frac{\partial^{k+1}}{\partial z^{k+1}} \Psi({\bf x},0)  \;\; . \nonumber \\
 & & 
\end{eqnarray}
For constant finite depth, 
\begin{eqnarray}
\label{eigen}
\Psi_n ({\bf x},t) = \frac{\cosh\left[ | {\bf k}_n| (z+d) \right]}{\cosh( |{\bf k}_n| d )} \exp(\imath {\bf k}_n \cdot {\bf x}) \;\; ,
\end{eqnarray}
\noindent where ${\bf k}_n=(k_x, k_y)$.    Equations \ref{eigen} and \ref{zvelo} are substituted into the free-surface boundary conditions \ref{dfsbc} and \ref{kfsbc}, and the resulting equations are solved using a pseudo-spectral method.   Details are provided in \mbox{\cite{dommermuth87}}.   Other examples of HOS formulations with applications are provided by \mbox{\cite{wu04}} and \mbox{\cite{blondel08}}.

\subsubsection{HOS smoothing} Smoothing is required in HOS simulations of broad-banded wave spectra.   Smoothing prevents the pileup of energy at high wavenumbers.     Filtering in wavenumber space is used as follows:
\begin{eqnarray}
\label{smooth_spectral}
F_{\rm ij} (F_c) =
\left\{
\begin{array}{lr}
1, & (\frac{k_{\rm i}}{F_c k_{\rm N_x}})^2+(\frac{k_{\rm j}}{F_c k_{\rm N_y}})^2 \le 1\\
   & \\
0, & (\frac{k_{\rm i}}{F_c k_{\rm N_x}})^2+(\frac{k_{\rm j}}{F_c k_{\rm N_y}})^2 > 1\\
\end{array} \;\; ,
\right.
\end{eqnarray}
\noindent where $k_{\rm i}$ and $k_{\rm j}$ are respectively the wavenumbers along the $x$ and $y-$axes, $k_{\rm N_x}$ and $k_{\rm N_y}$ are the corresponding Nyquist wavenumbers, and $0 < F_c \leq 1$ is the Fourier cut-off parameter.  Equation \ref{smooth_spectral} is applied to $\phi^s$ and $\eta$ every time step.

\subsubsection{HOS energy pumping}  As a result of smoothing, energy is not conserved in HOS simulations of free-surface waves.     The total energy as a function of time ($E(t)$) is
\begin{eqnarray}
\label{pump1}
E(t)=\int_{S_o} ds \, \phi^s \eta_t + \frac{1}{F_r^2} \int_{S_o} ds \, \eta^2 \;\; , 
\end{eqnarray}
\noindent where the first term on the right-hand side is the kinetic energy and the second term is the potential energy.  $S_o$ is the horizontal plane.  Due to the effects of smoothing, the total energy will decrease over time.  The total energy is conserved in the HOS simulations by rescaling the free-surface elevation and the surface potential at the end of every time step to generate new quantities.
\begin{eqnarray}
\label{pump2}
\eta^{\rm (NEW)} & = & S(t) \eta \nonumber \\
(\phi^s)^{\rm (NEW)} & = & S(t) \phi^s \;\; ,
\end{eqnarray}
\noindent where $S(t)$ is a scaling factor equal to the square root of the ratio of the current total energy to the initial total energy:
\begin{eqnarray}
\label{pump3}
S(t) =\left( \frac{E(t)}{E(0)} \right)^\frac{1}{2} \;\; .
\end{eqnarray}
As a result, energy is {\it pumped} into the free-surface waves.   Pumping nonlinear simulations of ocean waves  can be used to establish a $k^{-3}$ wavenumber dependence in wave spectra through the action of nonlinear wave interactions.    For example, pumped HOS simulations with third or higher order will fill in low-passed realizations of short-crested seas with a $k^{-3}$ power-law behavior corresponding to a saturated spectrum.  However, over long periods of time, energy will tend to pileup at high wavenumbers without an energy drain.  \mbox{\cite{nfa4}} provide additional details of energy pumping.

\subsubsection{HOS free-surface adjustment} Numerical simulations of nonlinear progressive waves are prone to developing spurious high-frequency standing waves unless the flow field is given sufficient time to adjust \mbox{\citep{dommermuth00}}.    An adjustment scheme allows the natural  development of nonlinear self-wave (locked modes) and inter-wave (free modes) interactions.  Linear Airy waves are adjusted to generate nonlinear waves in HOS simulations.   The nonlinear terms in HOS are isolated and slowly activated. 

We assign an adjustment factor $A(t)$ that slowly turns on nonlinearity as a function of time:
\begin{eqnarray}
\label{adjust1}
A(t) =\left\{ 
\begin{array}{ll}
\frac{1}{2}\left(1-\cos\left(\frac{2 \pi t}{T_a}\right)\right) &  {\rm for} \; t<T_a  \\  
0  & {\rm for} \; t \geq T_a
\end{array}
\right. \;\; ,
\end{eqnarray}
where $T_a$ is the adjustment time.

The following procedure is used to adjust the free-surface boundary conditions in HOS:
\begin{eqnarray}
\label{adjust2}
\frac{\partial \phi^s}{\partial t} & = & - \frac{1}{F_r^2} \eta +  \frac{2}{Re} (\frac{\partial^2 \phi}{\partial x^2}+ \frac{\partial^2 \phi}{\partial y^2} ) \nonumber \\
& + & [ \frac{1}{2} \left(1+\nabla_x \eta \cdot \nabla_x \eta \right) \frac{\partial \phi}{\partial z}^2  \nonumber \\
& - &  \frac{1}{2} \nabla_x \phi^s \cdot \nabla_x \phi^s ] A(t) \;\; {\rm on \; z= \eta} \nonumber \\
\frac{\partial \eta}{\partial t} & = &   W^{(1)}  + \frac{2}{Re} (\frac{\partial^2 \eta}{\partial x^2} + \frac{\partial^2 \eta}{\partial y^2} ) \nonumber \\
& + & [  (1+ \nabla_x \eta \cdot \nabla_x \eta )  \frac{\partial \phi}{\partial z}- W^{(1)}  \nonumber \\
& - &    \nabla_x \phi^s \cdot \nabla_x \eta ] A(t) \nonumber \\ & &  {\rm on \; z= \eta} \;\; ,
\end{eqnarray}
\noindent where $W^{(1)}$ is the leading-order component of the vertical velocity evaluated on the plane $z=0$:
\begin{eqnarray}
\label{adjust3}
W^{(1)}= \frac{\partial \phi}{\partial z}^{(1)}|_{z=0} \;\; .
\end{eqnarray}

\subsubsection{HOS time integration}  The linear terms are integrated analytically using an unrolling procedure.   First, the nonlinear terms are isolated as follows:
\begin{eqnarray}
\label{unroll1}
\frac{\partial \phi}{\partial t} & = & -g \eta + G(\eta,\phi) \nonumber \\
\frac{\partial \eta}{\partial t} & =  & \frac{\partial \phi}{\partial z} + F(\eta,\phi)  \;\; ,
\end{eqnarray}
\noindent where $F$ and $G$ are the nonlinear terms in the dynamic (\ref{dfsbc}) and kinematic (\ref{kfsbc}) free-surface boundary conditions, respectively.  For the sake of clarity, we have temporarily dropped our non-dimensional notation.

The Fourier decompositions of the free-surface elevation and the potential for a single mode are denoted as follows:
\begin{eqnarray}
\label{unroll2}
\phi & = & \hat{\phi}(t) \frac{\cosh(k ( z+d))}{\cosh(k d)} \exp(\imath {\bf k} \cdot {\bf x})+{\rm c.c.} \nonumber \\
\eta & = & \hat{\eta}(t) \exp(\imath {\bf k} \cdot {\bf x}) + {\rm c.c.}  \;\; ,
\end{eqnarray}
\noindent where the hat symbol denotes wavenumber space, and $k$ is the magnitude of the vector wavenumber $\bf k$.

Substitution of the preceding Equations \ref{unroll2} into \ref{unroll1} and taking the Fourier transform gives
\begin{eqnarray}
\label{unroll3}
\frac{\partial \hat{\phi}}{\partial t} & = & -g \hat{\eta} + \hat{G}(\eta,\phi) \nonumber \\
\frac{\partial \hat{\eta}}{\partial t} & =  &  k \tanh(k d) \hat{\phi} + \hat{F}(\eta,\phi)  \;\; .
\end{eqnarray}
The linear dispersion relation for this equation is
\begin{eqnarray}
\label{unroll4}
\omega^2 = k g \tanh( k d ) \;\; .
\end{eqnarray}

By definition, we let
\begin{eqnarray}
\label{unroll5}
\hat{\phi} & = & \frac{\imath g \hat{A}^+}{\omega} \exp(\imath \omega t)
                          -\frac{\imath g \hat{A}^-}{\omega} \exp(-\imath \omega t)  \nonumber \\
\hat{\eta} & = & \hat{A}^+ \exp(\imath \omega t)
                         +\hat{A}^- \exp(-\imath \omega t) \;\; ,
\end{eqnarray}
\noindent where $\hat{A}^+$ and $\hat{A}^-$ are the complex amplitudes associated with the positive and negative frequencies of the free-surface elevations, respectively.

Substitution of \ref{unroll5} into \ref{unroll3} gives the following evolution equations for the complex amplitudes:
\begin{eqnarray}
\label{unroll6}
\frac{\partial \hat{A}^+}{\partial t} & = & 
\left( \hat{F}(\eta,\phi) - \frac{\imath \omega}{g} \hat{G}(\eta,\phi) \right) \frac{\exp(-\imath \omega t)}{2} \nonumber \\
\frac{\partial \hat{A}^-}{\partial t} & = & 
\left( \hat{F}(\eta,\phi) + \frac{\imath \omega}{g} \hat{G}(\eta,\phi) \right) \frac{\exp(\imath \omega t)}{2} \;\; . \nonumber \\
\end{eqnarray}
These equations are integrated in time using a 4th-order Runge-Kutta scheme for each Fourier mode.

\subsection{\label{sec:vof}NFA formulation}

Consider the  immiscible turbulent flow at the interface between air and water with $\rho_a$ and $\rho_w$ respectively denoting the densities of air and water.   Similar to the potential-flow approaches, physical quantities are normalized by characteristic velocity ($U_o$), length ($L_o$),  time ($L_o/U_o$), density ($\rho_w$), and pressure ($\rho_w U_o^2$) scales.  

Let $\alpha$ denote the fraction of fluid that is inside a cell. By definition, $\alpha=0$ for a cell that is totally filled with air, and $\alpha=1$ for a cell that is totally filled with water.   In terms of $\alpha$, the normalized density is expressed as
\begin{eqnarray}
\label{alpha}
\rho(\alpha) & = & \lambda + (1 - \lambda) \alpha \;\; ,
\end{eqnarray}
\noindent where $\lambda = \rho_a/\rho_w$ is the density ratio between air and water.  

Let $u_i$ denote the normalized three-dimensional velocity field as a function of normalized space ($x_i$) and normalized time ($t$).  The conservation of mass is
\begin{eqnarray}
\label{mass}
\frac{\partial \rho}{\partial t} +\frac{\partial u_j \rho}{\partial x_j} = 0 \;\; .
\end{eqnarray}

\noindent For incompressible flow,
\begin{eqnarray}
\label{density}
\frac{\partial \rho}{\partial t} +u_j \frac{\partial  \rho}{\partial x_j} = 0 \;\; .
\end{eqnarray}

\noindent Subtracting Equation (\ref{density}) from (\ref{mass}) gives a solenoidal condition for the velocity:
\begin{eqnarray}
\label{solenoidal}
\frac{\partial u_i}{\partial x_i} = 0 \;\; .
\end{eqnarray}

\noindent Substituting Equation (\ref{alpha}) into (\ref{mass}) and making use of (\ref{solenoidal}), provides an advection equation for  the volume fraction:
\begin{eqnarray}
\label{vof}
\frac{\partial \alpha}{\partial t}+ \frac{\partial}{\partial x_j} \left(u_j \alpha \right)= 0 \;\; .
\end{eqnarray}

For an infinite Reynolds number, viscous stresses are negligible, and the conservation of momentum is
\begin{equation}
\label{momentum}
\frac{\partial u_i}{\partial t}+\frac{\partial}{\partial x_j} \left(u_j u_i \right)  =  -\frac{1}{\rho} \frac{\partial p}{\partial x_i} -\frac{p_s}{\rho} \frac{\partial H(\alpha)}{\partial x_i} - \frac{\delta_{i3}}{F_r^2}  \; ,
\end{equation}
\noindent where $F_r^2 = U_o^2/(g L_o)$ is the Froude number, and $g$ is the acceleration of gravity.  $p$ is the pressure and $p_s$ is a stress that acts normal to the interface.  $H(\alpha)$ is a Heaviside function, and $\delta_{ij}$ is the Kronecker delta function.   

The divergence of the momentum equations (\ref{momentum}) in combination with the solenoidal condition (\ref{solenoidal}) provides a Poisson equation for the dynamic pressure:
\begin{eqnarray}
\label{poisson} 
\frac{\partial}{\partial x_i} \frac{1}{\rho} \frac{\partial
p}{\partial x_i} = \Sigma \;\; ,
\end{eqnarray}
\noindent where $\Sigma$ is a source term.  The pressure is used to project the velocity onto a solenoidal field.  Details of the volume fraction advection, the pressure projection, and the numerical time integration are provided in \mbox{\cite{dommermuth07}} and \mbox{\cite{oshea08}}. Sub-grid scale stresses are modeled using an implicit model that is built into the treatment of convective terms.  The performance of the implicit SGS model is provided in \mbox{\cite{nfa3}}. 

\subsubsection{\label{sec:smooth_vof}NFA smoothing}  The free-surface boundary layer is not resolved in VOF simulations at high Reynolds numbers with large density jumps such as air and water.  Under these circumstances, the tangential velocity is discontinuous across the free-surface interface and the normal component is continuous.   As a result, unphysical tearing of the free surface tends to occur.   Favre-like filtering can be used to alleviate this problem by forcing the air velocity at the interface to be driven by the water velocity in a physical manner.    Consider the following projection,
\begin{eqnarray}
\label{smooth_velo}
\tilde{u}_i=
\left\{
\begin{array}{cl}
\left< \rho u_i  \right>/\left< \rho \right> & {\rm for} \; \alpha \le 0.5 \\
u_i  & {\rm for} \; \alpha > 0.5
\end{array}
\right. \;\; ,
\end{eqnarray}
where $\tilde{u}_i$ is the smoothed velocity field, $u_i$ is the unfiltered velocity field, $\rho$ is the density, and $\alpha$ is the volume fraction.  Brackets denote smoothing.
\begin{eqnarray}
\left< F({\bf x}) \right> = \int_{\rm v_\xi} W({\bf x_\xi}) F({\bf x}-{\bf x_\xi}) {\rm d v_\xi}  \;\; .
\end{eqnarray}
Here, $F(x)$ is a general function, ${\rm v}$ is a control volume that surrounds a cell, and $W(x)$ is a weighting function that neither overshoots or undershoots the maximum or minimum allowable density.  Due to the  high density ratio between water and air, equation~\ref{smooth_velo} tends to push the water-particle velocity into the air.   Once the velocity is filtered, we need to project it back onto a solenoidal field in the fluid volume (V).
\begin{eqnarray}
\label{project1}
u_i = \tilde{u}_i - \frac{1}{\rho} \frac{\partial \psi}{\partial x_i}  \;\; {\rm in \; V} \; ,
\end{eqnarray}
where $\psi$ is a potential function.     For an incompressible flow, we require that $u_i$ is solenoidal care of Equation~({\ref{solenoidal}).  Substituting (\ref{project1}) into (\ref{solenoidal}) gives a Poisson equation for $\psi$:
\begin{eqnarray}
\label{project2}
\frac{\partial}{\partial x_i} \frac{1}{\rho} \frac{\partial \psi}{\partial x_i} = \frac{\partial \tilde{u}_i}{\partial x_i}   \;\; {\rm in \; V} \; .
\end{eqnarray}
\noindent We typically apply the filtering every 20 to 80 time steps.  Details of the implementation of the preceding filter are provided in \mbox{\cite{nfa6}}.

\subsubsection{NFA time integration}  Based on \mbox{\cite{sussman03a}}, a second-order Runge-Kutta scheme is used to integrate with respect to time the field equations for the velocity field.  Here, we illustrate how a volume of fluid formulation is used to advance the volume-fraction function.  Similar examples are provided by \mbox{\cite{rider94}}.  During the first stage of the Runge-Kutta algorithm, a Poisson equation for the pressure
is solved:
\begin{eqnarray}
\label{eqn:poisson1}
\frac{\partial}{\partial x_i} \frac{1}{\rho(\alpha^k)} \frac{\partial P^*}{\partial x_i} =\frac{\partial}{\partial x_i} \left(
\frac{u^k_i}{\Delta t}+R_i \right) \;\; ,
\end{eqnarray}
where $R_i$ denotes the nonlinear convective, hydrostatic,  and atmospheric forcing terms in the momentum equations.
$u^k_i$ and $\rho^k$ are respectively the velocity components at time step $k$.  $\Delta t$ is the time step.  $P^*$ is the first prediction for the pressure field.

For the next step, this pressure is used to project the velocity onto a solenoidal field. The first prediction for the
velocity field ($u^*_i$) is
\begin{eqnarray}
\label{eqn:runge1}
u^*_i=u^k_i+\Delta t \left( R_i-\frac{1}{\rho(\alpha^k)}\frac{\partial P^*}{\partial x_i} \right) \;\; .
\end{eqnarray}
The volume fraction is advanced using a volume of fluid operator (VOF):
\begin{eqnarray}
\alpha^*=\alpha^{k}- {\rm VOF} \left( u^k_i,\alpha^k,\Delta t \right) \;\; .
\end{eqnarray}
A Poisson equation for the pressure is solved again during the second stage of the Runge-Kutta algorithm:
\begin{eqnarray}
\label{eqn:poisson2}
\frac{\partial}{\partial x_i} \frac{1}{\rho(\alpha^*)} \frac{\partial P^{k+1}}{\partial x_i}=\frac{\partial}{\partial x_i} \left(
\frac{u^*_i+u^k_i}{\Delta t}+R_i \right) \;\; .
\end{eqnarray}
$u_i$ is advanced to the next step to complete one cycle of the Runge-Kutta algorithm:
\begin{equation}
\label{eqn:runge2}
u^{k+1}_i=\frac{1}{2} \left( u^*_i + u^k_i +\Delta t \left( R_i -\frac{1}{\rho(\alpha^*)}\frac{\partial P^{k+1}}{\partial x_i}
\right) \right) \;\; , 
\end{equation}
and the volume fraction is advanced to complete the algorithm:
\begin{eqnarray}
\alpha^{k+1}=\alpha^k- {\rm VOF} \left( \frac{u^*_i+u^k_i}{2},\alpha^{k},\Delta t \right) \;\; .
\end{eqnarray}

For very large-scale simulations, with $\Delta t \ll 1$, the $\Delta t$ terms in equations \ref{eqn:poisson1} and \ref{eqn:poisson2} are eliminated to improve the numerical conditioning of the Poisson solver when there are divergence errors in the velocity.   Instead, the following projection operator is periodically applied to ensure that the velocity field is divergence free:
\begin{eqnarray}
\label{project_delt_1}
u_i=u_i^k-\frac{1}{\rho(\alpha^k)}\frac{\partial \psi}{\partial x_i} \;\; {\rm in \;\; V} \;\; ,
\end{eqnarray}
where $\psi$ is a potential function.   The preceding projection operator is similar to the smoothing algorithm in \S\ref{sec:smooth_vof}.   The resulting Poisson equation is
\begin{eqnarray}
\label{project_delt_2}
\frac{\partial}{\partial x_i} \frac{1}{\rho(\alpha^k)} \frac{\partial \psi}{\partial x_i} = \frac{\partial u^k_i}{\partial x_i}   \;\; {\rm in \; V} \; .
\end{eqnarray}
\noindent We typically apply the projection every 10 time steps whenever the smoothing algorithm is not applied.

In NFA, the free surface is reconstructed from the volume fractions using piece-wise linear polynomials.  The reconstruction is based on algorithms that are described by \mbox{\cite{gueyffier99}}. The surface normals are estimated using weighted central differencing of the volume fractions. A similar algorithm is described by \mbox{\cite{pilliod97}}.     The advection portion of the algorithm is operator split, and it is based on similar algorithms reported in \mbox{\cite{puckett97}}.   One difference between the present algorithm and earlier methods includes a special treatment to alleviate mass-conservation errors due to the presence of non-solenoidal velocity fields.

Let $F_i$ denote the flux through the faces of a cell: 
\begin{eqnarray}
F_i = A_i u_i  \;\; .
\end{eqnarray}
\noindent $A_i$ correspond to the surface areas that bound the cell.   Based on an application of Gauss's theorem to the volume integral of equation \ref{mass} and making use of equation \ref{solenoidal}:
\begin{eqnarray}
\label{eqn:diverge}
F^+_i-F^-_i = 0 \;\; ,
\end{eqnarray}
\noindent where $F^+_i$ is the flux on the positive i-th face of the cell and $F^-_i$ is the flux on the negative i-th face of the cell.   Due to numerical errors, equation \ref{eqn:diverge} is not necessarily satisfied.   Let $\cal E$ denote the resulting numerical error for any given cell.   For each cell whose flux is not conserved,  a correction is applied prior to performing the VOF advection.   For example,  the following reassignment of the flux along the vertical direction ensures that the redefined flux is conserved:
\begin{eqnarray}
\tilde{F}^+_3 & = & F^+_3 - \frac{{\cal E}}{2} \nonumber \\
\tilde{F}^-_3 & = & F^-_3  + \frac{{\cal E}}{2}  \;\; .
\end{eqnarray}
Based on this new flux, new face velocities are defined on the faces of the cell: 
\begin{eqnarray}
\tilde{u}_i & = &  \frac{\tilde{F}_i}{A_i}  \;\; .
\end{eqnarray}

Equation \ref{vof} is operator split.   A dilation term is added to ensure that the volume fraction remains between $0 \leq \alpha \leq 1$ during each stage of the splitting \mbox{\citep{puckett97,wey2010}}.  The resulting discrete set of equations for the first stage of the Runge-Kutta time-stepping procedure is provided below:    
{\small
\begin{eqnarray}
\label{operatorsplit}
\alpha^{(1)}      & = &  \alpha^k    \nonumber \\
& - & \frac{{\cal F}_1\left[ \left({u}^+_1\right)^k,\alpha^k,\Delta t \right] 
       -          {\cal F}_1\left[ \left({u}^-_1\right)^k,\alpha^k,\Delta t \right]}{\rm V} \nonumber \\
& + &  \Delta t  \, c_c \frac{\left({u}^+_1\right)^k-\left({u}^-_1\right)^k}{\Delta x_1}   \nonumber \\
\alpha^{(2)}   & = & \alpha^{(1)}  \nonumber \\  
& - & \frac{{\cal F}_2 \left[ \left({u}^+_2\right)^k,\alpha^{(1)} ,\Delta t \right]
       -          {\cal F}_2 \left[ \left({u}^-_2\right)^k,\alpha^{(1)} ,\Delta t \right]}{\rm V} \nonumber \\
& + & \Delta t  \, c_c \frac{\left({u}^+_2\right)^k-\left({u}^-_2\right)^k}{\Delta x_2}   \nonumber \\
\alpha^* & = &  \alpha^{(2)}  \nonumber \\
& - & \frac{{\cal F}_3 \left[ \left(\tilde{u}^+_3 \right)^k,\alpha^{(2)} ,\Delta t \right] 
      -           {\cal F}_3 \left[ \left(\tilde{u}^-_3 \right)^k,\alpha^{(2)} ,\Delta t \right]}{\rm V} \nonumber \\
& + & \Delta t \, c_c  \frac{\left(\tilde{u}^+_3\right)^k-\left(\tilde{u}^-_3\right)^k}{\Delta x_3} \;\; ,
\end{eqnarray}
}
${\cal F}_i$ denotes VOF advection  along the cartesian axes, ${\rm V}=\Delta x_1 \Delta x_2 \Delta x_3$ is the volume of the cell,  and $c_c$ is a coefficient that is designed to conserve flux and prevent overfilling or under filling of a cell \mbox{\citep{wey2010}}:
\begin{eqnarray}
c_c=
\left\{
\begin{array}{ll}
1 & {\rm for} \;  0.5 \leq \alpha^k \leq 1 \\
0 & {\rm for} \; 0 < \alpha^k \leq 0.5
\end{array} \;\; .
\right. 
\end{eqnarray}
Note that the order of the operator splitting is alternated from time step to time step to preserve second-order accuracy and to prevent any biasing.

\subsection{\label{sec:seaway}Seaway representation}  

A JONSWAP spectrum is used to initialize the HOS simulations that are assimilated into NFA.   The wavelength at the peak of the spectrum ($L_o$) is used to normalize length scales.   The velocity scale is $U_o=\sqrt{g L_o}$.   Based on these choices for $L_o$ and $U_o$, the Froude number equals one ($F_r=1$).  In normalized variables, the one-dimensional JONSWAP spectrum in wavenumber space is 
\begin{eqnarray}
\label{jonswap1}
S_{\rm J}(k)=\frac{\alpha_p}{2 k^3} \exp\left( -\frac{5}{4} \left(\frac{k_o}{k} \right)^2  \right) \gamma^a \;\; ,
\end{eqnarray}
\noindent where $\gamma$ controls the height of the spectral density relative to a Pierson-Moskowitz spectrum and
\begin{eqnarray}
\label{jonswap2}
a=\exp\left( -\frac{1}{2} \left( \frac{\sqrt{k}-\sqrt{k_o}} {\sigma_a \sqrt{k_o}} \right)^2  \right)  \;\; .
\end{eqnarray}
For a fully-developed wind-generated spectrum, $\alpha_p = 0.0081$.  $\sigma_a$ controls the widths of the left and right sides of the spectrum as follows:
\begin{eqnarray}
\label{jonswap3}
\sigma_a=\left\{
\begin{array}{c}
0.07, \; k < k_o\\
0.09,  \; k \geq k_o 
\end{array}
\right. \;\; .
\end{eqnarray}
A cosine spreading function is used to simulate a directional spectrum.
\begin{eqnarray}
\label{spread1}
D(\theta)= \frac{1}{2 \sqrt{\pi}} \frac{\Gamma(s+1)}{\Gamma(s+\frac{1}{2})} \cos^{2 s}( \frac{1}{2} (\theta -\theta_o) )  \;\; ,
\end{eqnarray}
\noindent where $s$ controls the amount of spreading and $\theta_o$ is the primary direction of the waves.  $\Gamma$ is the Gamma function. 

\subsection{\label{sec:profile}Vertical profiles of mean wind and wind drift}  

Log similarity profiles are used to specify the profiles of the mean wind and the mean wind drift.   For the flow in the air, 
\begin{equation}
\label{meanwind}
u(\zeta) = u_o+u_{\rm air} + \frac{u_*}{\kappa} {\rm ln} \frac{\zeta}{z_{\rm air}} \;\; {\rm for} \;  \zeta \ge z_{\rm air} \; ,
\end{equation}
and in the water,
\begin{equation}
\label{meandrift}
u(\zeta) = u_o -u_{\rm h_2o} - \frac{w_*}{\kappa} {\rm ln} \frac{-\zeta}{z_{\rm h_2o}} \;\; {\rm for} \;  \zeta \le -z_{\rm h2o} \; .
\end{equation}
$\zeta$ is a free-surface-following coordinate system  (see equation \ref{eq:profile1}).   $u_o$ is the wind drift on the free surface where $\zeta=0$.   $u_*$ and $w_*$ are the friction velocities in the air and the water, respectively.  $\kappa=0.4$ is the von K\'{a}rm\'{a}n constant.   $z_{\rm air}$ and $z_{\rm h2o}$ are the corresponding roughness heights.  $u_o+u_{\rm air}$ is the lower bound of the logarithmic profile in the air, and $u_o -u_{\rm h_2o}$ is the upper bound of the logarithmic profile in the water.   

The turbulent stresses are equal across the free-surface interface such that
\begin{equation}
\label{turbstress}
\rho_a (u_*)^2 = \rho_w (w_*)^2 \; .
\end{equation}

Linear profiles are used for the mean wind and mean wind drift close to the free surface.   For the mean wind,
\begin{equation}
\label{meanwindclose}
u(\zeta) = u_o+u_{\rm air} \frac{\zeta}{z_{\rm air}} \;\; {\rm for} \;  0 < \zeta < z_{\rm air} \; ,
\end{equation}
and for the mean wind drift,
\begin{equation}
\label{meandriftclose}
u(\zeta) = u_o +u_{\rm h_2o} \frac{\zeta}{z_{\rm h_2o}} \;\; {\rm for} \;   -z_{\rm h2o} < \zeta \leq 0 \; .
\end{equation}

\subsection{\label{sec:assimilate}Data assimilation}

As illustrated in Figure \ref{assimilate}, data is sequentially assimilated into NFA every N time steps.  Both the wavy and vortical portions of the flow are assimilated into NFA.   A separation of the flow into wavy and vortical components is similar to the Helmholtz decomposition that is used by  \mbox{\cite{dommermuth93}} in his studies of the interaction of a vortex pair with a free surface.    In the NFA data assimilation, energy cascades down from the lowest wavenumbers to the highest wavenumbers through nonlinear wave interactions where it is dissipated due the effects of wave breaking, and forced by the wind and the wind drift.   Vertical profiles of the mean wind and the mean wind-drift are assimilated into NFA to drive the vortical portion of the flow.    The turbulent fluctuations form naturally through the energy cascade, the generation of free-surface vorticity, and the breaking of waves.   

\begin{figure}[ht]
\begin{center}
\includegraphics[width=\linewidth]{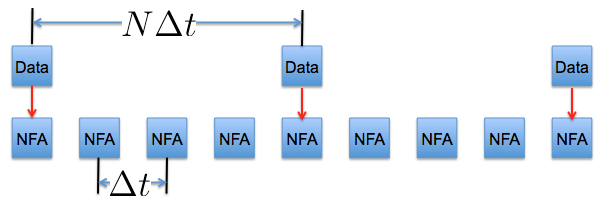}
\caption{\label{assimilate}  Sequential assimilation.}
\end{center}
\end{figure}

Here, we illustrate the assimilation of HOS simulations into NFA for the wavy portion of the flow (see \S\ref{sec:bvp}).   For the vortical portion of the flow, we assimilate analytical solutions based on similarity theory (see \S\ref{sec:profile}).  The procedure can be generalized to assimilate radar data into NFA for the wavy portion of the flow and velocimetry for the vortical portion of the flow.    

The HOS formulation is based on irrotational flow, \ie the wavy portion of the flow.   Depending on the type, radar measures the position of the ocean surface or the velocities on the ocean surface,  which can be directly related to the wavy portion of the flow.  HOS, like radar measurements, is bandwidth limited.   In the case of HOS, approximations, including the Taylor series approximation, the perturbation expansion, and the single-valued free surface, limit the relative difference of the maximum wavenumber to the wavenumber at the peak of spectrum to about two decades.   Radar measurements over a patch of the ocean surface have similar resolution limitations.   HOS simulations or radar measurements are assimilated into NFA to drive the lowest wave numbers in the NFA simulation.   The assimilation of the wavy flow is described in \S\ref{sec:wavy_flow}.

The vertical profiles of the mean vortical flow provide relevant statistics and are easier to measure and assimilate than the full three-dimensional, time-varying field.    First, the vortical portion of the flow is isolated from the wavy portion of the flow in the NFA simulation.   Then the profiles of the mean  vortical flow based on the NFA results are calculated and compared to data.   The differences between the numerical predictions and the data are used to nudge the numerical simulations toward the data.   The assimilation of the vortical flow is described in \S\ref{sec:vort_flow}.

\subsubsection{\label{sec:wavy_flow}Wavy portion of flow}

A properly posed free-surface problem requires the assimilation of two surface quantities for the wavy portion of the flow.    We assimilate the free-surface elevation and the normal component of velocity evaluated on the free surface.    As discussed earlier, these two quantities are assimilated differently depending on whether the wavenumber $k$ is above or below a cutoff wavenumber $k_c$.   The incremental changes in the free-surface elevation ($\Delta \eta$) and the normal velocity ($\Delta u_n$) are 
\begin{eqnarray}
\label{assimilate1}
\Delta \eta &  = & \Delta \eta^ L + \Delta \eta^H  \\
\label{assimilate2}
\Delta u_n & = &  \Delta u_n^L+\Delta u_n^H \;\; ,
\end{eqnarray}
\noindent where the superscript $L$ and $H$ symbols denote low pass and high pass filtering, respectively.   As shown in \S\ref{sec:low_wave}, incremental changes at low wave numbers are assimilated using a nudging technique.   As shown in \S\ref{sec:high_wave}, incremental changes at high wave numbers are estimated based on theoretical considerations of a saturated wave spectrum.

The change in the free-surface elevation is added to the old free-surface elevation to get the new position of the free surface:
\begin{eqnarray}
\label{assimilate3}
\eta^{\rm NEW}  & = & \eta^{\rm OLD} + \Delta \eta \;\; .
\end{eqnarray}
\noindent  As shown in \S\ref{sec:alpha}, the change in the free-surface elevation is also used to update the volume fraction:
\begin{eqnarray}
\label{assimilate5}
\alpha^{\rm NEW}  & = & U(\alpha^{\rm OLD}, \Delta \eta) \;\; ,
\end{eqnarray}
\noindent where $U$ is a function that preserves overturning waves, bubbles, and droplets that may be present in the volume fraction.   

The change in the normal component of velocity evaluated on the free surface is used to calculate new velocities throughout the entire domain: 
\begin{eqnarray}
\label{assimilate4}
u_i^{\rm NEW}  & = &  u_i^{\rm OLD} + {\cal P}(\Delta u_n) \;\; ,
\end{eqnarray}
\noindent where $\cal P$ is a projection operator that is described in \S\ref{sec:velocity}.   

\paragraph{\label{sec:transfer}Transfer functions.}  The volume fraction and water-particle velocities as calculated by NFA are not suitable for the assimilation of data.   As shown in \mbox{\cite{nfa5}}, height functions provided the best agreement between NFA predictions and experimental measurements of turbulent roughening of the free surface behind the transom stern of a model-scale ship.

We define the free surface in terms of height functions expressed in terms of the volume fraction.   The calculation is performed in three steps.   First, we define a height function  integrating the volume fraction from the bottom of the computational domain to the top and subtract out the water depth.  This provides an initial estimate of the free-surface elevation $\eta_1$:
\begin{equation}
\label{height1}
\eta_1(x,y,t)=\int_{-d}^h dz \alpha(x,y,z,t)-d \; ,
\end{equation}
where $d$ is the water depth, $h$ is the height of the air, and $\alpha$ is the volume fraction. Then the air pockets that are trapped beneath $\eta_1$ are added back to provide a water column without bubbles.
\begin{equation}
\label{height2}
\eta_2(x,y,t)=\eta_1(x,y,t)+\int_{-d}^{\eta_1(x,y,t)} dz (1-\alpha(x,y,z,t)) \; .
\end{equation}
Finally, the droplets above $\eta_2$ are subtracted out to provide a water column without bubbles and without droplets.
\begin{equation}
\label{height3}
\eta^{\rm NFA}(x,y,t)=\eta_2(x,y,t)-\int_{\eta_2(x,y,t)}^h dz \alpha(x,y,z,t) \; .
\end{equation}
We evaluate the surface water-particle velocities as predicted by NFA on $\eta^{\rm NFA}$.
\begin{equation}
\label{height4}
u^s_i(x,y,t)=\left. u_i \right|_{z=\eta^{\rm NFA}} \;\; .
\end{equation}
We define a unit normal $n^{\rm NFA}_i$ that points into the air in terms of the height function $\eta^{\rm NFA}$.  Then the normal component of velocity evaluated on the position of the free surface predicted by NFA is
\begin{equation}
\label{height5}
u_n^{\rm NFA}(x,y,t)= u^s_i n^{\rm NFA}_i \;\; .
\end{equation}

\paragraph{\label{sec:low_wave}Incremental changes at low wave numbers.} For $k \leq k_c$, the NFA free-surface elevations and the NFA normal component of velocity on the free surface are nudged toward their HOS counterparts as follows:
\begin{eqnarray}
\label{low1}
\Delta \eta^L   & = & - \beta N \Delta t (\eta^{\rm NFA} -\eta^{\rm HOS})^L \\
\label{low2}
\Delta u_n^L  & = &  -\beta N \Delta t (u_n^{\rm NFA} -u_n^{\rm HOS})^L \;\; ,
\end{eqnarray}
\noindent where $\beta=O(1)$ is a relaxation factor and recall that $N$ is the number of time steps between injections.  $\eta^{\rm HOS}$ is the position of the free surface as predicted by HOS, and $u_n^{\rm HOS}$ is the corresponding normal component of velocity.   In the context of Equations \ref{assimilate3} and \ref{assimilate4}, the differences between the NFA predictions and the HOS simulations as represented in the preceding equations are used to nudge the NFA predictions toward the HOS results for low wave numbers, $k \leq k_c$.

As a model of nudging, consider following difference equation:
\begin{eqnarray}
\label{low3}
\varphi^{n+1} = \varphi^{n}-\beta N \Delta t (\varphi^{n}-\varphi^*) \;\; ,
\end{eqnarray}
\noindent where $n$ denotes the number of the injection, $\varphi^{n+1}$ is the {\em new} variable, $\varphi^n$ is the {\em old} variable,  and $\varphi^*$ is the target variable.   
The general solution of \ref{low3} is 
\begin{eqnarray}
\label{low4}
\varphi^n = (\varphi^o - \varphi^*) (1-\beta N \Delta t)^n + \varphi^* \;\; ,
\end{eqnarray}
\noindent where $\varphi^o$ is the initial variable.    As $n \rightarrow \infty$, $\varphi^n \rightarrow \varphi^*$, \ie the target, as long as $\beta N \Delta t <1$.  As formulated, the difference between the current value of $\varphi^n$ and the target value $\varphi^*$ is used to nudge $\varphi$ toward the target solution.

\paragraph{\label{sec:high_wave}Incremental changes at high wave numbers.}   In the NFA data assimilation, energy cascades down from the lowest wavenumbers to the highest wavenumbers through the action of nonlinear wave interactions where it is dissipated due the effects of wave breaking, and forced by the wind and the wind drift.  Since the wavy portion of the flow forms naturally at high wavenumbers, the incremental changes at high wave numbers are zero and $\Delta \eta^H=\Delta u_n^H=0$.  Although we do not force the wavy portion of the flow at high wavenumbers in the present study, we provide here a formulation for completeness.

For $k>k_c$, the spectral densities of the free-surface elevation and normal velocity are calculated in the NFA data assimilation using the height function \ref{height3} and the normal component of velocity evaluated on the height function \ref{height5}.   We denote these one-dimensional spectral densities of the free-surface elevation and the normal velocity by  $S_\eta(k)$ and $S_u(k)$, respectively.  

For a fully saturated spectrum as $k \rightarrow \infty$, the spectral density of the free-surface elevation behaves as
\begin{eqnarray}
\label{high1}
S_{\eta_o}(k) \rightarrow \frac{\alpha_p}{2 k^3}  \;\; .
\end{eqnarray}
Similarly, the spectral density of the normal velocity on the free surface behaves as
\begin{eqnarray}
\label{high2}
S_{u_o}(k) \rightarrow \frac{\alpha_p}{2 k^2} \;\; .
\end{eqnarray}
Due to the effects of wave breaking and numerical dissipation, the wave energy in the NFA assimilation will decay between injections under most circumstances.   However, if the wave breaking is particularly energetic, the wave energy will actually grow at high wave numbers.   We increase or decrease the energy in the NFA assimilation at high wave numbers in a manner that is similar to the pumping that is used in the HOS simulations (see equations \ref{pump1}-\ref{pump3}).   

In wave number space, for $k > k_c$, we express the modal amplitudes of the increments of the free-surface elevation and normal velocity as follows:
\begin{eqnarray}
\label{high3}
\widetilde{\Delta \eta^H} & = & f_\eta(k) \widetilde{\eta^{\rm B}}(k,\theta)  \\
\label{high4}
\widetilde{\Delta u_n^H} & = & f_u(k) \widetilde{u_n^{\rm B}}(k,\theta) \;\; ,
\end{eqnarray}
\noindent where the tilde symbol denotes a Fourier transform in wave number space, $\widetilde{\eta^{\rm B}}$ and  $\widetilde{u_n^{\rm B}}$ are basis functions,  and $f_\eta$ and  $f_u$ are real factors that induce no change phase.  The wave-number vector expressed in terms of $\theta$ is ${\rm \bf{k}}=(k \cos(\theta), k \sin(\theta))$. The actual forms of basis functions will be provided later. 

The perturbations are added to the free-surface elevation and the normal velocity on the free surface to ensure that their respective spectrums are fully saturated.   This leads to the following quadratic equations for determining $f_\eta$ and  $f_u$:
\begin{eqnarray}
\label{high5}
f_\eta(k)^2 S_{\eta_p}(k)+ 2 f_\eta(k) S_{\eta_p \eta}(k) & & \nonumber \\ 
+S_\eta(k) -S_{\eta_o}(k) & = & 0 \\ 
\label{high6}
f_u(k)^2 S_{u_p}(k) + 2 f_u(k) S_{u_p u}(k) & & \nonumber \\ 
+S_u(k)-S_{u_o}(k) & =  & 0 \;\; . 
\end{eqnarray}
$S_{\eta_p}$ and $S_{u_p}$ are the spectral densities of the perturbations to the free-surface elevation and normal velocity, respectively.   Similarly, $S_{\eta_p \eta}$ and $S_{u_p u}$ are the spectral densities of the cross terms for the free-surface elevation and the normal velocity, respectively.

The solutions for $f_\eta$ and $f_u$ are
\begin{samepage}
\begin{eqnarray}
\label{high7}
f_\eta(k) =  \frac{{(S_{\eta_p \eta}^2 -S_{\eta_p}S_\eta + S_{\eta_p} S_{\eta_o})}^{1/2}-S_{\eta_p \eta}  }{S_{\eta_p}} \;\;\;\; & & \\
\label{high8}
f_u(k) =  \frac{{(S_{u_p u}^2 -S_{u_p}S_u + S_{u_p} S_{u_o})}^{1/2}-S_{u_p u}  }{S_{u_p}}  \;\; . & & 
\end{eqnarray}
\end{samepage}
If the arguments to the square roots are negative, we set  $f_\eta$ and $f_u$ as follows:
\begin{samepage}
\begin{eqnarray}
\label{high9}
f_\eta(k) =  \frac{S_{\eta_o} -S_\eta}{2 S_{\eta_p \eta}}  \;\;\;\; & & \\
\label{high10}
f_u(k) =  \frac{S_{u_p}-S_{u_o}}{2 S_{u_p u}}   \;\; . & & 
\end{eqnarray}
\end{samepage}
We further limit the maximum allowable change that can occur at high wave numbers by enforcing $|f_\eta | \leq f_{\rm max}$ and $|f_u | \leq  f_{\rm max}$.    We typically set  $f_{\rm max}=0.2$.

The basis functions $\eta^{\rm B}$ and $u_n^{\rm B}$ are filtered along the linear dispersion curve in the direction of the wind.   To perform this filtering, we let
\begin{eqnarray}
\label{high11}
\widetilde{\eta^{\rm NFA}}(k,\theta,t) =  \tilde{\eta}^+(k,\theta) \exp(\imath \omega t) \;\;\;\;\; & & \nonumber \\
                                                       + \tilde{\eta}^-(k,\theta)\exp(-\imath \omega t) \;\; & &  \\
\label{high12}
\widetilde{u_n^{\rm NFA}}(k,\theta,t) = \tilde{u}^+(k,\theta) \exp(\imath \omega t)    \;\;\;\;\;\; & &  \nonumber \\ 
                                                     + \tilde{u}^-(k,\theta) \exp(-\imath \omega t)  \;\;  ,
\end{eqnarray}
where the plus and minus superscripts indicate modal amplitudes associated with the positive and negative frequencies, respectively, and $\omega$ is the wave frequency based on the linear dispersion relationship \ref{unroll4}.

Two successive time steps are used to solve for modal amplitudes of the positive and negative frequencies as shown below:
\begin{eqnarray}
\label{high13}
\tilde{\eta}^+ = \frac{\imath( \widetilde{\eta^{\rm NFA}}^{N-1}-\widetilde{\eta^{\rm NFA}}^N\exp(\imath \omega \Delta t))}{2 \sin(\omega \Delta t)} \;\;\;\;\;\;\;\;\;\;  & &  \nonumber \\
\tilde{\eta}^- = \frac{\imath ( -\widetilde{\eta^{\rm NFA}}^{N-1}+\widetilde{\eta^{\rm NFA}}^N\exp(-\imath \omega \Delta t))}{2 \sin(\omega \Delta t)}  \;\;\;\;  & & \\
\label{high14}
\tilde{u}^+ = \frac{\imath( \widetilde{u_n^{\rm NFA}}^{N-1}-\widetilde{u_n^{\rm NFA}}^N\exp(\imath \omega \Delta t))}{2 \sin(\omega \Delta t)}  \;\;\;\;\;\;\;\;\; & & \nonumber \\
\tilde{u}^- = \frac{\imath ( -\widetilde{u_n^{\rm NFA}}^{N-1}-\widetilde{u_n^{\rm NFA}}^N\exp(-\imath \omega \Delta t))}{2 \sin(\omega \Delta t)} \;\; , & &
\end{eqnarray}
where the superscript $N$ and $N-1$ respectively denote the time steps.   Given the modal amplitudes of the positive and negative frequencies, the basis functions are formulated as follows:
\begin{eqnarray}
\label{high15}
\widetilde{\eta^{\rm B}}(k,\theta) = g(\theta) \eta^+(k,\theta)  + g(\theta-\pi) \eta^-(k,\theta)  \;\;\; & & \\
\label{high16}
\widetilde{u_n^{\rm B}}(k,\theta) = g(\theta) u^+(k,\theta) + g(\theta-\pi) u^-(k,\theta) \;\; , &&
\end{eqnarray}
where the filter $g$ is defined below:
\begin{eqnarray}
\label{high17}
g(\theta) = \left\{
\begin{array}{ll} 
0  \;\; , \;\; & |\theta-\theta_o| > \Delta \theta  \\
& \\
1 \;\; , \;\;  & |\theta-\theta_o| \leq \Delta \theta \;\; .
\end{array} \right.
\end{eqnarray}
Recall that $\theta_o$ is the direction of the wind (see equation \ref{spread1}).   $\Delta \theta$ is the width of the sector that is centered over the linear dispersion curve.   Typically, $\Delta \theta \leq \pi/4$.

If the filtering along the linear dispersion curve based on equations \ref{high11} and \ref{high12} is not accurate enough, spurious standing waves will form.   Under these circumstances, a least-squares analysis over many time steps can be used to calculate the basis functions with more accuracy.

The basis functions are designed to ensure that the incremental changes at high wave numbers occur in the direction of the wind.   If the basis functions had been set to the unfiltered free-surface elevation and normal velocity, standing waves would grow un-physically.  As formulated, standing waves are free to form under the action of wave breaking, but once they are formed, the standing waves are not forced at high wave numbers.

\paragraph{\label{sec:alpha}Total free-surface increment.}   The increment in the free-surface elevation needs to be converted into a change in the volume fraction.   The conversion should preserve any wave overturning that is present in the old volume fraction.    The change in the free-surface elevation is used to update the volume fraction by defining a pseudo velocity:
\begin{eqnarray}
W(x,y,z)=\frac{\Delta \eta(x,y)}{\tau}  \;\; ,
\end{eqnarray}
\noindent where $\Delta \eta(x,y)$ is the free-surface increment and $\tau$ is a pseudo time.  

The resulting pseudo velocity $W$ is constant in time and along the z axis.  Based on a Courant constraint, a pseudo time step is chosen such that $\Delta \tau<\min(\Delta z)/\max(W)$, where $\Delta z$ is the grid spacing along the z axis.  Setting $\tau=1$, we let $\Delta \tau = 1/N_\tau$, where here $N_\tau$ is an integer such that  $N_\tau>\max(\Delta \eta)/\min(\Delta z)$.   

The volume fraction is updated by integrating the following equation:
\begin{eqnarray}
\frac{\partial \alpha}{\partial \tau} + \frac{\partial W \alpha}{\partial z} = 0 \;\; ,
\end{eqnarray}
\noindent where at $\tau=0$, $\alpha=\alpha^{\rm OLD}$.   

Similar to Equation \ref{operatorsplit}, VOF reconstruction and advection are used to integrate the preceding equation in $N_\tau$ steps:
\begin{eqnarray}
\alpha^{\rm n+1} =  \alpha^{n} 
-  \frac{{\cal F}_3 \left[ W,\alpha^{n} ,\Delta \tau \right] 
      -           {\cal F}_3 \left[ W,\alpha^{n} ,\Delta \tau \right]}{\rm V} \nonumber \\ 
      \;\; {\rm for} \;\; n = 0,\ldots,N_\tau,
\end{eqnarray}
\noindent where $\alpha^0=\alpha^{\rm OLD}$ and $\alpha^{\rm NEW}=\alpha^N$.

\paragraph{\label{sec:velocity}Total velocity increment.}  The increment in the normal component of velocity evaluated on the position of the free surface is used to provide an Neumann condition for a projection operator.   Since the wavy portion of the flow is irrotational, we define velocity potentials in the water and air that are respectively denoted by $\phi^{\rm H_2O}$ and $\phi^{\rm Air}$.   The following boundary-value problem is solved in the water:
\begin{eqnarray}
\label{eq:water}
\nabla^2 \phi^{\rm H_2O} & = & 0   \;\; {\rm for} \;\;  -d \leq z \leq \eta^{\rm NEW} \nonumber \\
\frac{\partial \phi}{\partial n}^{\rm H_2O} & = & \Delta u_n  \;\; {\rm on}  \;\; z = \eta^{\rm NEW}  \nonumber \\
\frac{\partial \phi}{\partial z}^{\rm H_2O} & = & 0  \;\; {\rm on}  \;\; z = -d \;\; .
\end{eqnarray}
Here, $\vec{n}$ is the unit normal to $\eta^{\rm NEW}$ that points from the water into the air.  Similarly, the boundary-value problem in the air is
\begin{eqnarray}
\label{eq:air}
\nabla^2 \phi^{\rm Air} & = & 0    \;\; {\rm for} \;\;  \eta^{\rm NEW}  \leq z \leq h \nonumber \\
\frac{\partial \phi}{\partial n}^{\rm Air} & = & \Delta u_n \;\; {\rm on}  \;\; z = \eta^{\rm NEW}  \nonumber \\
\frac{\partial \phi}{\partial z}^{\rm Air} & = & 0   \;\; {\rm on}  \;\; z = h  \;\; .
\end{eqnarray}
\noindent Equations \ref{eq:water} and \ref{eq:air} are solved using the method of fractional areas.  Details associated with the calculation of the area fractions are provided in \mbox{\cite{sussman2000}} along with additional references. 

Given the velocity potentials in the water and air, the velocity increment is projected down into the water and up into the air as follows:
\begin{eqnarray}
\label{eq:unew}
u^{\rm NEW}_i = u^{\rm OLD}_i + 
\left\{
\begin{array}{ll} 
\frac{\partial \phi} {\partial x_i}^{\rm H_2O} \;\; &  -d \leq z \leq \eta^{\rm NEW}  \\
& \\
\frac{\partial \phi}{\partial x_i}^{\rm Air} \;\;  & \eta^{\rm NEW}  \leq z \leq h
\end{array}
\right.  . 
\nonumber \\
\end{eqnarray}
\noindent As constructed, the normal component of the velocity increment across the air-water interface is continuous whereas the tangential component is discontinuous.   The density-weighted velocity smoothing that is discussed in \S\ref{sec:smooth_vof} mitigates this effect.

\subsubsection{\label{sec:vort_flow}Vortical portion of flow}

\indent The total velocity field ($u_i$) is separated into wavy and vortical portions as follows:
\begin{eqnarray}
\label{eq:decomp}
u_i = w_i+v_i \;\; ,
\end{eqnarray}
where $w_i$ and $v_i$ respectively denote the wavy and vortical portions of the flow.   A boundary-value problem is solved to get the wavy portion of the flow, and then the difference between the total velocity field and wavy velocity field is used to derive the vortical velocity field.   

The wavy portion of flow is irrotational in the air and the water, and we define velocity potentials in the water ($\Phi^{\rm H_2O}$) and air ($\Phi^{\rm Air}$) similar to Equations \ref{eq:water}-\ref{eq:unew}, which are used to calculate the {\it increment} of the normal component of velocity evaluated on the free surface.      The wavy portion of the flow is prescribed in terms of the normal component of velocity evaluated on the free surface (see Equation \ref{height5}):
\begin{eqnarray}
\label{eq:neumann}
\begin{array}{rll}
w_i n^{\rm NFA}_i =  & u^{\rm NFA}_n & \;\; {\rm on} \;\; z=\eta^{\rm NFA}  \\
 & \\
v_i n^{\rm NFA}_i  =  & 0            &  \;\; {\rm on} \;\; z=\eta^{\rm NFA} \;\; .
\end{array}
\end{eqnarray}
Here, $n^{\rm NFA}_i$ is the unit normal to $\eta^{\rm NFA}$ that points from the water into the air, and the normal component of the vortical portion of the flow is zero on free surface by construction.   In the water,
\begin{eqnarray}
\label{eq:wavy_water}
\nabla^2 \Phi^{\rm H_2O} & = & 0  \;\; {\rm for} \;\;  -d \leq z \leq \eta^{\rm NFA} \nonumber \\
\frac{\partial \Phi}{\partial n}^{\rm H_2O} & = & u^{\rm NFA}_n \;\; {\rm on}  \;\; z = \eta^{\rm NFA}  \nonumber \\
\frac{\partial \Phi}{\partial z}^{\rm H_2O} & = & 0  \;\; {\rm on}  \;\; z = -d \;\; ,
\end{eqnarray}
and in the air,
\begin{eqnarray}
\label{eq:wavy_air}
\nabla^2 \Phi^{\rm Air} & = & 0    \;\; {\rm for} \;\;  \eta^{\rm NFA}  \leq z \leq h \nonumber \\
\frac{\partial \Phi}{\partial n}^{\rm Air} & = & u^{\rm NFA}_n \;\; {\rm on}  \;\; z = \eta^{\rm NFA}  \nonumber \\
\frac{\partial \Phi}{\partial z}^{\rm Air} & = & 0   \;\; {\rm on}  \;\; z = h  \;\; .
\end{eqnarray}
The velocity field of the wavy portion of the flow is specified in terms of the divergence of $\Phi^{\rm H_2O}$ and  $\Phi^{\rm Air}$:
\begin{eqnarray}
\label{eq:wavy}
w_i =
\left\{
\begin{array}{ll} 
\frac{\partial \Phi}{\partial x_i}^{\rm H_2O} \;\; &  -d \leq z \leq \eta^{\rm NFA}  \\
& \\
\frac{\partial \Phi} {\partial x_i}^{\rm Air} \;\;  & \eta^{\rm NFA}  \leq z \leq h
\end{array}
\right.  , 
\end{eqnarray}
whereupon the vortical velocity field is $v_i=u_i-w_i$.

The incremental change in the vortical velocity field is expressed in terms of its mean and fluctuating components as a function of their distance to the free surface.    First, we define a free-surface-following coordinate system:
\begin{eqnarray}
\label{eq:profile1}
\zeta =
\left\{
\begin{array}{ll} 
\frac{h (z-\eta^{\rm NFA})}{(h-\eta^{\rm NFA})} \;\;  & \eta^{\rm NFA}  \leq z \leq h \\ 
& \\
\frac{d (z-\eta^{\rm NFA}) }{(d+\eta^{\rm NFA}) } \;\; &  -d \leq z \leq \eta^{\rm NFA} 
\end{array}
\right.  . 
\end{eqnarray}
With respect to this coordinate system, the mean and fluctuations of the vortical velocity field are defined as 
\begin{eqnarray}
\label{eq:profile2}
v_i(x,y,\zeta,t) = \left< v_i(\zeta,t) \right> + v'_i(x,y,\zeta,t) \;\; ,
\end{eqnarray}
where angle brackets and single primes respectively denote mean and fluctuations.  The vertical profile of the wind corresponds to $\zeta > 0$, and the vertical profile of the wind-drift currents corresponds to $\zeta \le 0$.   

The mean is calculated using spatial averaging in the horizontal plane:
\begin{eqnarray}
\left< F(\zeta,t) \right> = \frac{1}{L W} \int\int dx dy F(x,y,\zeta,t) \;\; ,
\end{eqnarray}
where $L$ and $W$ respectively denote the length and width of the data assimilation.   $F$ is a three-dimensional function.   We note that $\left< w_1(0,t) \right>$ and $\left< w_2(0,t) \right>$  are the horizontal components of the Stokes drift evaluated on the free surface.

The incremental changes in the vortical velocity field are expressed in terms of the mean and the fluctuations:
\begin{eqnarray}
\Delta v_i = \left<   \Delta v_i \right>  + \Delta v'_i  \;\; .
\end{eqnarray}
As shown in \S\ref{sec:mean}, incremental changes in the mean portion of the vortical velocity field are assimilated using a nudging technique.   As shown in \S\ref{sec:fluctuations}, incremental changes in the fluctuating portion of the vortical velocity field can be estimated based on theoretical considerations of fully-developed turbulence.  

The incremental change in the vortical velocity field is added to the old vortical velocity field to get the new vortical velocity field:
\begin{eqnarray}
v_i^{\rm NEW} = v_i^{\rm OLD}+\Delta v_i + V_i \;\; .
\end{eqnarray}
$V_i$ ensures that velocity increments are solenoidal with zero Neumann boundary conditions on free surface care of Equation \ref{eq:neumann}.    $V_i$ is formulated in terms of gradients of scalar potentials in the water ($\psi^{\rm H_2O}$) and the air  ($\psi^{\rm Air}$).   The boundary-value problem in the water is
\begin{eqnarray}
\label{eq:water_vort}
\nabla^2 \psi^{\rm H_2O} & = & -\frac{\partial \Delta v_i}{\partial x_i}   \;\; {\rm for} \;\;  -d \leq z \leq \eta^{\rm NEW} \nonumber \\
\frac{\partial \psi}{\partial n}^{\rm H_2O} & = & -\Delta v_i n_i  \;\; {\rm on}  \;\; z = \eta^{\rm NEW}  \nonumber \\
\frac{\partial \psi}{\partial z}^{\rm H_2O} & = & 0  \;\; {\rm on}  \;\; z = -d \;\; .
\end{eqnarray}
Here, $n_i$ is the unit normal to $\eta^{\rm NEW}$ that points from the water into the air.  Similarly, the boundary-value problem in the air is
\begin{eqnarray}
\label{eq:air_vort}
\nabla^2 \psi^{\rm Air} & = & -\frac{\partial \Delta v_i}{\partial x_i}     \;\; {\rm for} \;\;  \eta^{\rm NEW}  \leq z \leq h \nonumber \\
\frac{\partial \psi}{\partial n}^{\rm Air} & = & -\Delta v_i n_i \;\; {\rm on}  \;\; z = \eta^{\rm NEW}  \nonumber \\
\frac{\partial \psi}{\partial z}^{\rm Air} & = & 0   \;\; {\rm on}  \;\; z = h  \;\; .
\end{eqnarray}
\noindent Equations \ref{eq:water_vort} and \ref{eq:air_vort}, similar to Equations \ref{eq:water} and \ref{eq:air},  are solved using the method of fractional areas.   The final expression for $V_i$ is
\begin{eqnarray}
\label{eq:unew_vort}
V_i=
\left\{
\begin{array}{ll} 
\frac{\partial \psi}{\partial x_i}^{\rm H_2O} \;\; &  -d \leq z \leq \eta^{\rm NEW}  \\
& \\
\frac{\partial \psi}{\partial x_i}^{\rm Air} \;\;  & \eta^{\rm NEW}  \leq z \leq h
\end{array}
\right.  . 
\end{eqnarray}
Equations \ref{eq:water}-\ref{eq:unew} and \ref{eq:water_vort}-\ref{eq:unew_vort} can be linearly superposed in the water and the air to simultaneously solve for the total velocity increment for the wavy and vortical velocity fields, \ie two boundary-value problems instead of four.

\paragraph{\label{sec:mean}Mean wind and wind-drift profiles.} The mean increment in the vortical portion of the velocity field is imposed using nudging:
\begin{eqnarray}
\label{betavortical}
\lefteqn{\left< \Delta v_i(\zeta,t) \right> =} && \nonumber \\
&& -\beta N \Delta t \left( \left< v_i^{\rm NFA}(\zeta,t) \right> -  v^{\rm LOG}_i(\zeta,t) \right) \; , \;\; 
\end{eqnarray}
where $\beta$ is a relaxation factor, $N$ is the number of time steps between injections, and $\Delta t$ is the time step.  $v^{\rm LOG}_i(\zeta,t)$ is the measured mean profiles of the wind for $\zeta>0$ and  the wind-drift current for $\zeta \leq 0$.     Based on theoretical considerations, $v^{\rm LOG}_i(\zeta,t)$ has log-like behavior for equilibrium conditions.

We anticipate that the measured profiles of the wind and the wind drift in equation \ref{betavortical} will be based on temporal averaging.   In this case, the temporal measurements in the field are related to spatial averaging in the data assimilations through a Galilean transformation.   \mbox{\cite{dommermuth:2002}} use a similar formulation to perform wake relaxation to study the formation of pancake eddies in a stratified fluid.  In fact, the nudging approach that is used in this paper generalizes the wake-relaxation technique to unsteady flows.

\paragraph{\label{sec:fluctuations}Fluctuations in wind and wind-drift profiles.} The turbulent fluctuations form naturally through the energy cascade, the generation of free-surface vorticity, and the breaking of waves.   Since the turbulent fluctuation form naturally, the incremental changes of the turbulent fluctuations are zero and $\Delta v'_i=0$.  Although we do not force the turbulent fluctuations in the present study, we provide here a formulation for completeness.

The fluctuations are rescaled using a scaling factor.
\begin{eqnarray}
\Delta v'_i = f(\zeta) v'_i(x,y,\zeta,t)
\end{eqnarray}
The scaling factor is
\begin{eqnarray}
f(\zeta) = \frac{v^*(\zeta,t)}{{\left< v'_i(\zeta,t) v'_i(\zeta,t) \right>}^{1/2}}-1
\end{eqnarray}
where $v^*$ is the rms velocity of the fluctuations as a function of the distance to the free surface.  We further limit the maximum allowable change that can occur for the fluctuations by enforcing $|f(\zeta) | \leq f_{\rm max}$.    We typically set  $f_{\rm max}=0.2$.   The spectral content of the turbulent fluctuations could also be forced to give a $k^{-5/3}$ power-law behavior based on Kolmogorov's theory of turbulence. 

\section{Results}

Data assimilations of breaking waves in equilibrium with the wind are used here to investigate the structures of the upper oceanic boundary layer (OBL) and the lower marine atmospheric boundary layer (ABL).  NFA is adapted to HOS simulations of a JONSWAP  (\ref{jonswap1}-\ref{jonswap3}) spectrum to study the effects of wave breaking for short-crested seas.   The peak enhancement factor is $\gamma=6$.   The angular spreading is  $s=50$ with $\theta_o=0$ (see equation \ref{spread1}).   The results are converted to dimensional units by choosing $L_o=100 {\rm m}$ for the length of wave at the peak of the spectrum.   The significant wave height is $H_s=3.66 \; {\rm m}$.    The HOS simulations use $N_x \times N_y = 256 \times 64=16,384$ de-alaised Fourier modes with a fourth-order approximation.  The length (L), width (W), and depth (d) of the HOS simulations are respectively 500 m, 125 m, and 50 m.  The period of adjustment is $T_a=31.93 \; {\rm s}$ (see \ref{adjust1}).   For reference, the wave period at the peak of the spectrum is $T_o=8.00 \; {\rm s}$.  The HOS simulations are run for $99.8 \; {\rm s}$  to adjust the waves before being assimilated.  The HOS are injected just before the Kurtosis of the free-surface elevation reaches its maximum value.   The HOS simulations are smoothed every time step with $F_c=0.9$ (see \ref{smooth_spectral}), and each HOS simulation uses energy pumping to maintain the total energy (see \ref{pump1}-\ref{pump3}).   

Coarse and medium-sized assimilations are performed.  The lengths (L), widths (W), depths of water (d), and heights of air (h) of the data assimilations are respectively 500 m, 125 m, 50 m, and 50 m.    The number of grid points along the $x$, $y$, and $z$-axes are respectively 4096, 1024, and 512 for the coarse-sized assimilation and 8192, 2048, and 1024 for the medium-sized assimilation.   The total numbers of grid points are approximately 2.15  and 17.2 billion grid points for the coarse and medium-sized assimilations, respectively.  The coarse and medium data assimilations are performed with respectively 12.2 and 6.10 cm resolution.  Grid stretching is used to cluster points near the free surface along the z axis.   The durations of the coarse and medium assimilations are 249.5 and 50.89 seconds, and the number of time steps for the coarse and medium data assimilations are 125,000 and 51,000, respectively.

Data are assimilated every 0.07983 and 0.03992 seconds for respectively the coarse and medium-sized assimilations.  The HOS simulations and log profiles are injected into the NFA simulations every $N=40$ time steps for each assimilation.  The high data rate is required to prevent ringing.  The HOS data  is assimilated for wave numbers $k_c \leq 0.70 \; {\rm rad/m}$,  which corresponds to a wavelength of 8.976 m (see Equations \ref{low1}-\ref{low2}).   The cutoff is chosen to match \underline{Wa}ve and Surface Current \underline{Mo}nitoring \underline{S}ystem (WaMoS) measurements of the ocean surface \mbox{\citep{lund2012}}.   Density-weighted velocity smoothing is applied every 100 and 140 time steps for the coarse and medium-sized assimilations, respectively (\ref{smooth_velo}-\ref{project2}).     The relaxation factor is $\beta=2$ for each assimilation (see equations \ref{low1}, \ref{low2}, and \ref{betavortical}).  

The wind speed at 10 meters height is $U_{10}=11.1 \; {\rm m/s}$.  For reference, the phase speed at the peak of the spectrum is $c_o=12.5 \; {\rm m/s}$ such that  $U_{10}/c_o=0.89$. The friction velocities in the atmosphere and the ocean are $u_*=0.814 \; {\rm m/s}$ and $w_* = 2.83 \;  {\rm cm/s}$, respectively.  The wind drift on the ocean surface is $u_o=31.3 \; {\rm cm/s}$, which is 2.82\% of the wind speed at 10 m.  The roughness heights in the air and the water are respectively $z_{air}=10.0 \; {\rm cm}$ and $z_{h_2o}=40.0 \; {\rm cm}$.   $u_{air}= 1.44 \; {\rm m/s}$ and $u_{h_2o}= 3.13 \;  {\rm cm/s}$ are used to specify the profiles of the mean wind and mean wind drift  (see equations \ref{meanwind} and \ref{meandrift}).  As constructed, the profile in the water has \mbox{5.99 cm/s} return flow at the bottom.  The Stokes drift at z=0 is $U_s \approx 10. \; {\rm cm/s}$.  The turbulent Langmuir number is $La_{tur} = (w_*/U_s)^{1/2}=0.53$.    The height of the critical layer where the wind speed is equal to the phase speed of the wave at the peak of the spectrum is $z_c=19.6 \; {\rm m}$.  The ratio of the density of the air to the density of the water is 0.001207.  There is no stratification, and Coriolis effects are not considered.

The coarse and medium data assimilations ran for 520 and 740 wall-clock hours, and they generated 112 and 275 TB of data, respectively.  The core solver is written in Fortran 90 and communication between processors is performed using MPI. 

\subsection{\label{sec:windrows}Windrows}

Figures \ref{fig:windrow1}a and b are views looking down on windrows forming and waves breaking for the coarse and medium assimilations.   These figures and subsequent figures include links to animations in the figure captions.   The animations, if they are included, are highlighted in blue.  The figures in the electronic version of this paper can be zoomed to view details that would not otherwise be visible.  Similarly, the animations can be viewed in high definition using the HD options that are available on YouTube.

The waves and the wind are moving from left to right in Figures  \ref{fig:windrow1}a and b.  Lagrangian particles are used to illustrate the formation of windrows under the action of wave breaking and the formation of swirling jets.  The particles are initially uniformly distributed.   The free surface in the upper panel is shaded with the water-particle velocity in the direction of the wind.  The results are shown at $t=190 \; {\rm s}$ and $t=50.9 \; {\rm s}$ for the coarse and medium assimilations, respectively.  

Particles surf breaking waves.   Particles spill over the front and out the sides of breaking waves.     The streaks in the velocity are due to the formation of swirling jets.   The direction of the fluid velocity within the jets is downwind.    The particles spill over the fronts and sides of breaking waves in the same regions where swirling jets form because the velocities are lower there than at neighboring points on the fronts of breaking waves.    As a result of this action, the particles are aligned with the swirling jets.   Contrary to Langmuir's original hypothesis \mbox{\citep{langmuir1938}}, windrows do not form due to flow converging transverse to the wind on the free surface due to the effects of Langmuir cells.    For fully-developed seas, windrows form under the action of breaking waves and the formation of swirling jets.

Figures \ref{fig:windrow2}a and b are perspective view of windrows forming.   The results for the coarse and medium-sized data assimilations are shown at the same time instance at time $t=7.90 \; {\rm s}$.   As before, Lagrangian particles are used to illustrate the formation of windrows under the action of wave breaking and the formation of swirling jets.  The shedding of droplets into the atmosphere is visible in the animations when viewed at the highest resolution that is permitted.  The droplets have a whirling orbit that indicates that they are being entrained by swirling jets that are being shed into the atmosphere.   The patterns of the Lagrangian particles are similar for the coarse and medium assimilations at this early time.

Figures \ref{fig:windrow3} show the formation of windrows along with the three components of vorticity for the coarse and medium assimilations.      The x-component of vorticity is in line with the wind.   The y and z-components of the vorticity are transverse to the wind in the horizontal and vertical planes, respectively.  The results are shown at time instances that are at the ends of the assimilations.    The vorticity is plotted on a surface that is 20 cm below the free surface in a free-surface-following coordinate system.  The x-component of vorticity is associated with the swirling portion of the swirling jets.    The y-component of vorticity is related to surfing and flow separation off the backs of breaking waves.   The z-component of vorticity is associated with the jet portion of the swirling jets.  These results are best viewed in the animations using the 4K ProRes format.

The swirling jets are bounded by two bands of the z-component of vorticity.    The upper band is positive and the lower band is negative giving a jet flow that is downwind.    The diameters of the swirling jets are less than 80 cm.   The origins of the swirling jets are the uneven fronts of breaking waves where the y-component of vorticity is positive.   The swirling jets are stretched out over the wakes of the spilling breaking waves. The trailing edges of the wakes are often marked by bands of negative y-component of vorticity that are transverse to the wind.  The wake vorticity is rapidly absorbed into the background shear.  The swirling portion of the flow is visible  in the x-component of vorticity as streaks with either sign that are about 10 to 20 meters long and less than 40 to \mbox{80 cm} in diameter.  The streaks in the x-component of vorticity are most evident in the medium-sized assimilation at time instant when the breaking is stronger.  The swirling jets are strongest during active spilling in regions that are behind the fronts of breaking waves.   The lateral spacing of the strongest swirling jets is similar to the Langmuir cells that form immediately beneath them, and unlike the lateral spacing of the particles, does not vary much over the course of the assimilation. 

The Lagrangian particles surf the fronts of the breaking waves.  The surfing scrubs the free surface clean of particles as the particles line up along the fronts of breaking waves.    Particles spill over the fronts of breaking waves or are swept to the corners of the breaking fronts at the same points where swirling jets form because the water-particle velocities at those points are less than neighboring points on the breaking front.   Floating matter and swirling jets are collinear due to this effect of surfing.   The windrows get longer with each successive passing of a breaking wave.  Remnants of swirling jets are also reenergize by breaking waves.  However, not every swirling jet has particles that are collinear with it.   Also, the spacings of the streaks of particles transverse to the wind tends to get wider with time with a corresponding increase in the density of the particles.   The streaks of particles are longer than the surfing effects that form them. 
\begin{figure*}
\begin{center}
\begin{tabular}{cc}
\includegraphics[width=0.45\linewidth]{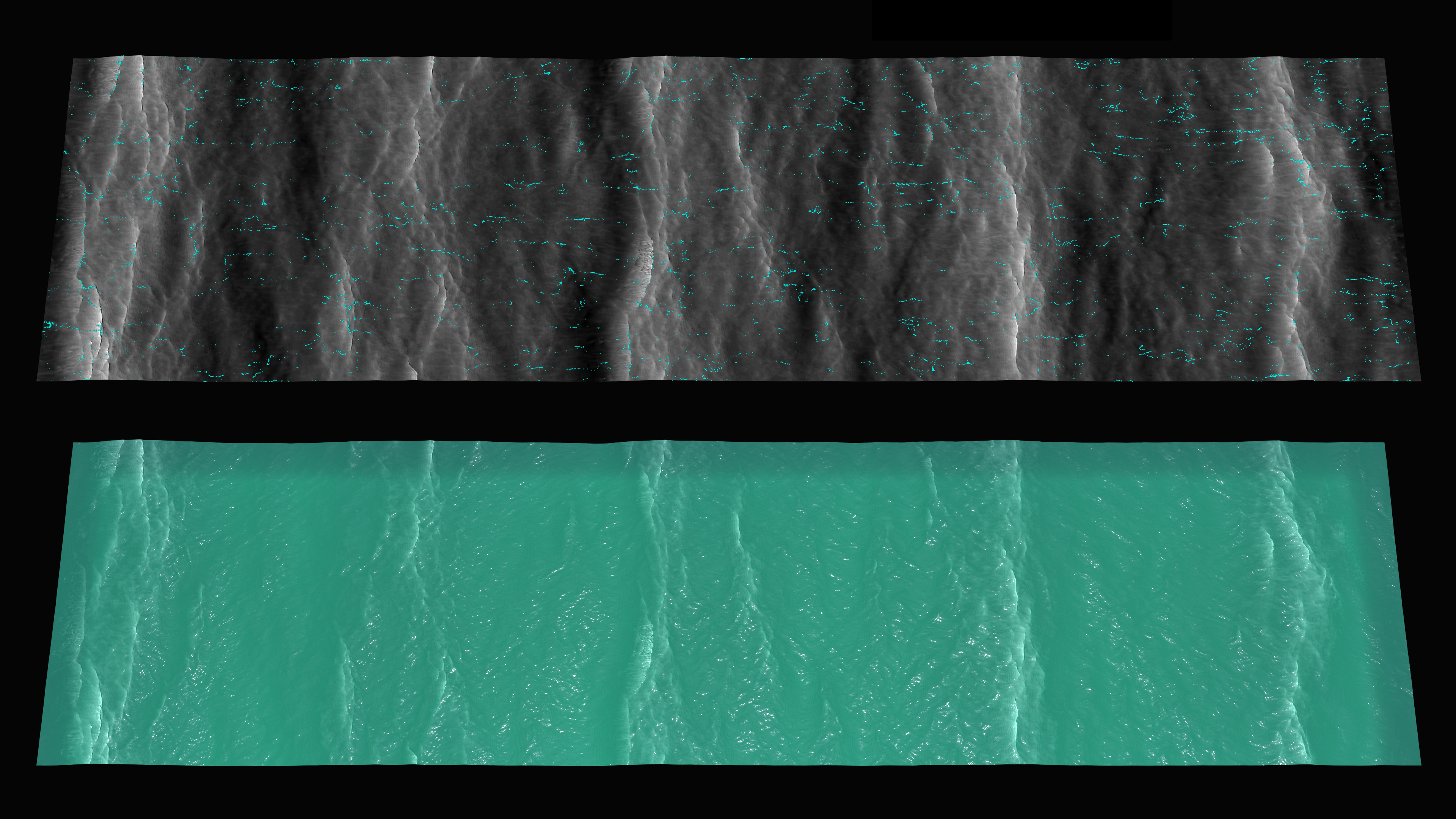} &
\includegraphics[width=0.45\linewidth]{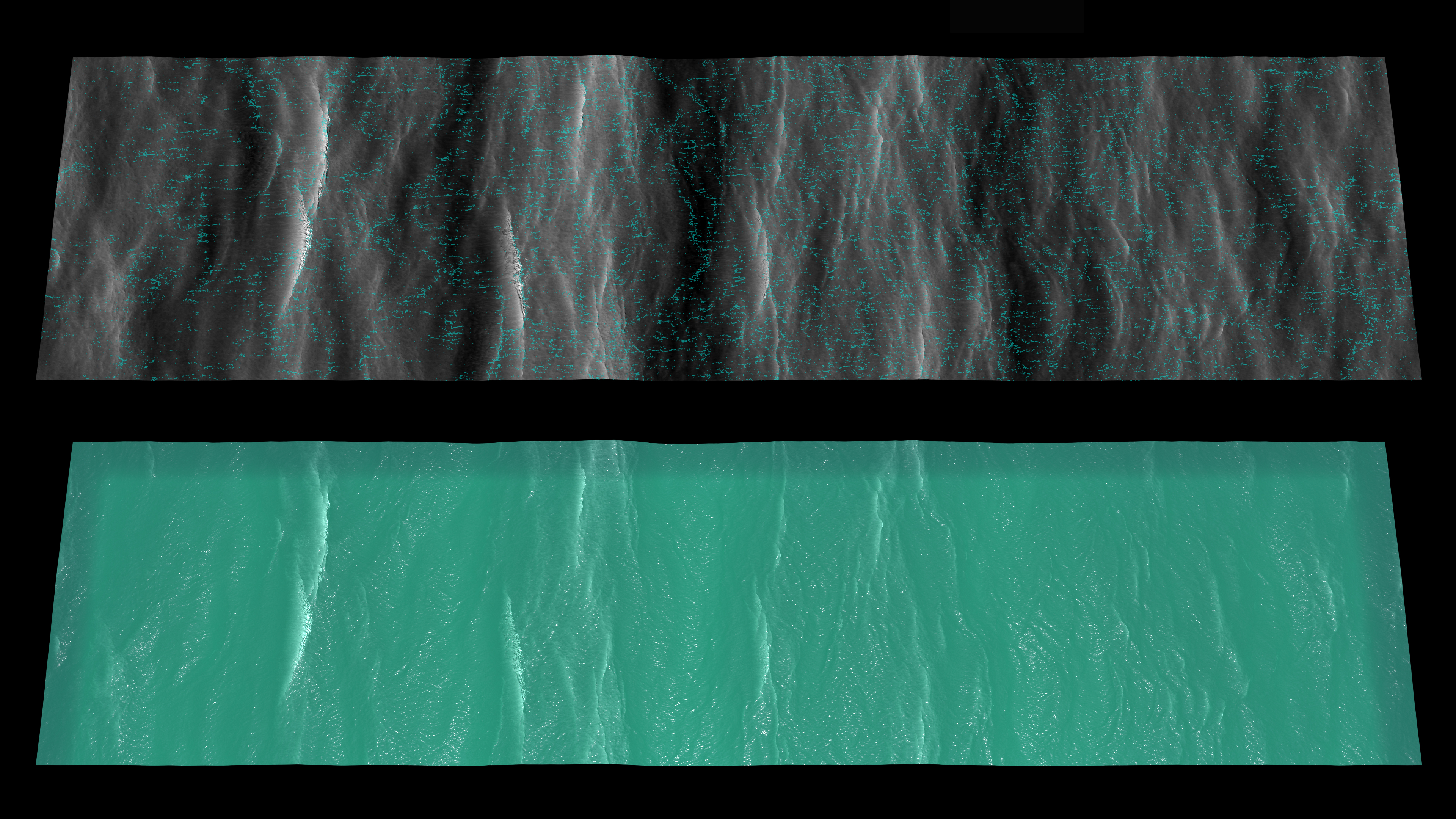} \\
(a) & (b)
\end{tabular}
\end{center}
\caption{\label{fig:windrow1} View looking down on windrows forming and wave breaking.   The results are shown for the (a) coarse-sized assimilation and  the (b) medium-sized assimilation.  The green particles (part a) and the blue particles (part b) are Lagrangian markers that are constrained to the free surface.  The diameters of the particles are 80 and 40 cm for respectively the coarse and medium assimilations.  The free surfaces in the  upper panels in each figure are shaded  with the component of the water-particle velocity in the direction of the wind.  The lower and upper limits of the grey scales are respectively 1.88 m/s and 11.59 m/s.   The grey scales of values outside that range are saturated.    The lower panels in the figures are ray-traced imagery of the wave breaking without the particles.  The Animations of these results are available at \mbox{\cite{daac5}} \href{http://youtu.be/IhrHRJSYfPE}{view looking down on windrows forming and waves breaking (coarse)} and \mbox{\cite{daac11}} \href{http://youtu.be/rUoSb6BWy4I}{view looking down on windrows forming and waves breaking (medium)}.   Animations of just the windrows forming are available at \mbox{\cite{daac6}} \href{http://youtu.be/Pfa54wLAXVY}{view looking down on windrows forming (coarse)} and \mbox{\cite{daac12}} \href{http://youtu.be/Sbn23QN3jD4}{view looking down on windrows forming (medium)}.}
\end{figure*}

\begin{figure*}
\begin{center}
\begin{tabular}{cc}
\includegraphics[width=0.45\linewidth]{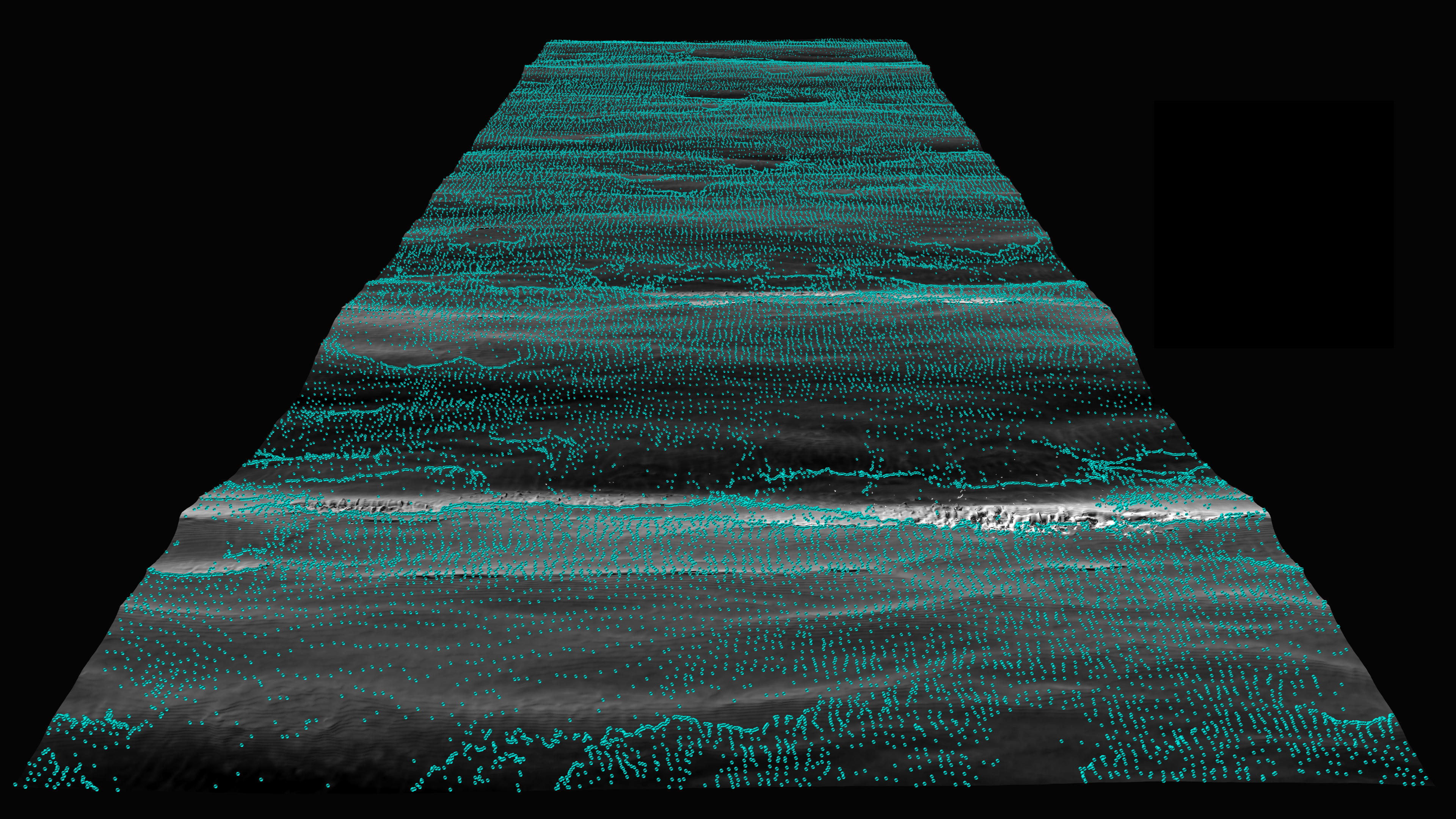} &
\includegraphics[width=0.45\linewidth]{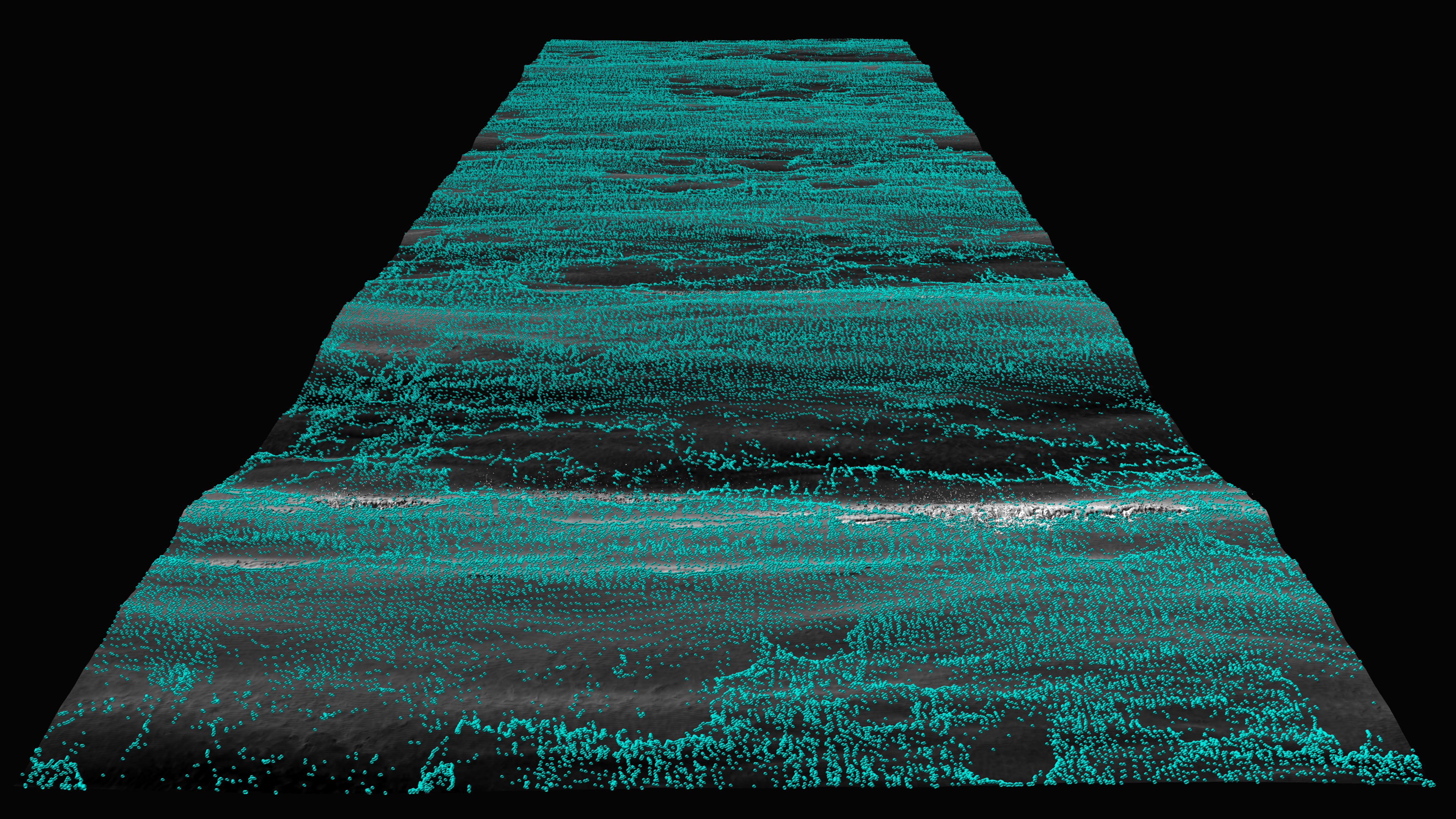} \\
(a) & (b)
\end{tabular}
\end{center}
\caption{\label{fig:windrow2} Perspective view of windrows forming.   The results are shown for (a) the coarse-sized assimilation and  (b) the  medium-sized assimilation.  The blue particles are Lagrangian markers that are constrained to the free surface.  The diameters of the particles are 80 and 40 cm for respectively the coarse and medium assimilations.  The free surfaces are shaded  with the component of the water-particle velocity in the direction of the wind.  The lower and upper limits of the grey scales are respectively 1.88 m/s and 11.59 m/s.   The grey scales of values outside that range are saturated.     Animations of these results are available at \mbox{\cite{daac3}} \href{http://youtu.be/6r9JbrykUcE}{perspective view of windrows forming (coarse)} and \mbox{\cite{daac9}} \href{http://youtu.be/JnLGxS3c4dA}{perspective view of windrows forming (medium)}.}
\end{figure*}

\begin{figure*}
\begin{center}
\begin{tabular}{c}
\includegraphics[trim=90mm 0mm 95mm 0mm,clip=true,angle=0,width=0.80\linewidth]{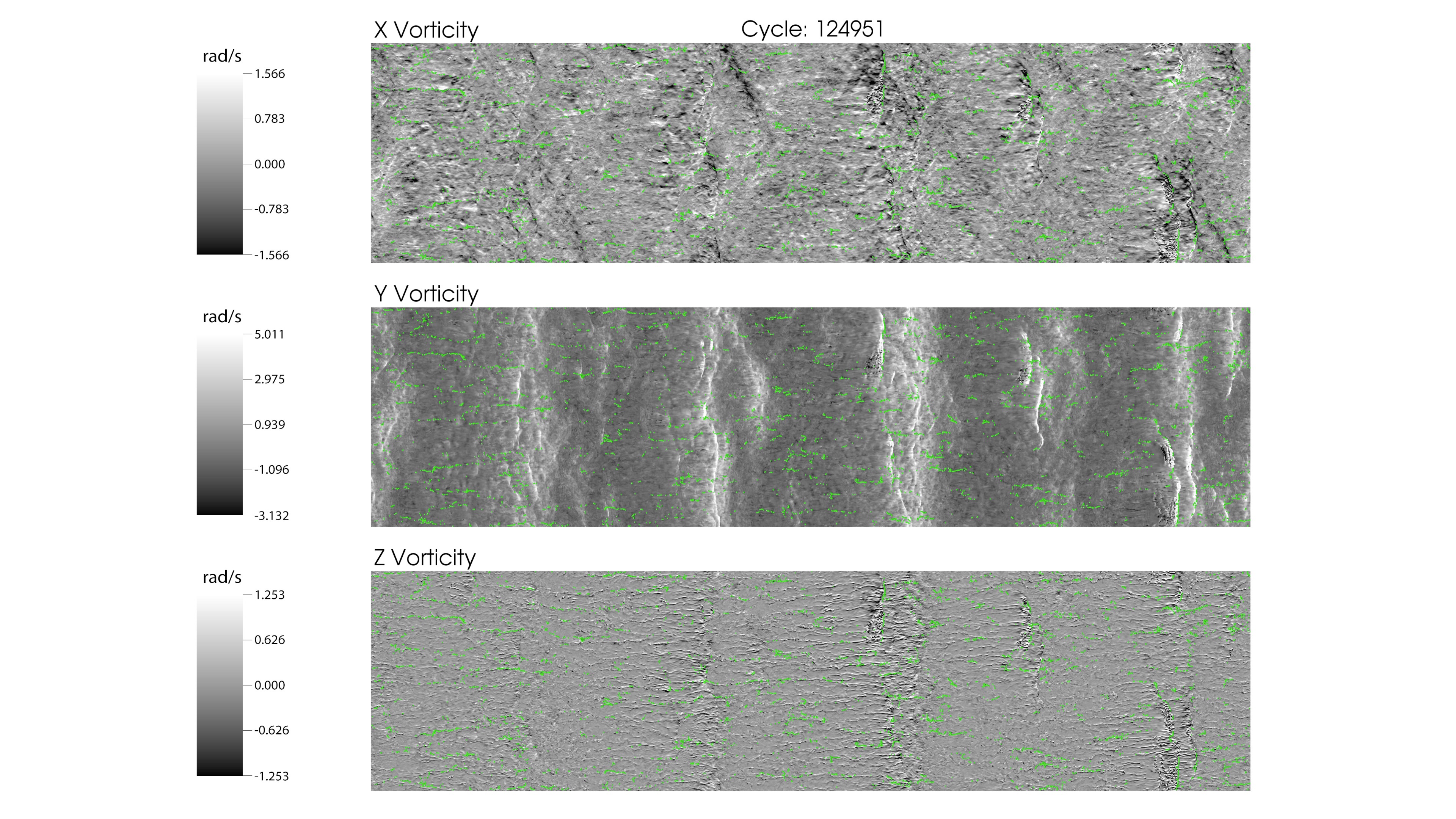} \\
(a) \\
\includegraphics[trim=90mm 0mm 95mm 0mm,clip=true,angle=0,width=0.84\linewidth]{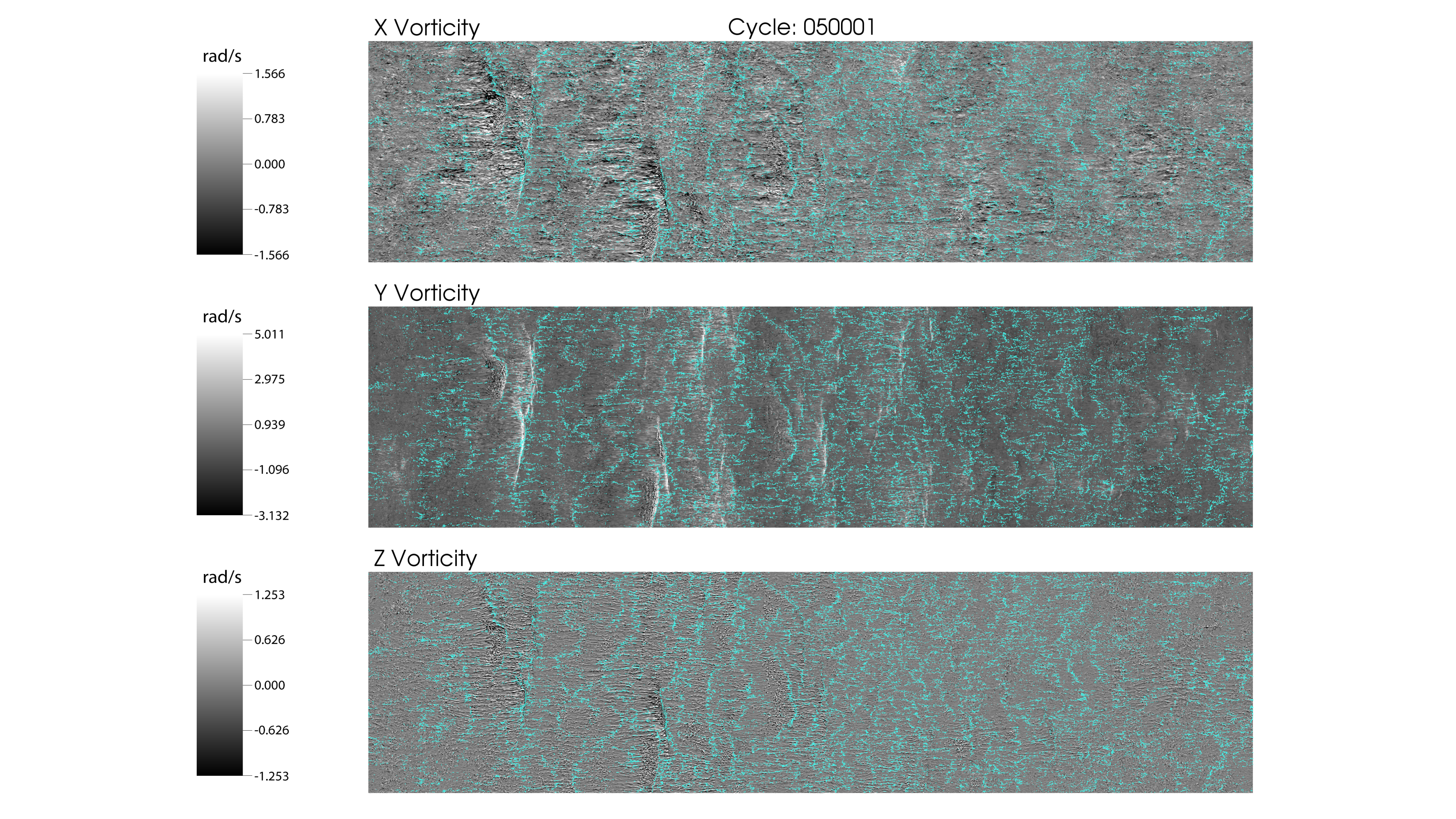} \\
(b)
\end{tabular}
\end{center}
\caption{\label{fig:windrow3} The formation of windrows due to surfing and swirling jets.    The results are shown for the (a) coarse-sized  assimilation and  the (b) medium-sized assimilation.   The green particles (part a) and the blue particles (part b) are Lagrangian markers that are constrained to the free surface.  The diameters of the particles are 80 and 40 cm for respectively the coarse and medium assimilations.  The three components of vorticity are shown on a surface that is 20 cm beneath the free surface in a free-surface-following coordinate system.   Animations of these results are available at \mbox{\cite{daac1}} \href{http://youtu.be/Mg8iJR-\_C48}{windrows forming under the action of breaking waves (coarse)} and \mbox{\cite{daac8}} \href{http://youtu.be/DR5eNIQ2ZDY}{windrows forming under the action of breaking waves (medium)}.}
\end{figure*}

\subsection{\label{sec:langmuirstreaks}The structure of Langmuir cells and wind streaks}

Figure \ref{fig:centerplane} shows the three components of the turbulent fluctuations on a centerplane cut.  The results are based on the coarse assimilation.  The turbulent fluctuations are calculated by subtracting out the wavy portion of the flow using equations \ref{eq:decomp} to \ref{eq:wavy}.   Once the wavy portion of the flow is eliminated, the vertical profiles of the mean vortical flow are also subtracted out.   A zero Neumann boundary condition is enforced on the free surface for the vertical profiles of the  mean  vortical flow to ensure that the mean wind and the mean wind drift are parallel to the free surface.  The boundary-value problem that is solved to subtract out the mean wind and the mean wind drift is similar to equations \ref{eq:water_vort} to \ref{eq:unew_vort} with zero Neumann conditions.   Two different color scales are used for the flow in the air and the water.    The colors of values outside the ranges of the scales are saturated. The figures in the electronic version of this paper can be zoomed to view details that would not otherwise be visible. The caption provides a link to an animation that shows the vortical flow developing as a function of time.

The flows in the water and the air are initially seeded with random velocity fluctuations.   Flow separation occurs at the free surface.    The turbulent diffusion is much greater in the air than it is in the water.    The turbulent fluctuations in the air are over ten times greater than those in the water.   Additional details of the turbulent mixing are provided in \S \ref{sec:mixing}.  The inclined stripes of negative and positive contours in the streamwise ($u$) and spanwise ($v$) velocities in the atmosphere are similar to stripes in the spanwise velocity of Figure 12 of  \mbox{\cite{marusic2012}}.

Figures \ref{fig:langmuir}a-i show the structure of Langmuir cells.  The results are based on the coarse assimilation.   The three components of velocity are shown at various depths below the free surface from $\zeta=0 \; {\rm m}$ to  $\zeta=-12.8 \; {\rm m}$, where $\zeta=z+\eta(x,y)$ is a free-surface-following coordinate system.   The results are time-averaged over two wave periods of the wave at the peak of the spectrum.  The results are shown at time $t=249.5 \; {\rm s}$ in Figure \ref{fig:langmuir}.   The duration of the animations is about 14 wave periods or 111.8 seconds.  

The u and v-components of velocity are streaky near the free surface.  There is some cross hatching in the w-component near the free surface that may be due to back scatter late in the simulation as discussed in \S \ref{sec:growth} (see Figures \ref{fig:langmuir}a-c). Without cross hatching, the w-component of velocity is smoother than the horizontal components of velocity.   Streaks form in the w-component of velocity at deeper depths as the swirling jets are tilted (see Figures \ref{fig:langmuir}d-h).   As the streaks form in the w-component of velocity, the v-component of velocity becomes less streaky.   Close to the free-surface, the streaks in the horizontal component of velocity are due to the meandering of the swirling jets.     Additional details of the meandering are provided in \S \ref{sec:profiles}.  

The vertical bands in the u and w-components of velocity are due to passage of waves.   The u and w-components of velocities interweave and form Y-junctions, which is especially evident in the animations.   For the interweaving in the u-component of velocity, see Figures \ref{fig:langmuir}a-g.  For the interweaving in the w-component of velocity, see Figures  \ref{fig:langmuir}d-h.  The positive and negative streaks in the u and w-components of velocities are 180 degrees out-of-phase between bands of streaks.    Streaks of the same sign are sometimes connected with wisps that reach either one below or one above to connect to streaks in neighboring bands due to a helical flow.      The interweaving of streaks and the formation of Y-junctions leading to mixing in the lateral direction is likely related to the meandering in the cross drift and crosswind that is discussed \S \ref{sec:profiles}. 

 The lateral spacing of the streaks gets greater at deeper depths.  For example, compare the fine-scale streaks in the u-component of velocity at $z=-20 \; {\rm cm}$ to the streaks at $z=-3.2 \; {\rm m}$.    At $z=-3.2 \; {\rm m}$, the spacing is about 11 m.    The Langmuir cells are not fully formed at the deeper depths as Figures \ref{fig:langmuir}h and i show.    At $z=-12.8 \; {\rm m}$,  there are downwind jets in the u-component of velocity and downwelling jets in the w-component of velocity.   The downwelling jets are similar to those that had been observed in \mbox{\cite{sullivan2007}}.   The v-component of velocity shows a quadrupole structure at $z=-12.8 \; {\rm m}$ due to pairing of swirling jets.  \mbox{\cite{walker2009}} (\href{http://news.bbc.co.uk/earth/hi/earth\_news/newsid\_8035000/8035593.stm}{BBC video}) shows similar pairing of swirling jets in the crest of plunging wave (Source BBC News/bbc.co.uk - \textcopyright 1999 BBC).   The large-scale structures at depth are due to the tilting of swirling jets that had been formed at the free surface.  The jet features in the u and w-component of velocity are due to the jet portion of the flow in swirling jets.    The quadrupole features in the v-component of velocity are due to the swirling portion of the flow in swirling jets.

Figures \ref{fig:windstreaks1}a-i show the structure of wind streaks.  The results are based on the coarse assimilation.  The three components of velocity are shown at various heights above the free surface from $\zeta=0 \; {\rm m}$ to  $\zeta=12.8 \; {\rm m}$, where $\zeta=z+\eta(x,y)$ is a free-surface-following coordinate system.   The results are time-averaged over two wave periods of the wave at the peak of the spectrum.   The results are shown at time $t=249.5 \; {\rm s}$ in Figure \ref{fig:windstreaks1}.  The duration of the animations is 14 wave periods.

As is the case for the Langmuir cells, the u and v-components of velocity for the wind streaks are streaky near the free surface.  There is some cross hatching in the w-component near the free surface that may be due to back scatter late in the simulation as discussed in \S \ref{sec:growth}. Without cross hatching, the w-component of velocity is smoother than the horizontal components of velocity.   The wind streaks are much finer close to the free surface than they are away from the free surface, especially for the w-component of velocity.   The animations show that the wind streaks close to the free surface have a higher frequency content than those higher above the free surface.   This is also evident in the plots of the crosswind meandering in \S \ref{sec:profiles}. The effects of the waves diminish as the height above the free surface increases, where the wind speed is higher.   The longest wind streaks at the highest elevation at $z=12.8 \; {\rm m}$ are \mbox{500 m}, \ie as long as the computational domain.    At the higher elevations, the wind streaks in the u and w-components of velocity with the same sign are connected by wisps, similar to the streaks beneath the free surface. 

Figure \ref{fig:windstreaks2} shows wind streaks based on radar processing using SuFMos \mbox{\citep{horstmann2014}}.  As \mbox{\cite{dankert2005}} discuss, radar backscatter of the ocean surface is used to measure wind and  wave fields.   For the X-band radar that is used with SuFMos, the radar back scatter is sensitive to surface roughness with length scales that are order 3 cm.   Such short waves would respond differently to wind gusts and lulls, which may help to explain the preponderance of positive streaks in Figure \ref{fig:windstreaks2} (compare Figures  \ref{fig:windstreaks2}d and i).  Even so, the streaks in Figure \ref{fig:windstreaks2} are strikingly similar to the wind streaks in Figure \ref{fig:windstreaks1}i.  In particular, the length scales are comparable.   Comparing the animations of the streaks in the radar measurements to the streaks in the results of the data assimilations, also show many similarities.

\begin{figure*}
\begin{center}
\begin{tabular}{c}
\includegraphics[trim=0mm 0mm 0mm 0mm,clip=true,angle=0,width=\linewidth]{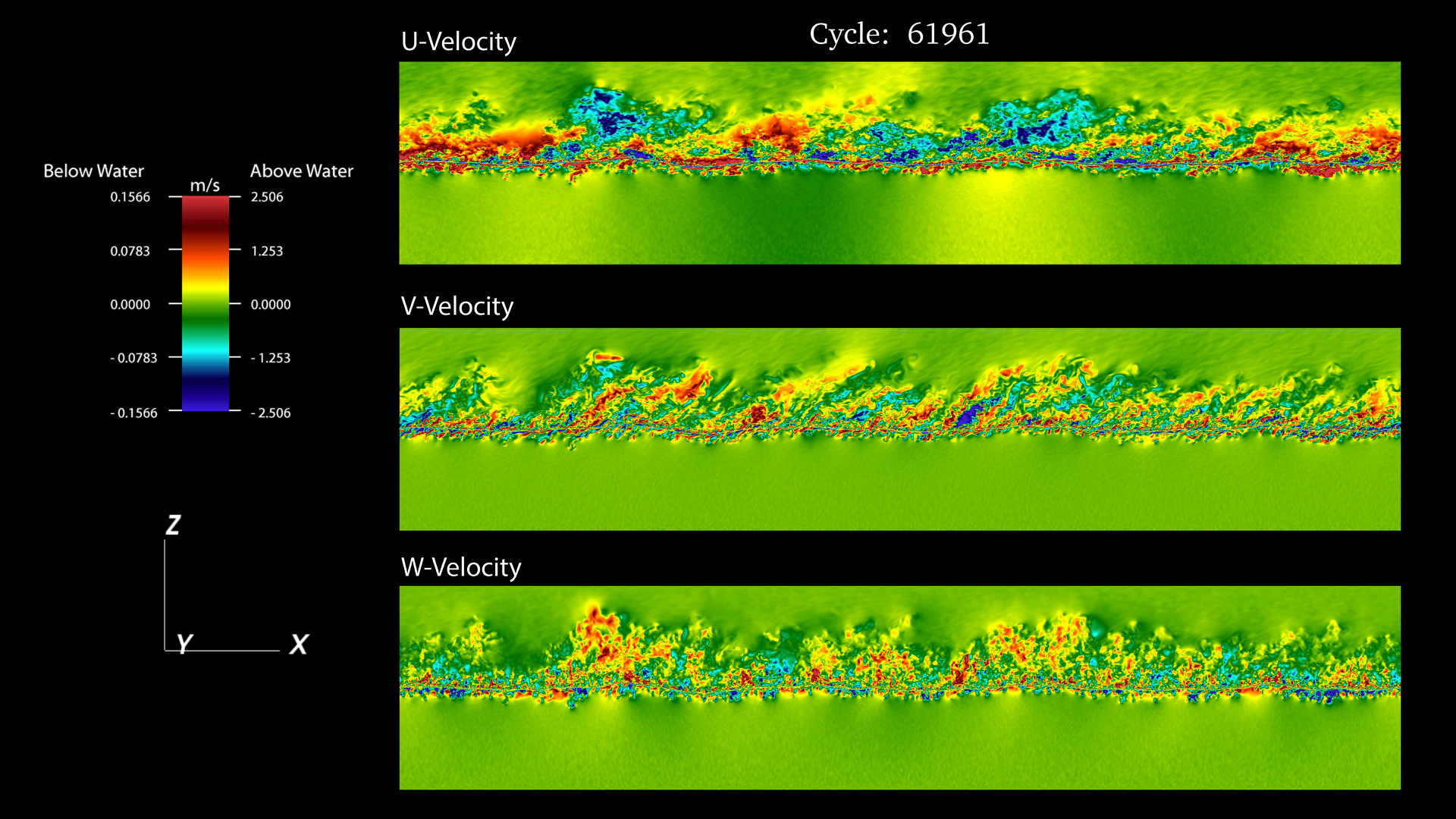}
\end{tabular}
\end{center}
\caption{\label{fig:centerplane} Centerplane cuts of the turbulent fluctuations.    An  animations of this result is available at \mbox{\cite{daac2}} \href{http://youtu.be/t7noqS2H\_RM}{centerplane cuts of velocity}.}
\end{figure*}

\begin{figure*}
\begin{center}
\begin{tabular}{ccc}
\includegraphics[width=0.30\linewidth]{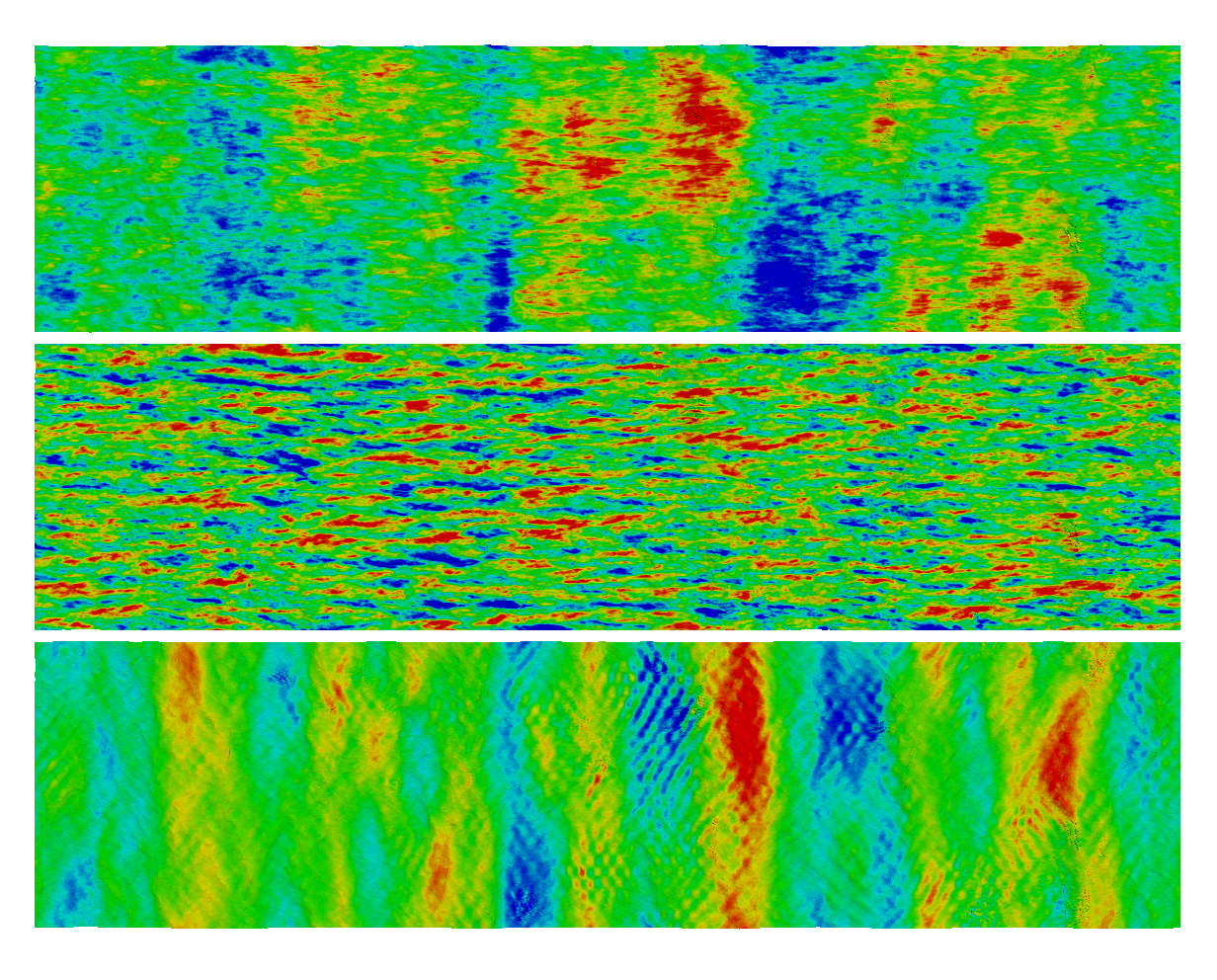} &
\includegraphics[width=0.30\linewidth]{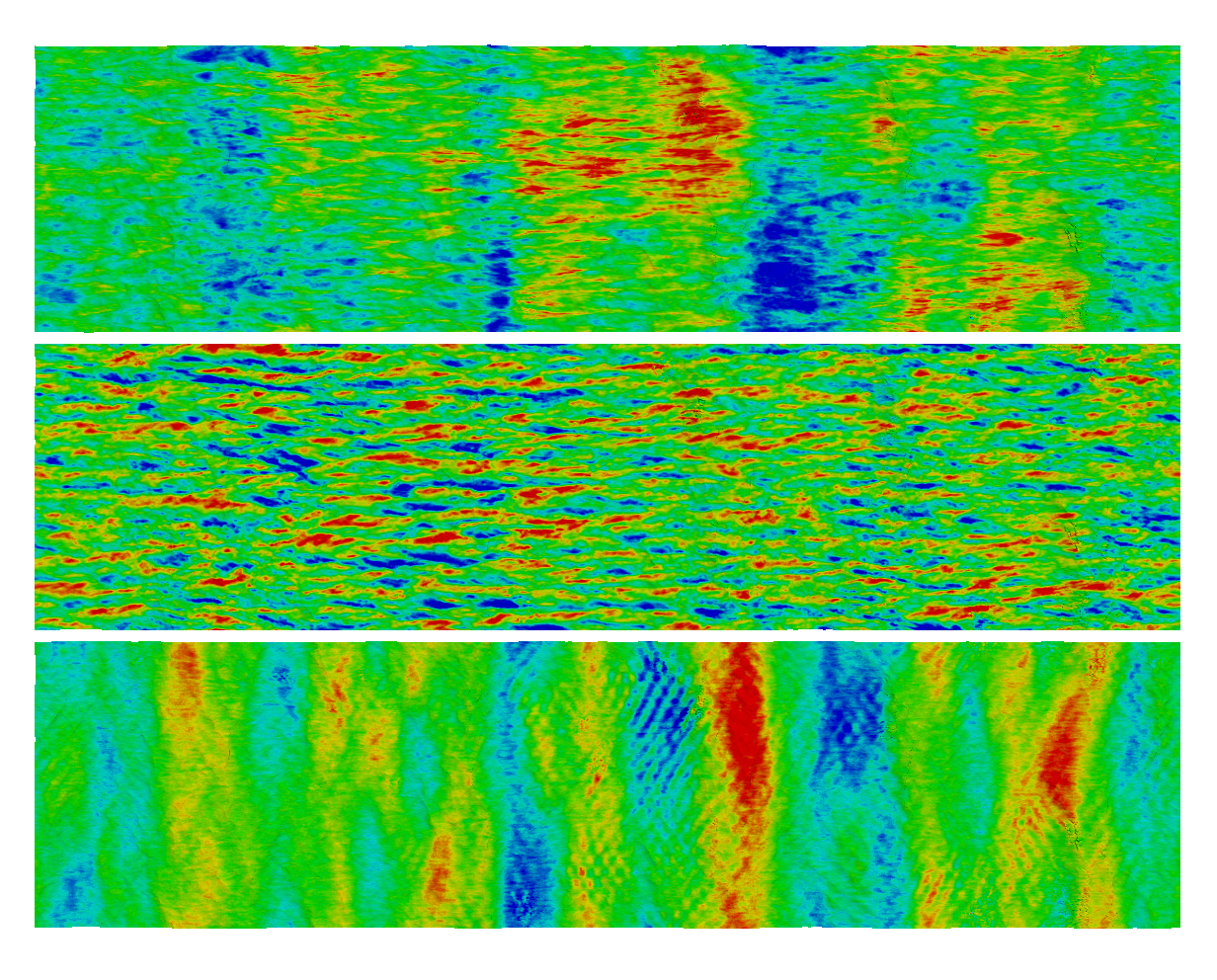} &
\includegraphics[width=0.30\linewidth]{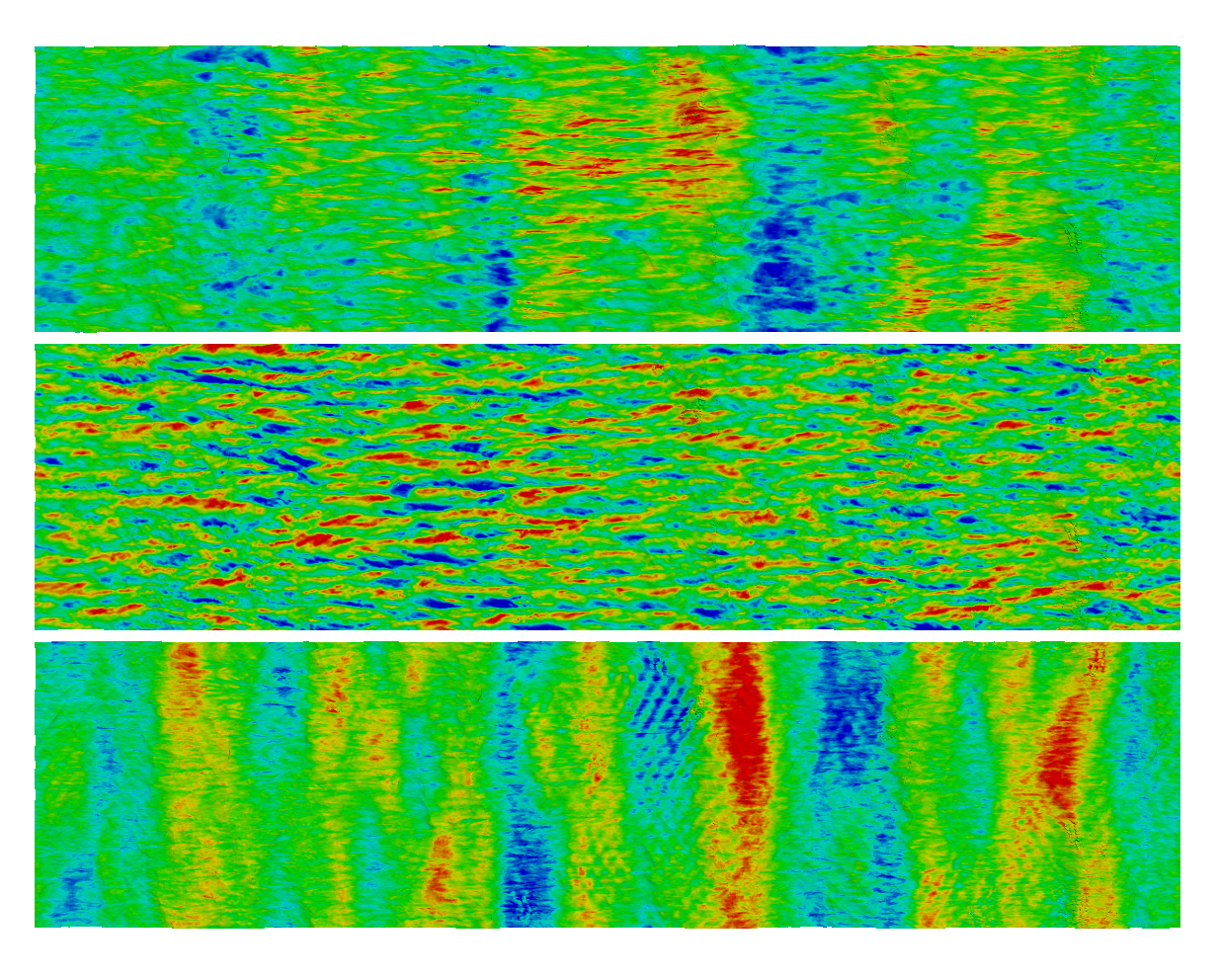} \\
(a) & (b) & (c) \\
\includegraphics[width=0.30\linewidth]{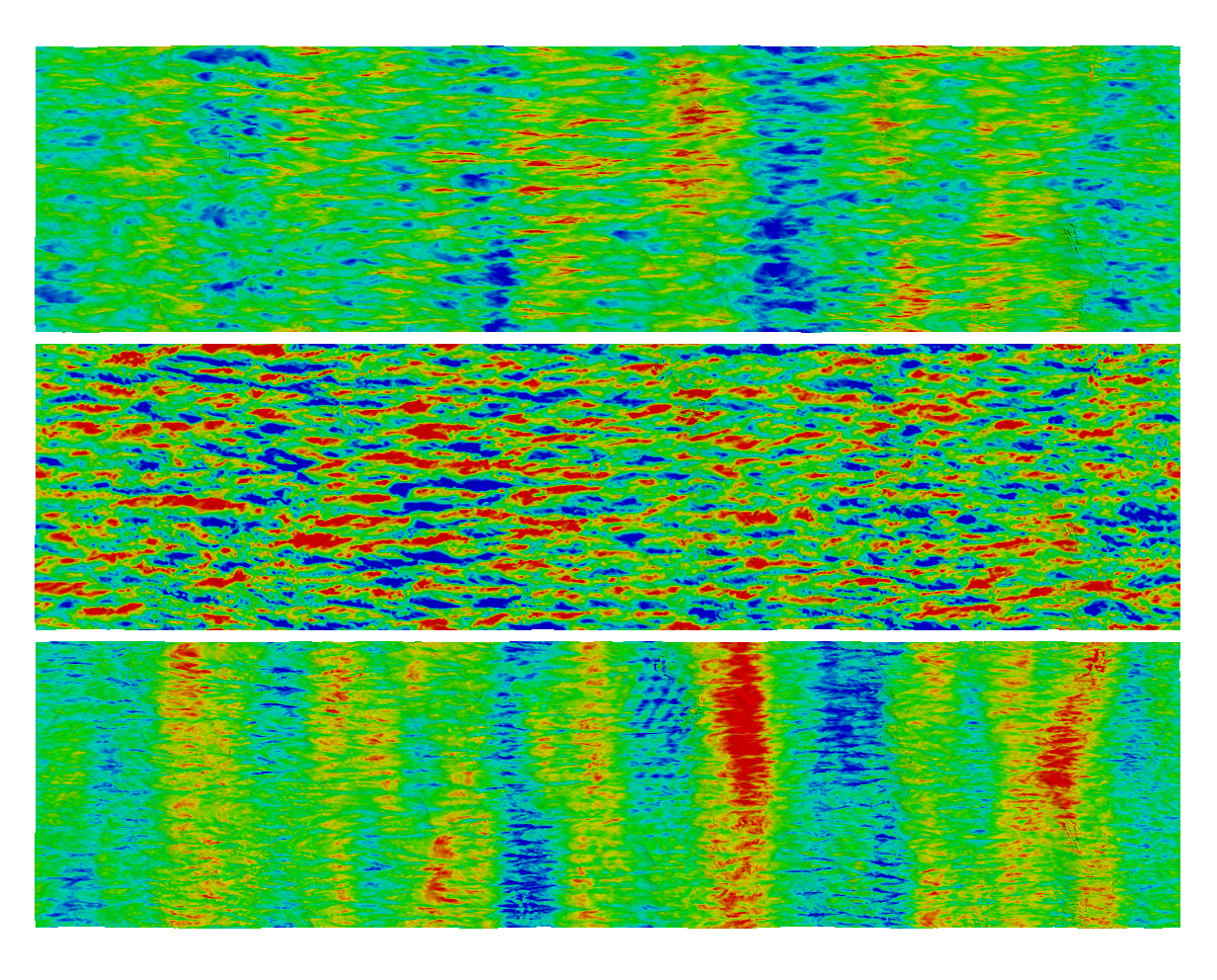} &
\includegraphics[width=0.30\linewidth]{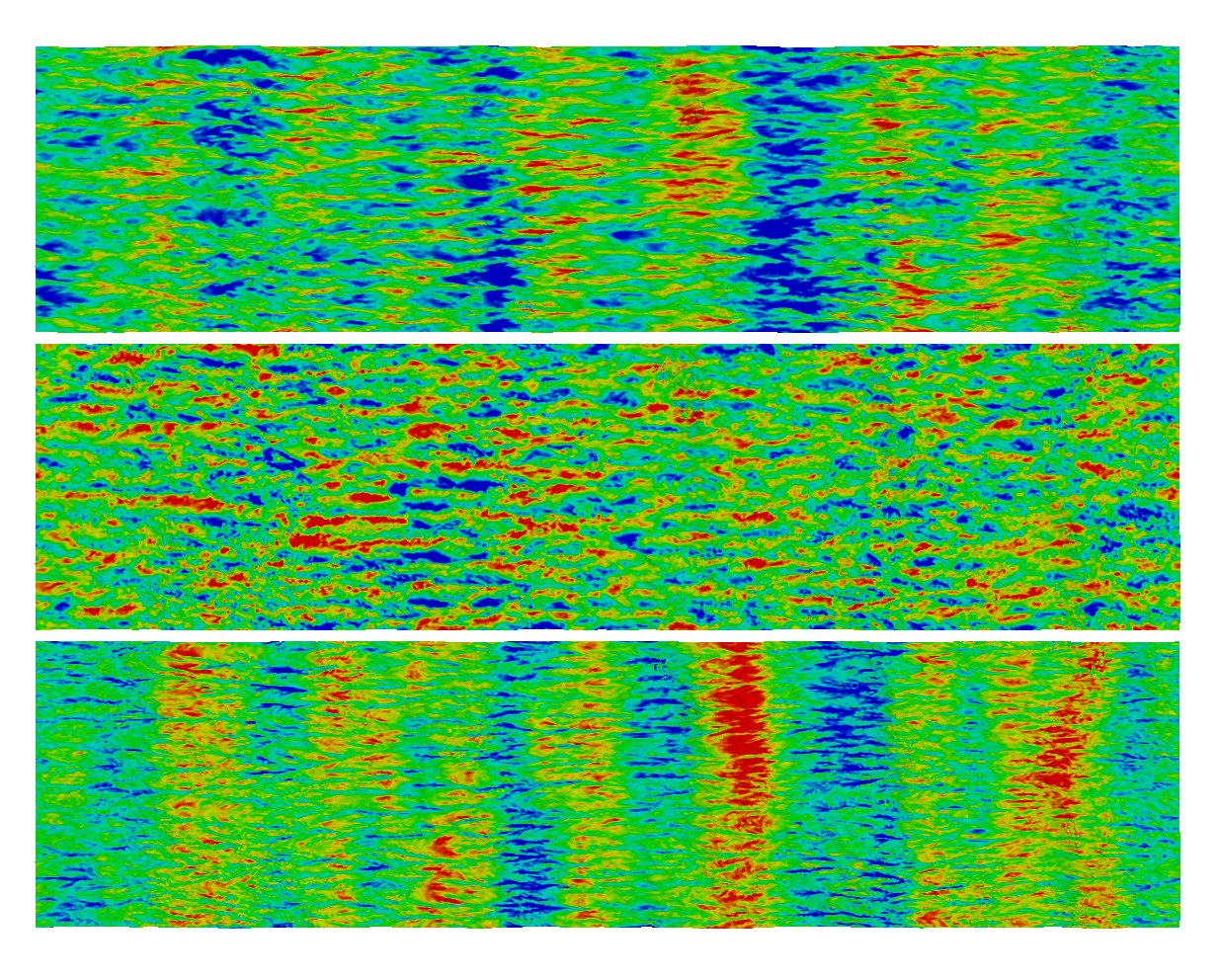} &
\includegraphics[width=0.30\linewidth]{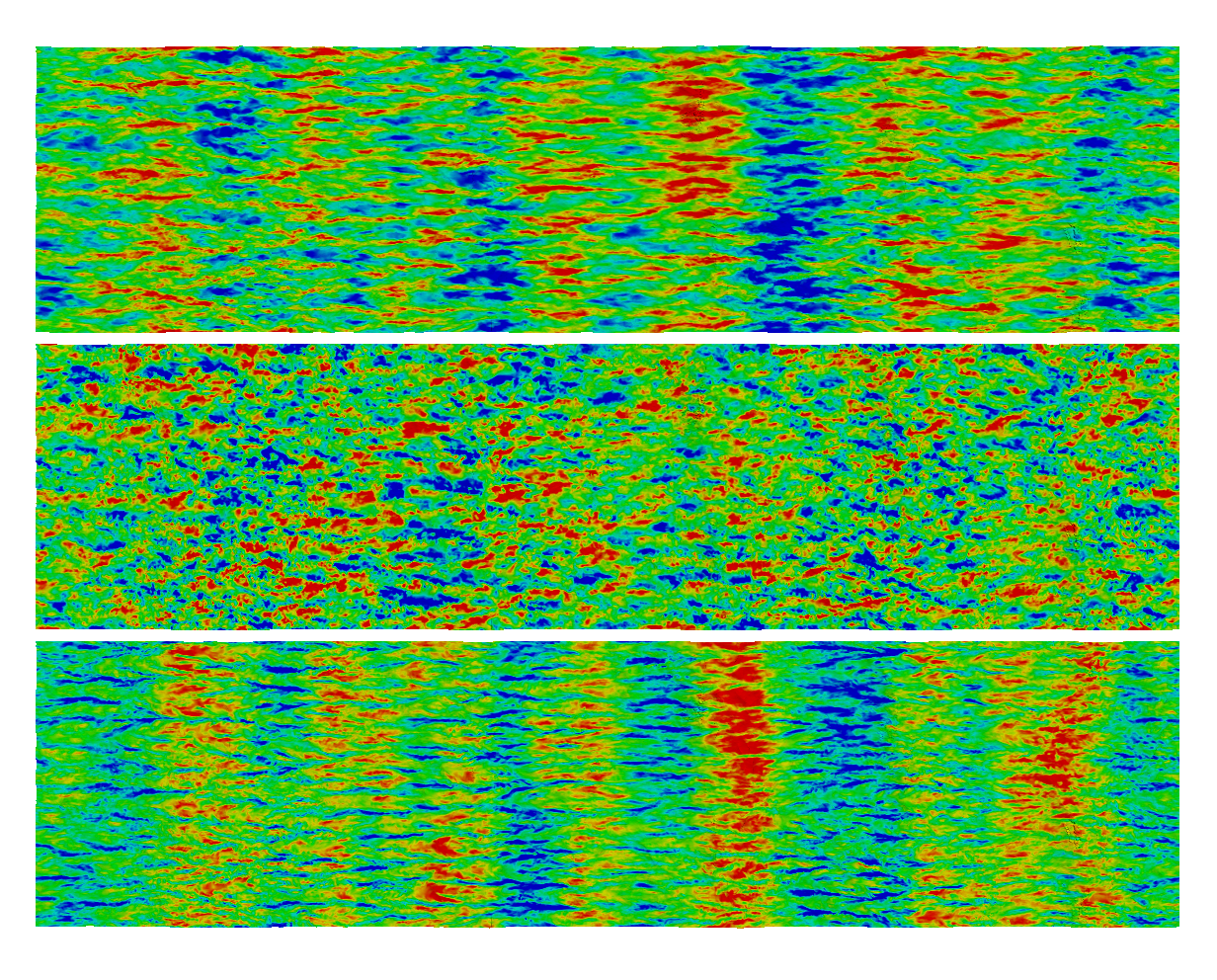} \\
(d) & (e) & (f) \\
\includegraphics[width=0.30\linewidth]{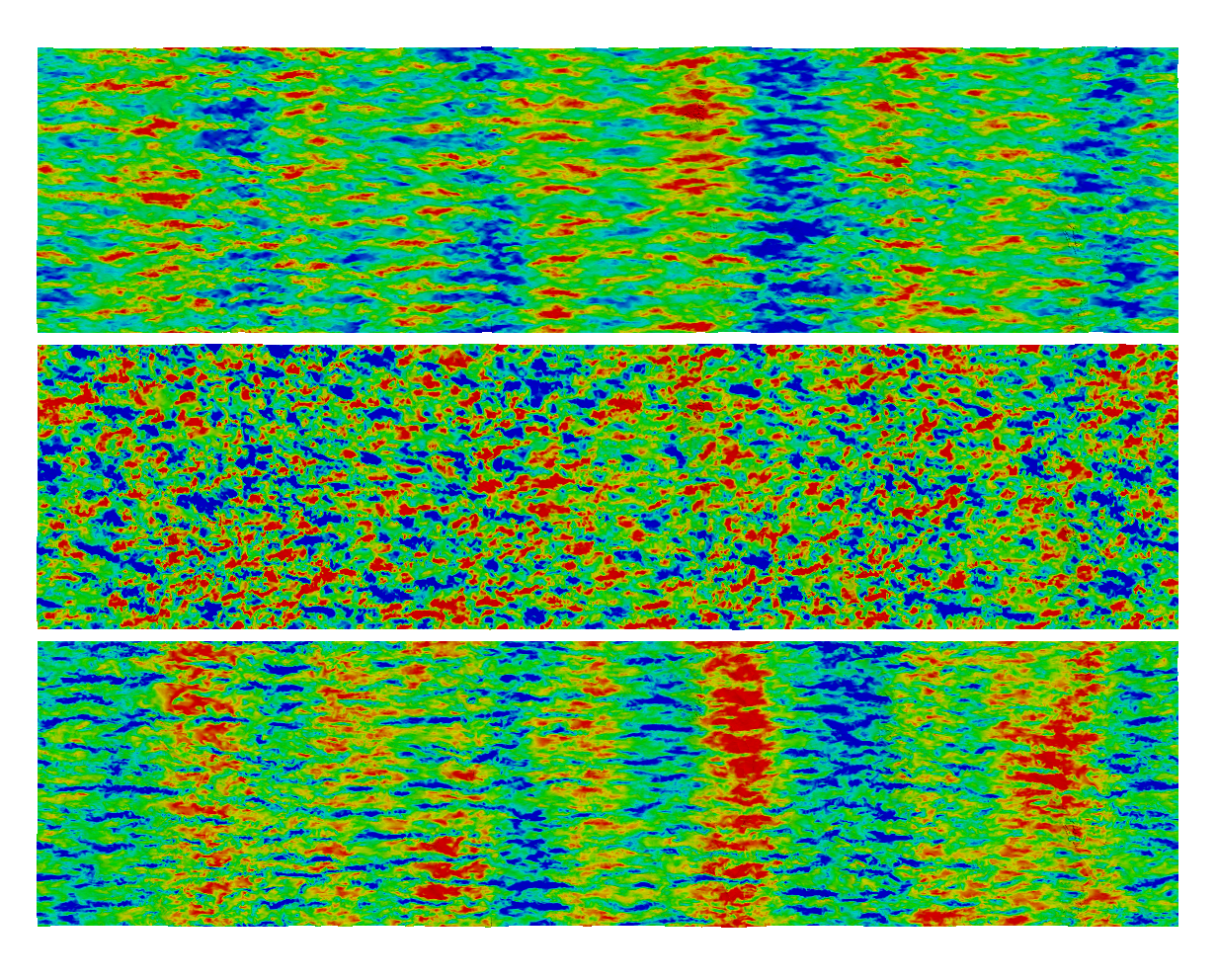} &
\includegraphics[width=0.30\linewidth]{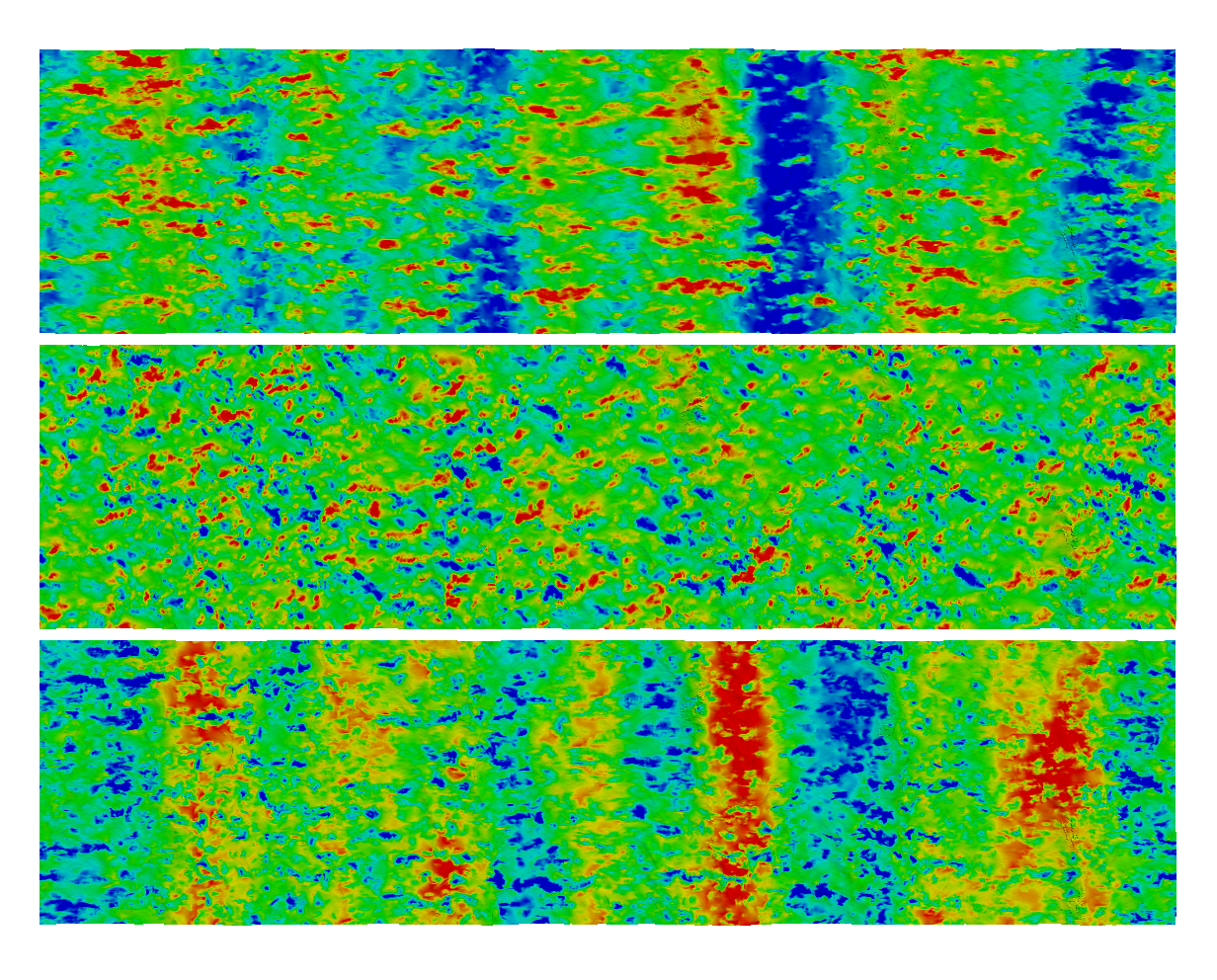} &
\includegraphics[width=0.30\linewidth]{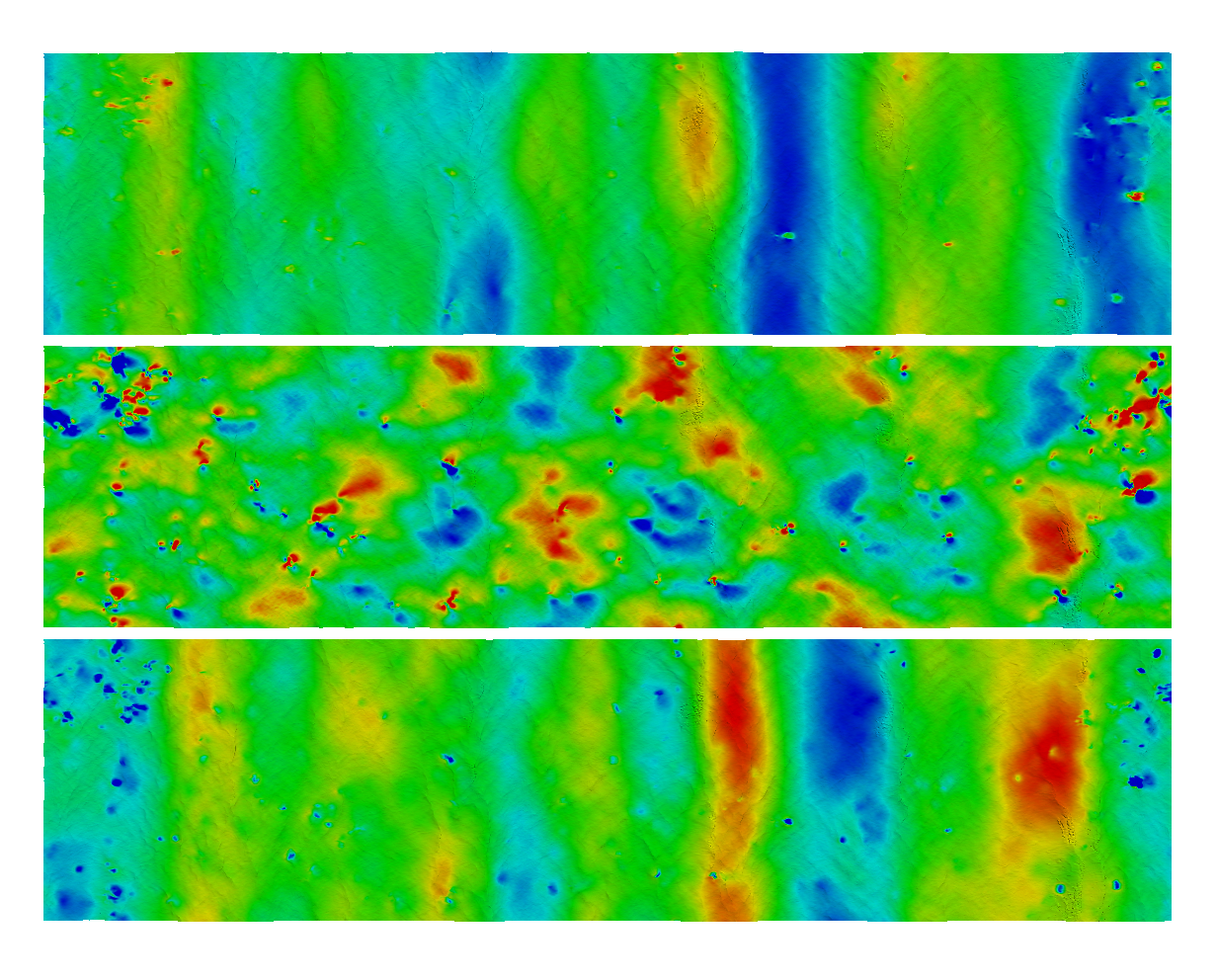} \\
(g) & (h) & (i) \\
\end{tabular}
\end{center}
\caption{\label{fig:langmuir} The structure of Langmuir cells at (a) \mbox{z=0}, (b) \mbox{z=-10 cm}, (c) \mbox{z=-20cm}, (d) \mbox{z=-40cm}, (e) \mbox{z=-80cm}, (f) \mbox{z=-1.6m}, (g) \mbox{z=-3.2m}, (h) \mbox{z=-6.4m}, and (i) \mbox{z=-12.8m}.  The $u$, $v$, and $w$ velocities are respectively shown in the top, middle, and bottom panels of each figure.  Animations of these results are available at (a) \mbox{\cite{velo1}} \href{http://youtu.be/gz8bAryvkUQ}{\mbox{z=0}}; (b) \mbox{\cite{velo2}} \href{http://youtu.be/0EOpjjBl3M4}{\mbox{z=-10cm}}; (c) \mbox{\cite{velo3}} \href{http://youtu.be/hJO8qjd6Rlw}{\mbox{z=-20cm}};  (d) \mbox{\cite{velo4}} \href{http://youtu.be/yNI2TlpCQnM}{\mbox{z=-40cm}}; (e) \mbox{\cite{velo5}} \href{http://youtu.be/9KSXAFmPbG8}{\mbox{z=-80cm}}; (f) \mbox{\cite{velo6}} \href{http://youtu.be/dCvulCJbDNY}{\mbox{z=-1.6m}}; (g) \mbox{\cite{velo7}} \href{http://youtu.be/iR1mPPKRoz8}{\mbox{z=-3.2m}}; (h) \mbox{\cite{velo8}} \href{http://youtu.be/CQje2hMpMs0}{\mbox{z=-6.4m}}; and (i) \mbox{\cite{velo9}} \href{http://youtu.be/TvD1SEX4H14}{\mbox{z=-12.8m}}.  The velocity ranges are 
(a) \mbox{-31.3cm/s $\leq u \leq$ 37.6cm/s}, \mbox{$|v| \leq$ 18.8cm/s}, \mbox{$|w| \leq$ 15.7cm/s};
(b) \mbox{-31.3cm/s $\leq u \leq$ 37.6cm/s}, \mbox{$|v| \leq$  18.8cm/s}, \mbox{$|w| \leq$ 15.7cm/s};
(c) \mbox{-31.3cm/s $\leq u \leq$ 37.6cm/s}, \mbox{$|v| \leq$  18.8cm/s}, \mbox{$|w| \leq$ 15.7cm/s}; 
(d) \mbox{-25.1cm/s $\leq u \leq$ 31.3cm/s}, \mbox{$|v| \leq$ 12.5cm/s}, \mbox{$|w| \leq$ 15.7cm/s};
(e) \mbox{-18.7cm/s $\leq u \leq$ 25.1cm/s}, \mbox{$|v| \leq$  12.5cm/s}, \mbox{$|w| \leq$  15.7cm/s}; 
(f) \mbox{$|u| \leq$ 18.8cm/s}, \mbox{$|v| \leq$ 9.4cm/s}, \mbox{$|w| \leq$  15.7cm/s};
(g) \mbox{-15.7cm/s $\leq u \leq$ 18.8cm/s}, \mbox{$|v| \leq$ 6.3cm/s}, \mbox{$|w| \leq$ 12.5cm/s}; 
(h) \mbox{-9.4cm/s $\leq u \leq$ 12.5cm/s}, \mbox{$|v| \leq$ 6.3cm/s}, \mbox{$|w| \leq$ 9.4cm/s}; and
(i) \mbox{-6.3cm/s $\leq u \leq$ 9.4cm/s}, \mbox{$|v| \leq$ 1.6cm/s}, \mbox{$|w| \leq$ 6.3cm/s}. 
The colors for values outside the ranges are saturated. Mean values are subtracted out.}
\end{figure*}

\begin{figure*}
\begin{center}
\begin{tabular}{ccc}
\includegraphics[width=0.30\linewidth]{./figures30th/structures/velo_1.png} &
\includegraphics[width=0.30\linewidth]{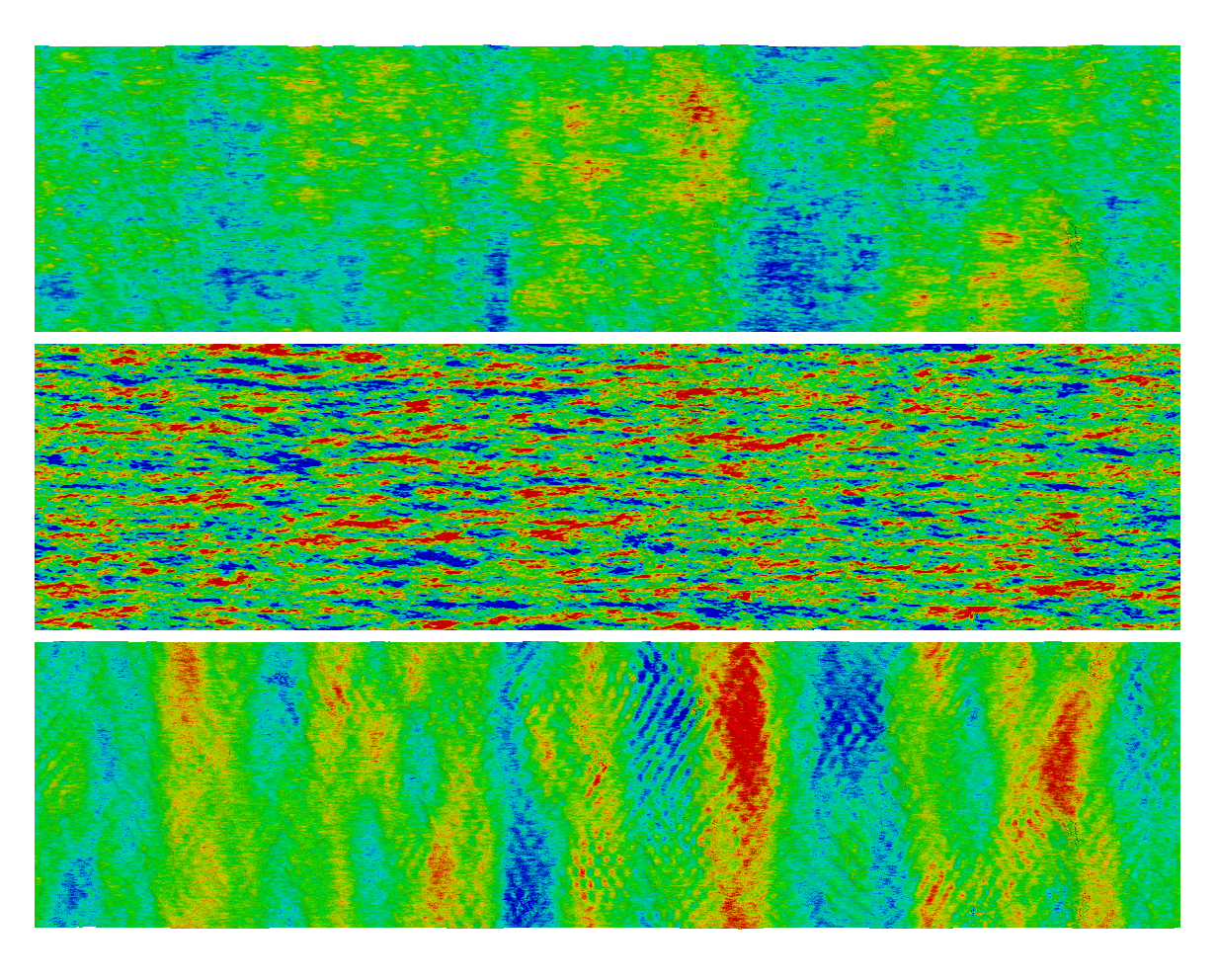} &
\includegraphics[width=0.30\linewidth]{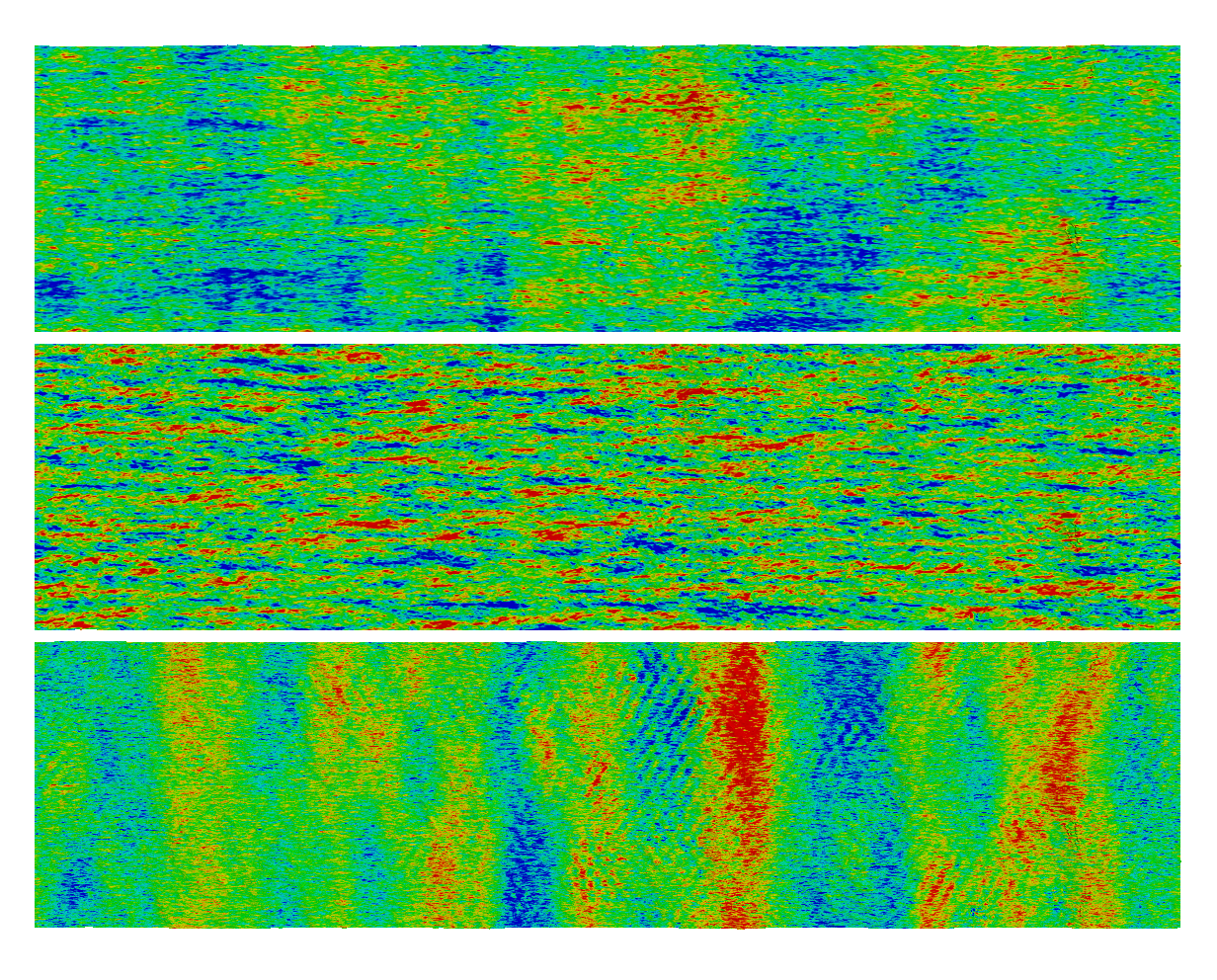} \\
(a) & (b) & (c) \\
\includegraphics[width=0.30\linewidth]{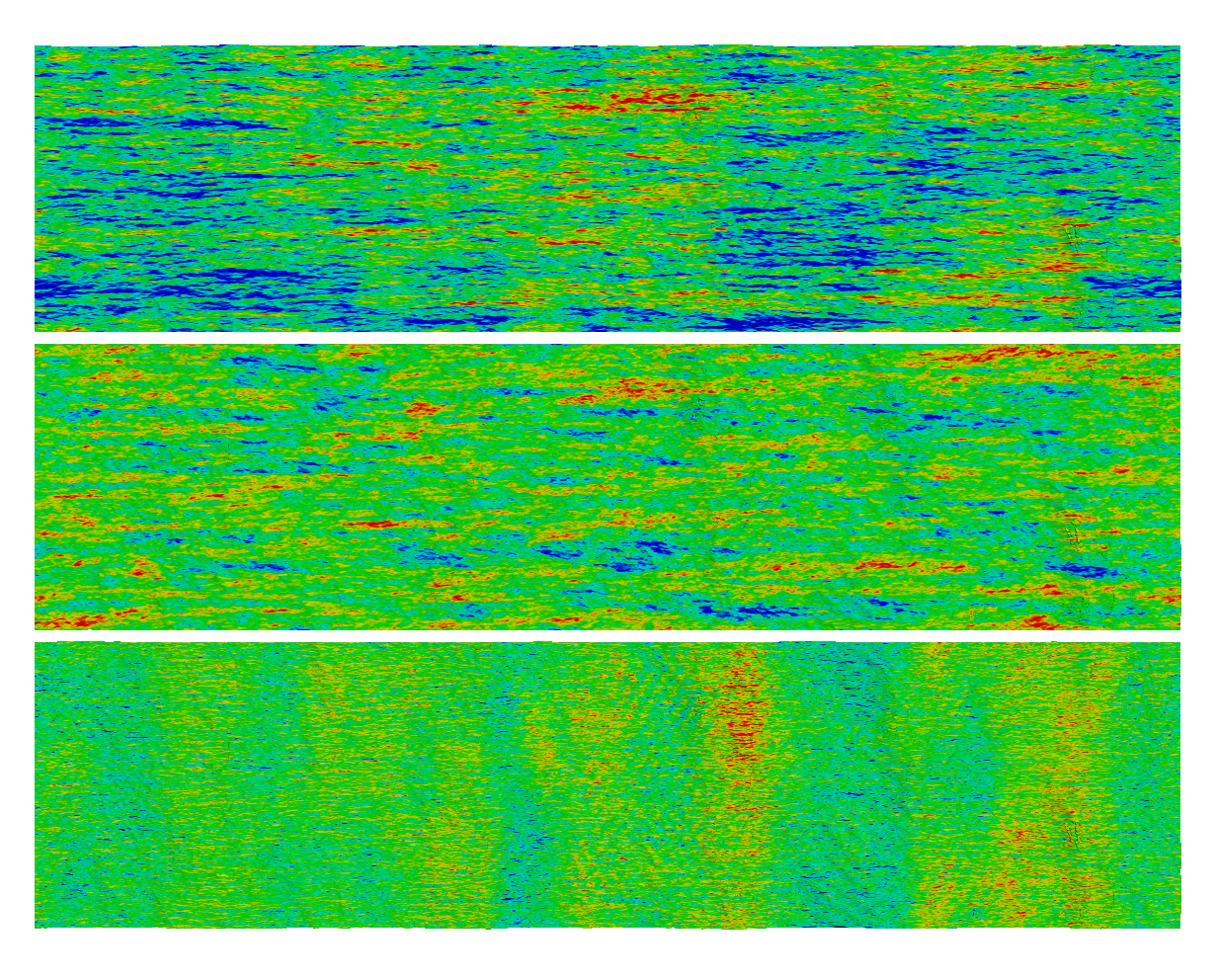} &
\includegraphics[width=0.30\linewidth]{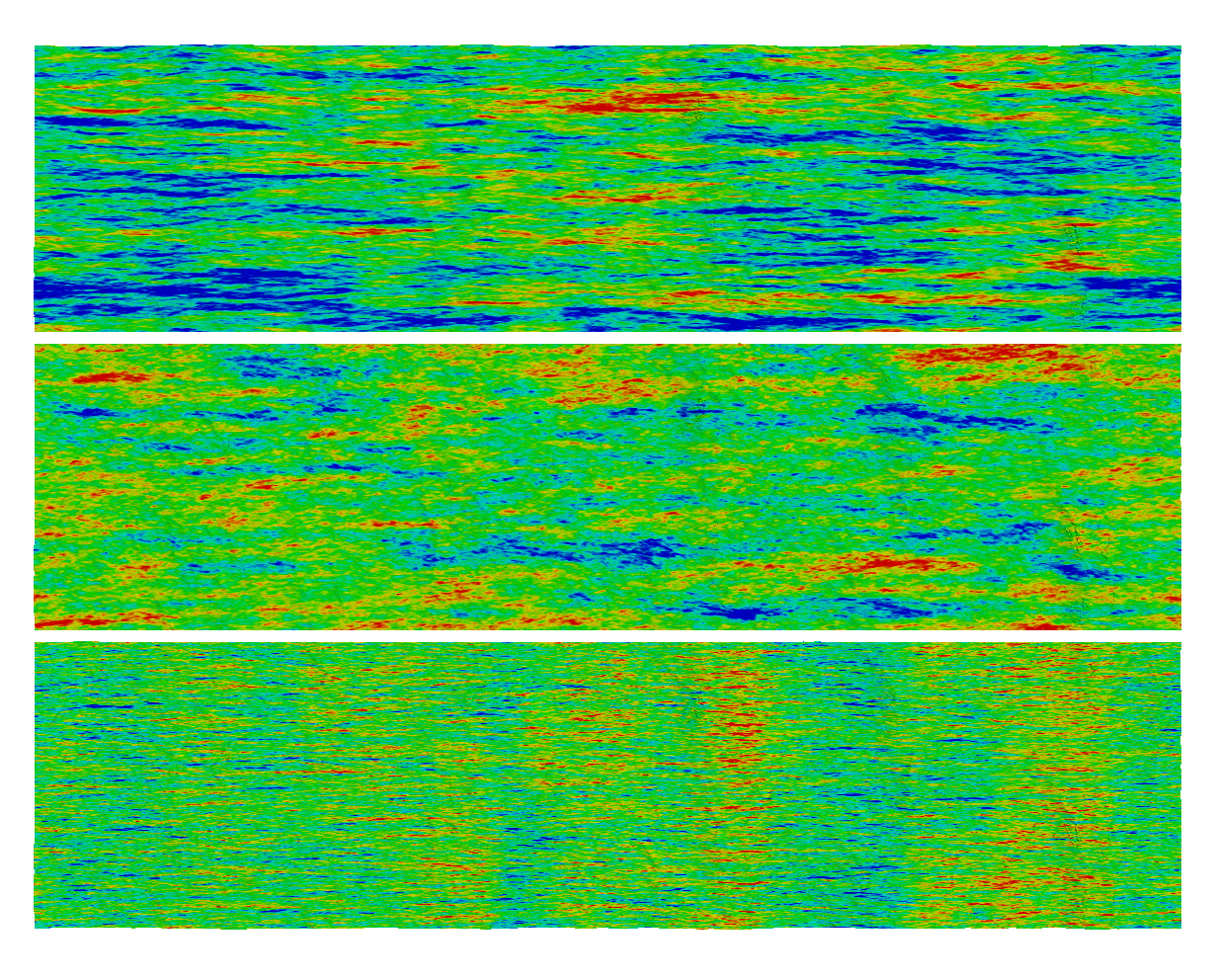} &
\includegraphics[width=0.30\linewidth]{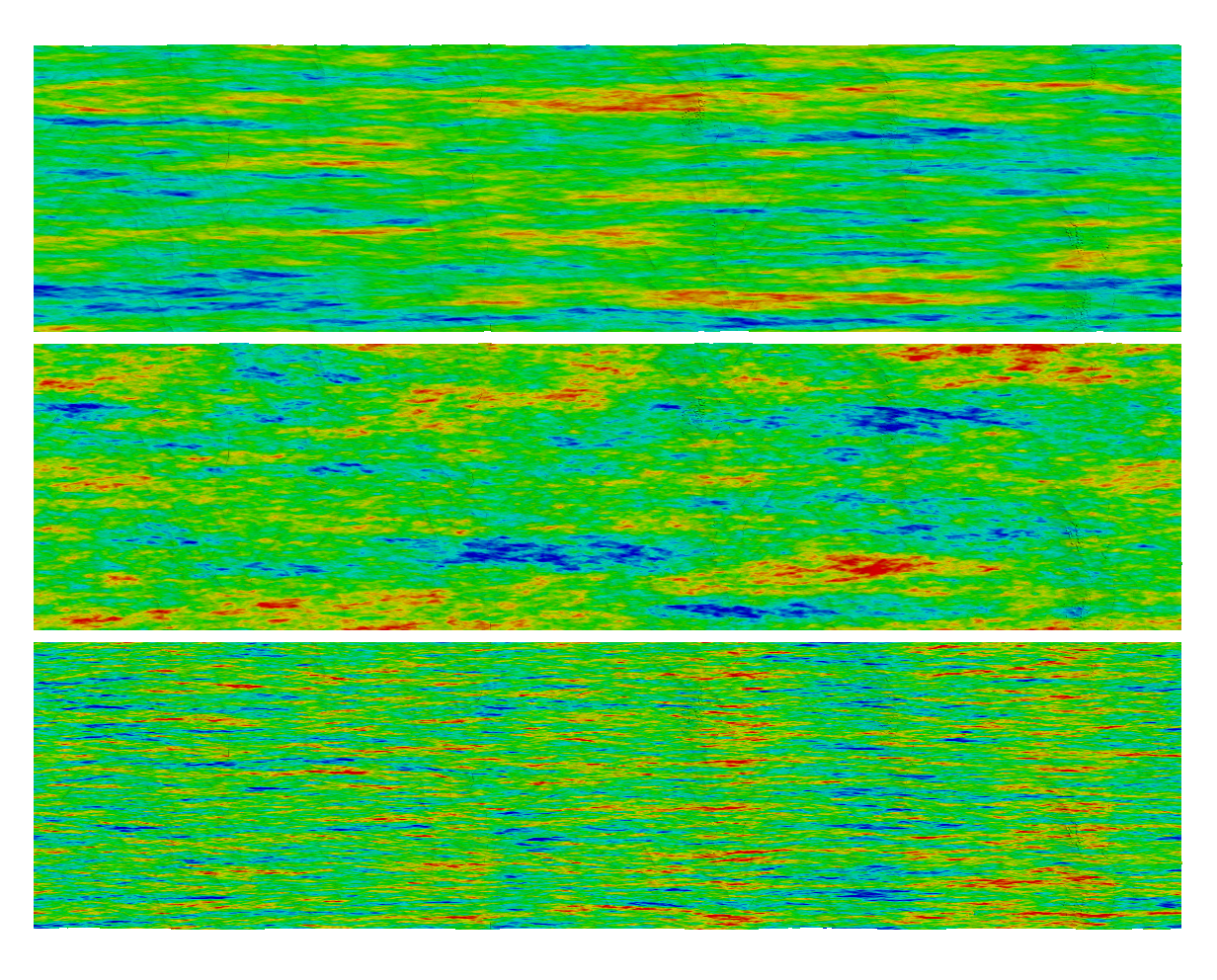} \\
(d) & (e) & (f) \\
\includegraphics[width=0.30\linewidth]{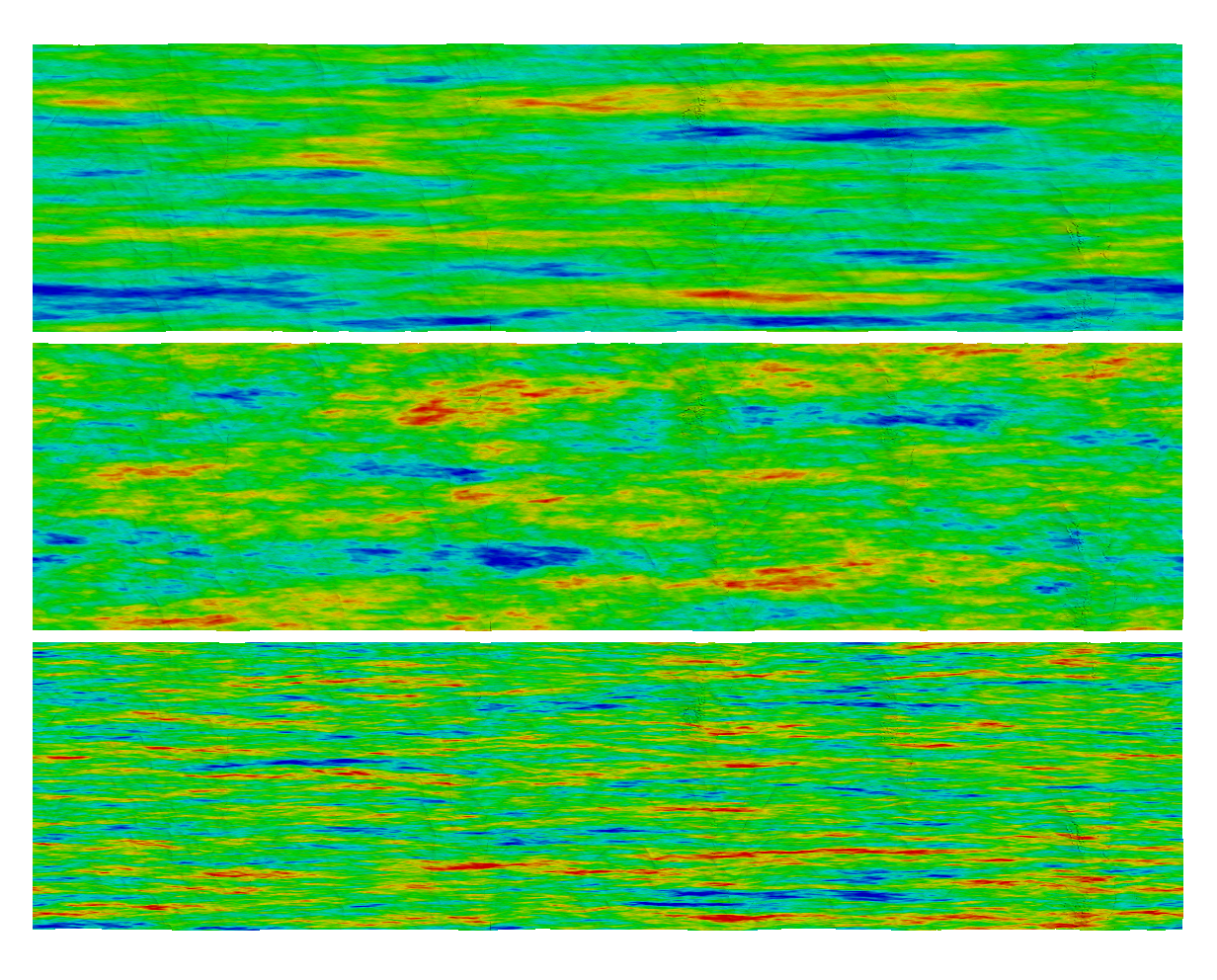} &
\includegraphics[width=0.30\linewidth]{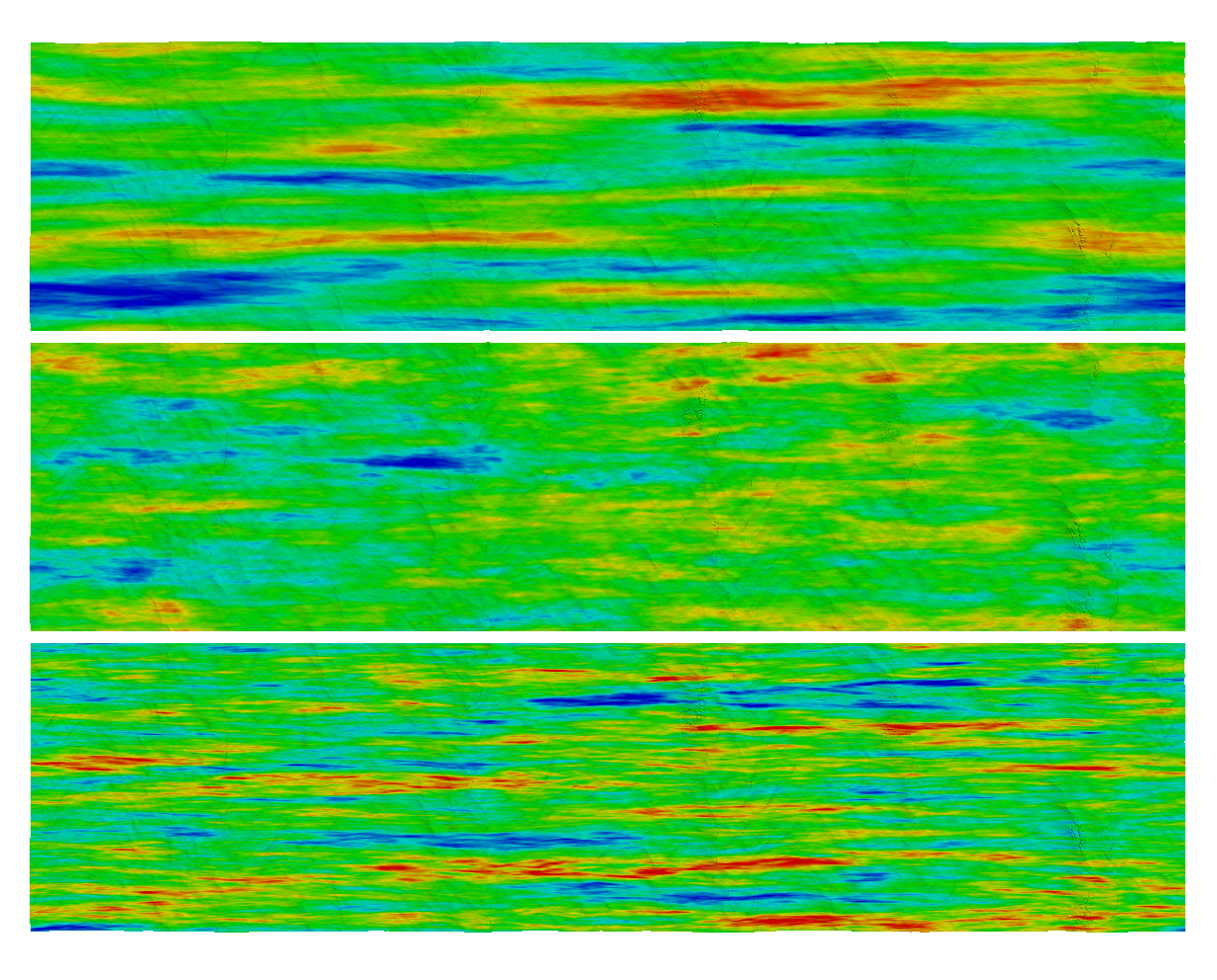} &
\includegraphics[width=0.30\linewidth]{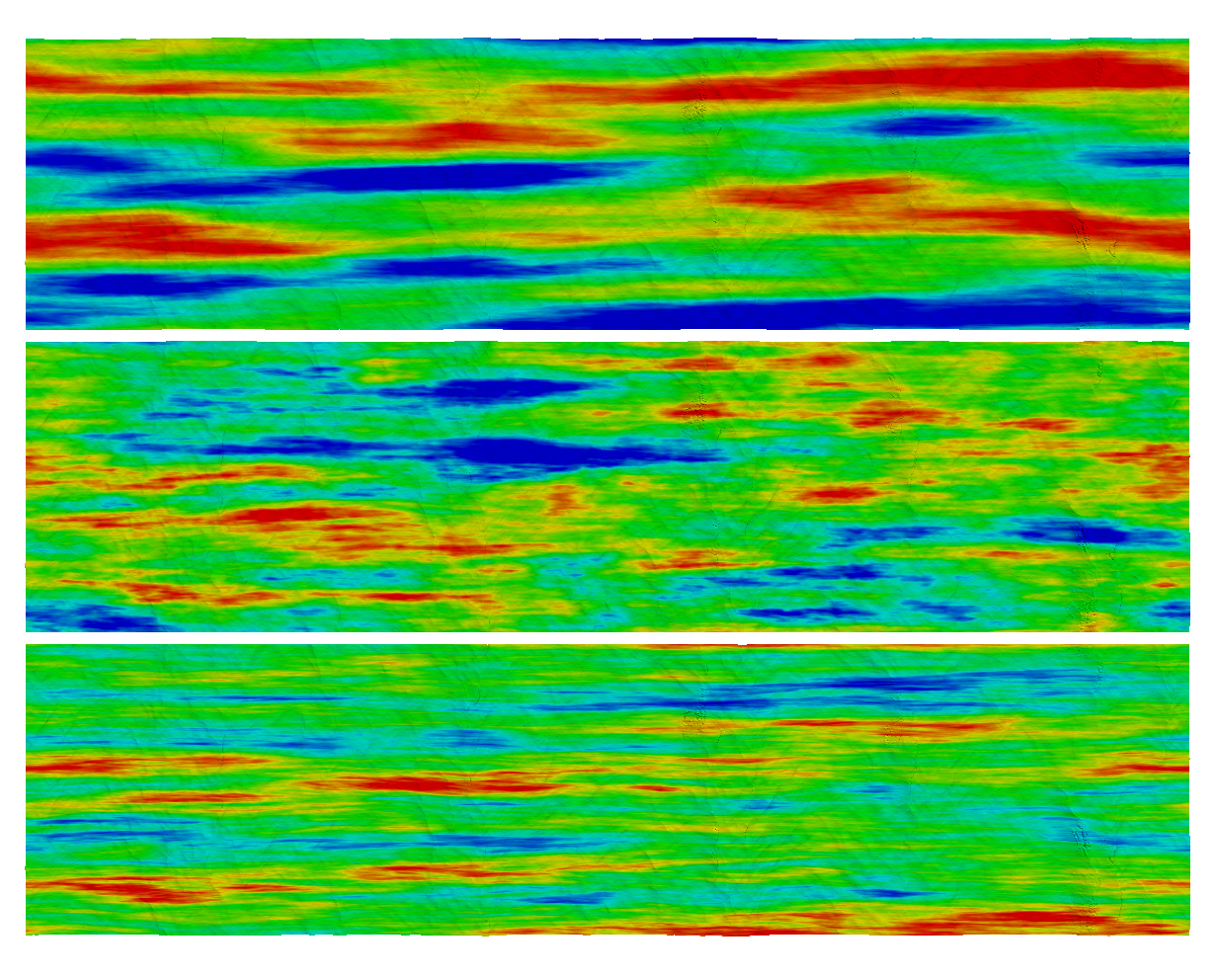} \\
(g) & (h) & (i) \\
\end{tabular}
\end{center}
\caption{\label{fig:windstreaks1} The structure of wind streaks at (a) \mbox{z=0}, (b) \mbox{z=10 cm}, (c) \mbox{z=20cm}, (d) \mbox{z=40cm}, (e) \mbox{z=80cm}, (f) \mbox{z=1.6m}, (g) \mbox{z=3.2m}, (h) \mbox{z=6.4m}, and (i) \mbox{z=12.8m}. The $u$, $v$, and $w$ velocities are respectively shown in the top, middle, and bottom panels of each figure.  Animations of these results are available at (a) \mbox{\cite{velo1}} \href{http://youtu.be/gz8bAryvkUQ}{\mbox{z=0}};  (b) \mbox{\cite{velo10}} \href{http://youtu.be/t-zYofbawcw}{\mbox{z=10cm}}; (c) \mbox{\cite{velo11}} \href{http://youtu.be/xRxjqpdB2P8}{\mbox{z=20cm}};  (d) \mbox{\cite{velo12}} \href{http://youtu.be/vCRokJx\_Emw}{\mbox{z=40cm}};  (e) \mbox{\cite{velo13}} \href{http://youtu.be/-f6CfR\_lx9s}{\mbox{z=80cm}}; (f) \mbox{\cite{velo14}} \href{http://youtu.be/YbfKpxd3VYA}{\mbox{z=1.6m}}; (g) \mbox{\cite{velo15}} \href{http://youtu.be/Z8JRTIJcX-c}{\mbox{z=3.2m}}; (h) \mbox{\cite{velo16}} \href{http://youtu.be/r8cwATgsMf4}{\mbox{z=6.4m}}; and (i) \mbox{\cite{velo17}} \href{http://youtu.be/zpjbr947YKg}{\mbox{z=12.8m}}. The velocity ranges are 
(a) \mbox{-31.3cm/s $\leq u \leq$ 37.6cm/s}, \mbox{$|v| \leq$ 18.8cm/s}, \mbox{$|w| \leq$ 15.7cm/s};
(b) \mbox{-47.0cm/s $\leq u \leq$ 62.6cm/s}, \mbox{$|v| \leq$ 18.8cm/s}, \mbox{$|w| \leq$ 15.7cm/s};
(c) \mbox{-47.0cm/s $\leq u \leq$ 62.6cm/s}, \mbox{$|v| \leq$ 25.1cm/s}, \mbox{$|w| \leq$ 15.7cm/s};
(d) \mbox{-62.6cm/s $\leq u \leq$ 93.9cm/s}, \mbox{$|v| \leq$ 47.0cm/s}, \mbox{$|w| \leq$ 31.3cm/s};
(e) \mbox{-62.6cm/s $\leq u \leq$ 93.9cm/s}, \mbox{$|v| \leq$ 47.0cm/s}, \mbox{$|w| \leq$ 31.3cm/s};
(f) \mbox{-1.09m/s $\leq u \leq$ 1.25m/s}, \mbox{$|v| \leq$ 47.0cm/s}, \mbox{$|w| \leq$ 31.3cm/s};
(g) \mbox{-1.09m/s $\leq u \leq$ 1.41m/s}, \mbox{$|v| \leq$ 47.0cm/s}, \mbox{$|w| \leq$ 31.3cm/s};
(h) \mbox{-1.09m/s $\leq u \leq$ 1.25m/s}, \mbox{$|v| \leq$ 47.0cm/s}, \mbox{$|w| \leq$ 31.3cm/s}; and
(i) \mbox{$|u| \leq$ 78.3cm/s}, \mbox{$|v| \leq$ 25.1cm/s}, \mbox{$|w| \leq$ 31.3cm/s}.
The colors for values outside the ranges are saturated. Mean values are subtracted out.}
\end{figure*}

\begin{figure}
\begin{center}
\begin{tabular}{c}
\includegraphics[trim=30mm 18mm 35mm 38mm,clip=true,angle=0,width=\linewidth]{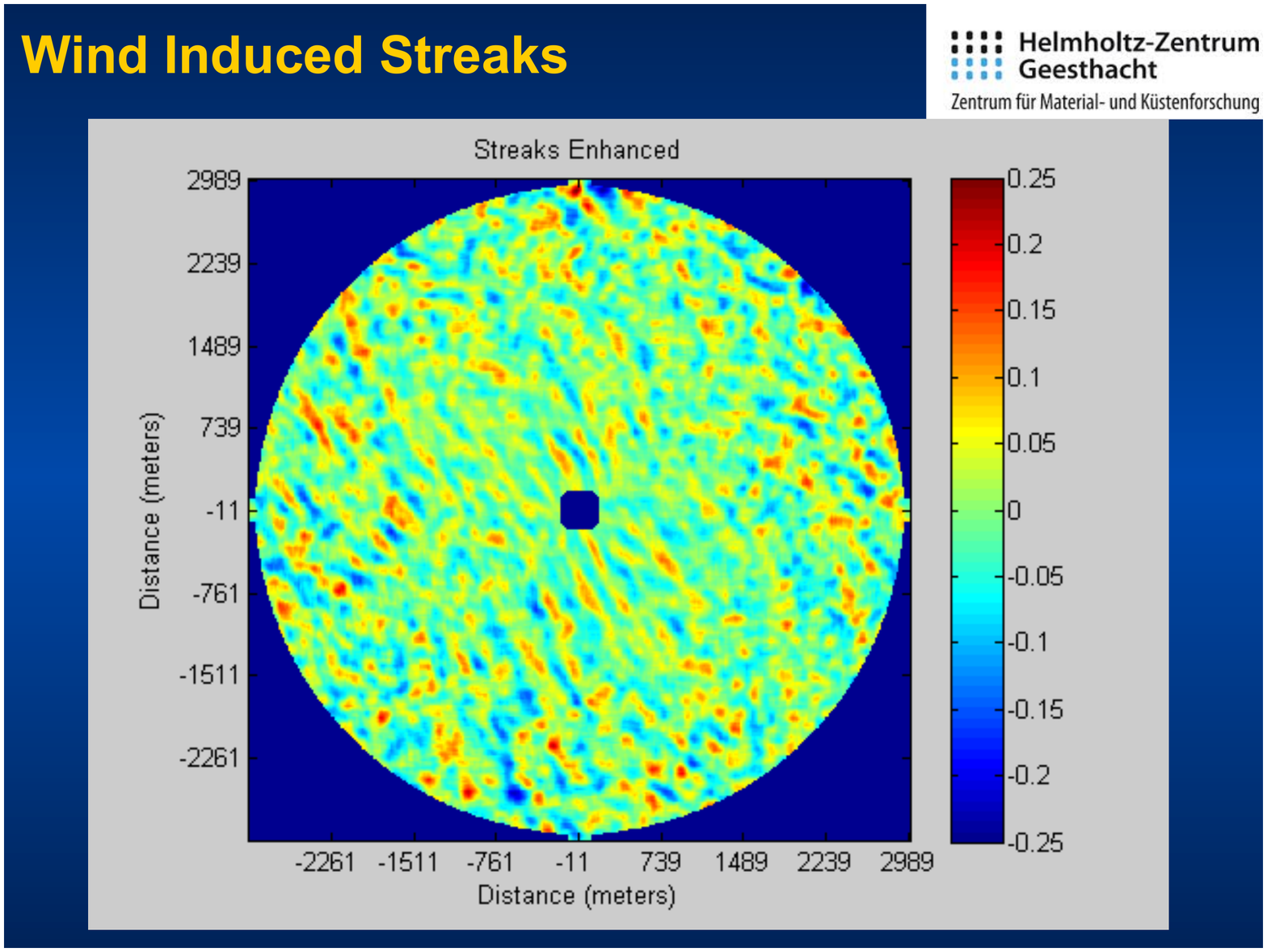}
\end{tabular}
\end{center}
\caption{\label{fig:windstreaks2} Wind-induced streaks observed using an X-band marine radar operating at grazing incidence.   The image shows the normalized mean of  approximately 60 seconds of radar data (approximately 30 individual images) after removing the range and azimuth dependence.  This figure is courtesy of Dr.\  Jochen Horstmann, Helmhotz-Zentrum Geesthacht, Germany \mbox{\citep{horstmann2014}}.}
\end{figure}

\subsection{\label{sec:profiles}Vertical profiles of mean quantities}

Figures \ref{fig:crosswind}a and b show the vertical profiles of the crosswind meandering and the vertical streaming in the lower ABL.   The vertical profiles are expressed in terms of a free-surface-following coordinate system based on equation \ref{eq:profile1}.   Mean quantities are calculated based on equation \ref{eq:profile2}.  The mean quantities are calculated in terms of the total velocity field, including the wavy and vortical portions of the flow.    The results are based on the coarse assimilation.    The animations are particularly useful for observing the frequency content and the vertical diffusion.    

The crosswind meandering and the vertical streaming develop as the vortical portion of the flow diffuses upward.   Based on Figure  \ref{fig:crosswind}a, the angle of attack of the wind changes as function of time and distance above the free surface.  The frequency content is higher closer to free surface than it is away from the free surface.   The meandering of the crosswind extends to the top of the domain at $z=50 \; {\rm m}$.  In Figure  \ref{fig:crosswind}b,  the spatially-averaged mean velocity of the wind in the vertical direction is directed downward toward the free surface 4 to \mbox{5 m} above the free surface, and a sharp gradient in the mean velocity is located within one meter of the free surface. The two peaks are associated with making the wind parallel to the free surface.   Vertical streaming is evident in the animation of Figure  \ref{fig:crosswind}b, especially toward the end.  The meandering of the crosswind and the vertical streaming are very large-scale effects that are difficult to quantify using field measurements.

Figures \ref{fig:crossdrift}a and b show the vertical profiles of the cross drift meandering and the vertical streaming in the upper OBL.   The calculation of the mean quantities in the OBL is similar to the calculation in the ABL.   As for the crosswind, the mean cross drift and vertical streaming in the OBL are calculated in terms of the total velocity field, including the wavy and vortical portions of the flow. The results are based on the coarse assimilation.   

The meandering of the cross drift in combination with the meandering of the crosswind may promote angular spreading of the waves.   The angular spreading may be self-regulating because as the waves spread, the formation of large-scale vortices in the atmosphere and the ocean would be less coherent.  Spreading due to the formation of large-scale vortices occurs in combination with spreading due to nonlinear wave interactions.  As discussed in the preceding section, the cross drift and crosswind meandering are likely related to the formation of Y-junctions and interweaving of streaks.   The meandering of the cross drift penetrates down to $z=12.5 \; {\rm m}$ within 4 minutes and is still deepening at the end of the assimilation.   The large variations in the vertical streaming in the OBL that are observed in the animation are  due to forcing the wind drift to be parallel to the free surface.

Figure \ref{fig:stokesdrift} shows the Stokes drift in the water and the air.  The calculation of the Stokes drift is based on the wavy portion of the flow using results from the coarse assimilation.   A link to an animation is provided in the figure caption.   Based on the animation, the variation of the Stokes drift as a function of time is minimal.  The Stokes drift is symmetric across the free surface.  The amplitude of the Stokes drift is much less than the water-particle velocity in the crest of a breaking wave, which is equal to the phase speed.

\begin{figure*}
\begin{center}
\begin{tabular}{cc}
\includegraphics[trim=14mm 65mm 22mm 65mm,clip=true,angle=0,width=0.45\linewidth]{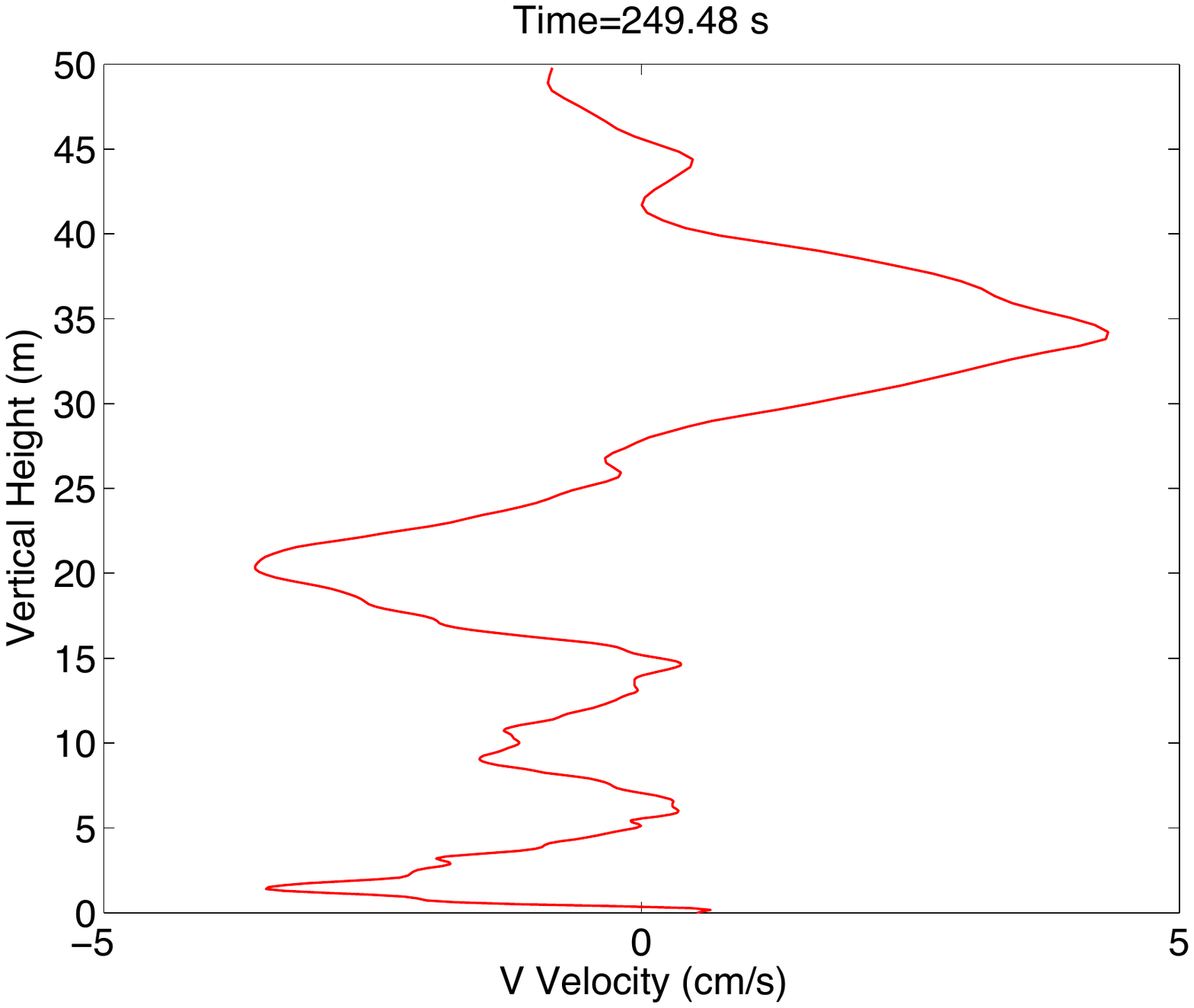}  &
\includegraphics[trim=14mm 65mm 22mm 65mm,clip=true,angle=0,width=0.45\linewidth]{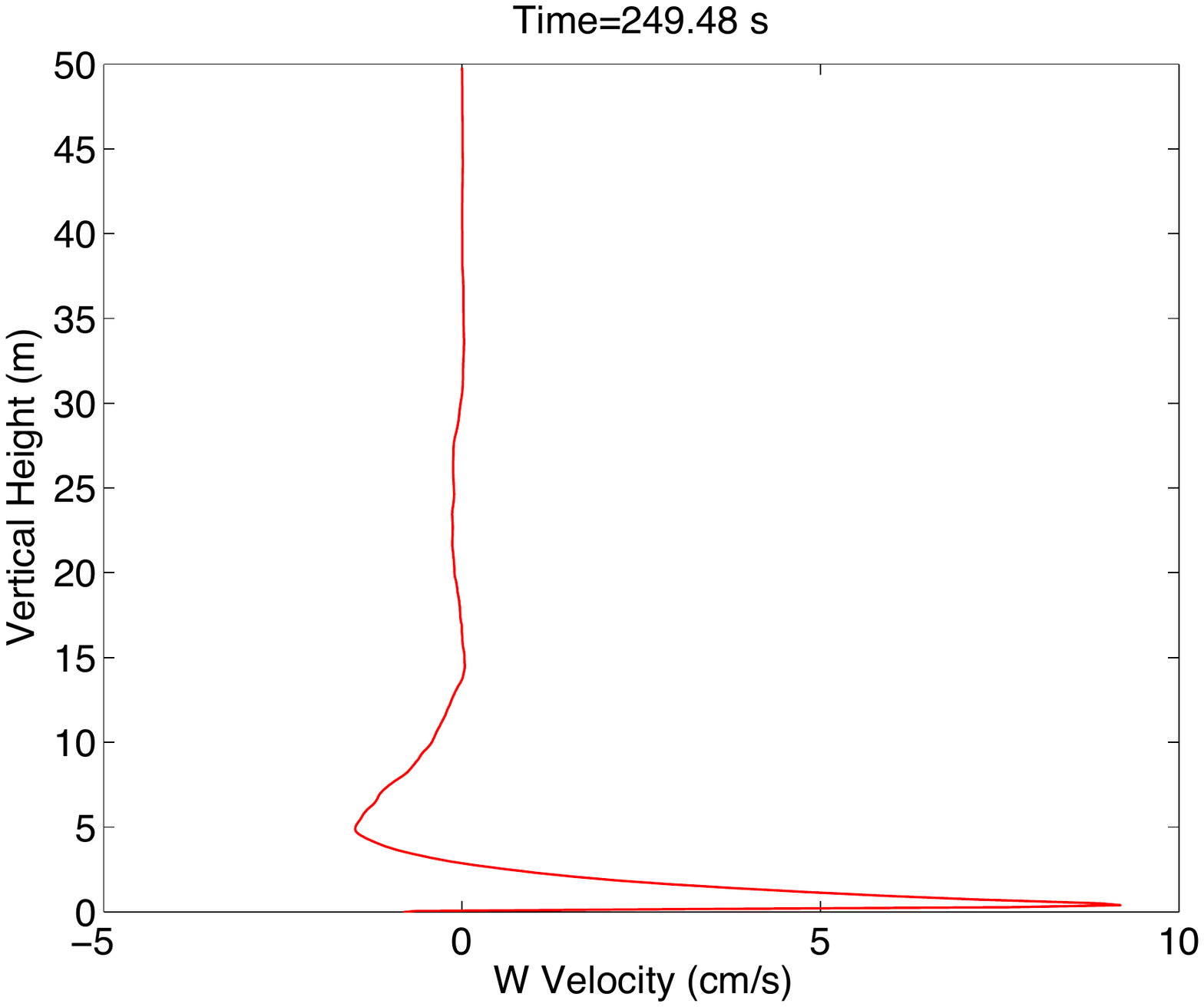} \\
(a) & (b)
\end{tabular}
\end{center}
\caption{\label{fig:crosswind} Crosswind meandering and vertical streaming in lower atmosphere.    (a) crosswind.  (b) vertical streaming.  Animations of these results are available at \mbox{\cite{crosswind}} \href{http://youtu.be/SZmlWlb4AQw}{crosswind meandering in the lower atmosphere} and \mbox{\cite{streamingair}} \href{http://youtu.be/7HUz0tagF-A}{vertical streaming flow in the lower atmosphere}.}
\end{figure*}

\begin{figure*}
\begin{center}
\begin{tabular}{cc}
\includegraphics[trim=14mm 65mm 22mm 65mm,clip=true,angle=0,width=0.45\linewidth]{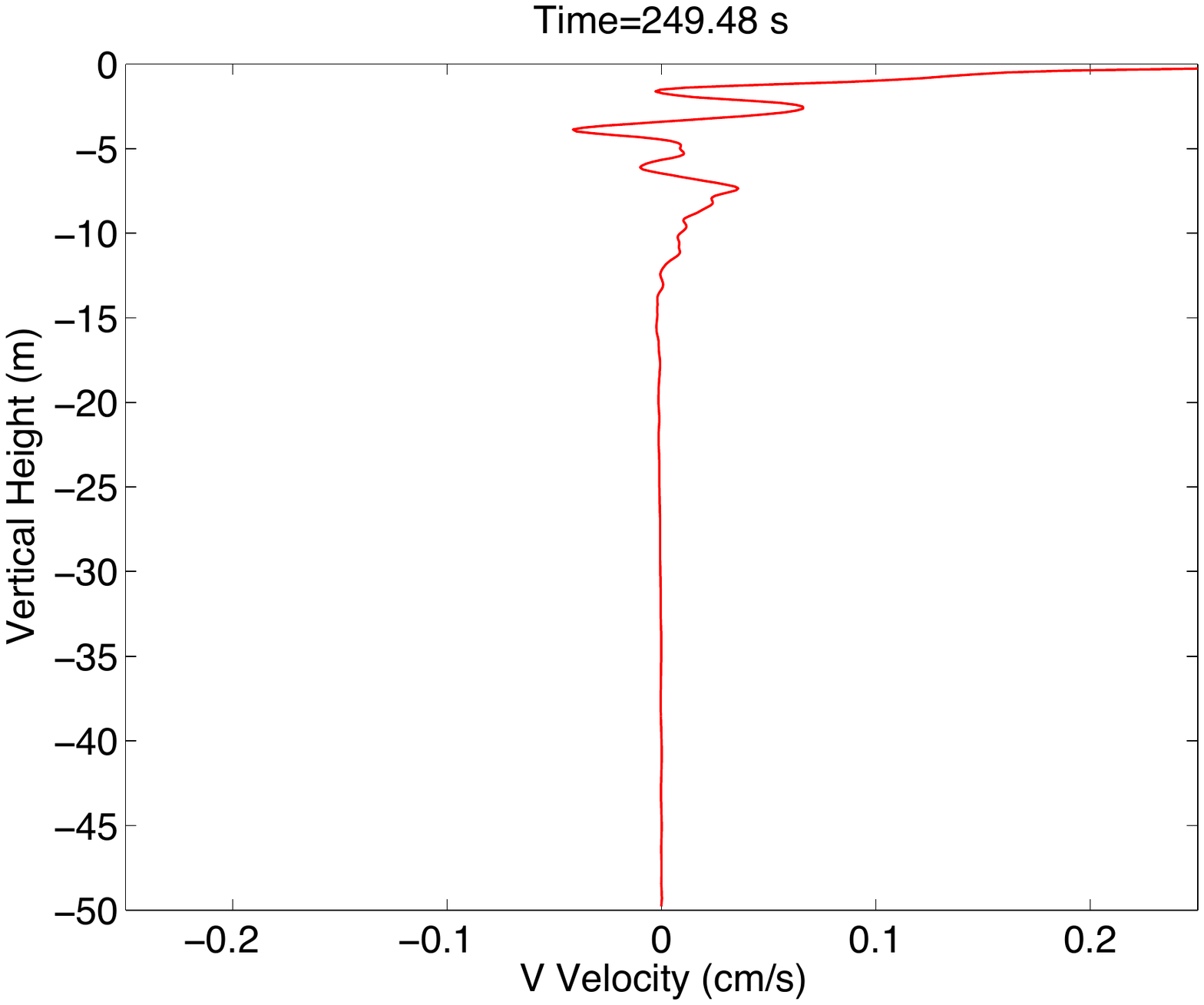} & 
\includegraphics[trim=14mm 65mm 22mm 65mm,clip=true,angle=0,width=0.45\linewidth]{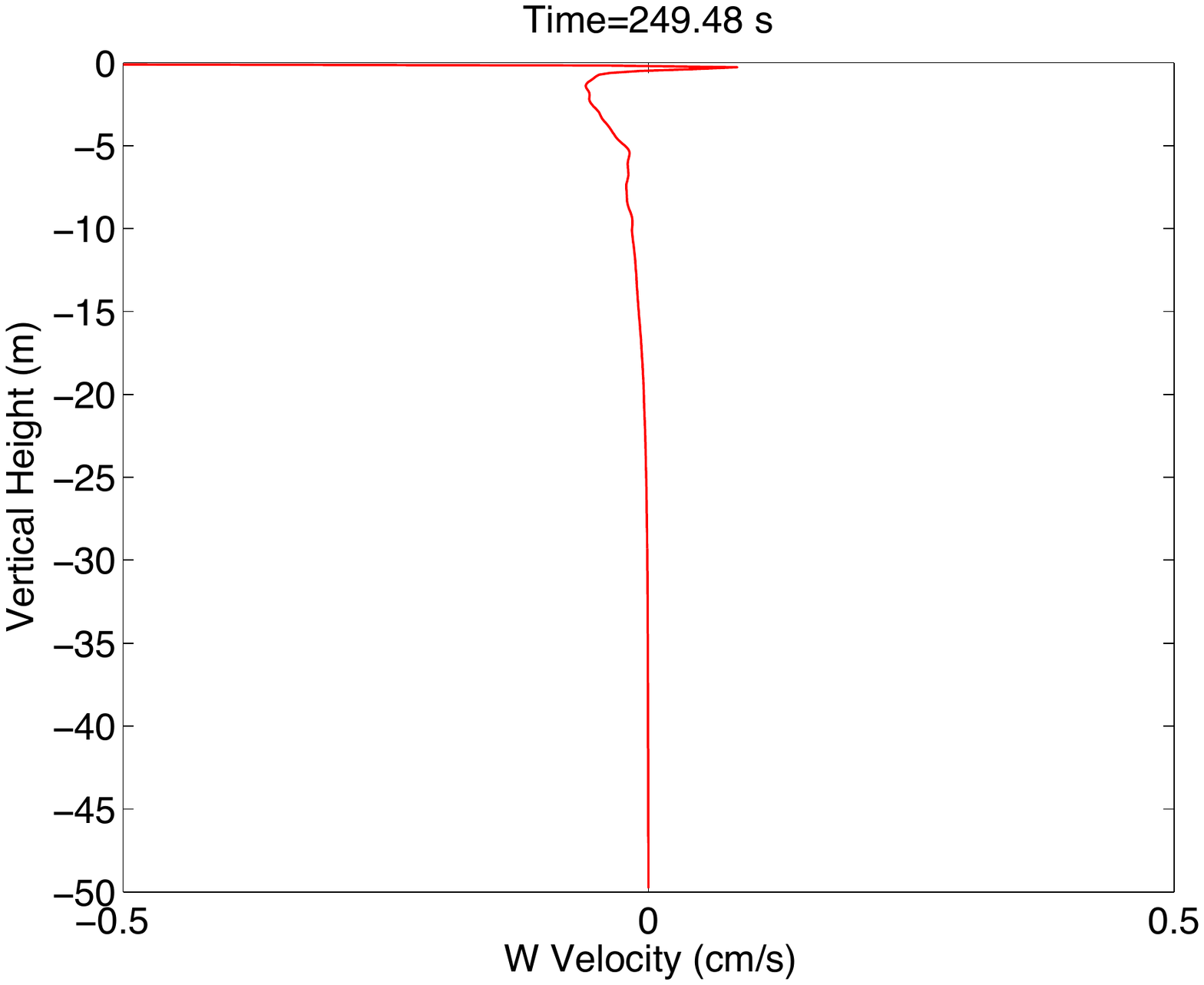} \\
(a) & (b)
\end{tabular}
\end{center}
\caption{\label{fig:crossdrift} Cross drift meandering and vertical streaming in upper ocean.    (a) cross drift.  (b) vertical streaming.  Animations of these results are available at  \mbox{\cite{crossdrift}} \href{http://youtu.be/JeT4oMrCSss}{cross drift meandering in the upper ocean} and \mbox{\cite{streamingh2o}} \href{http://youtu.be/b7CmVseM3sk}{vertical  streaming flow in the upper ocean}.}
\end{figure*}

\begin{figure}
\begin{center}
\begin{tabular}{c}
\includegraphics[trim=14mm 65mm 22mm 65mm,clip=true,angle=0,width=\linewidth]{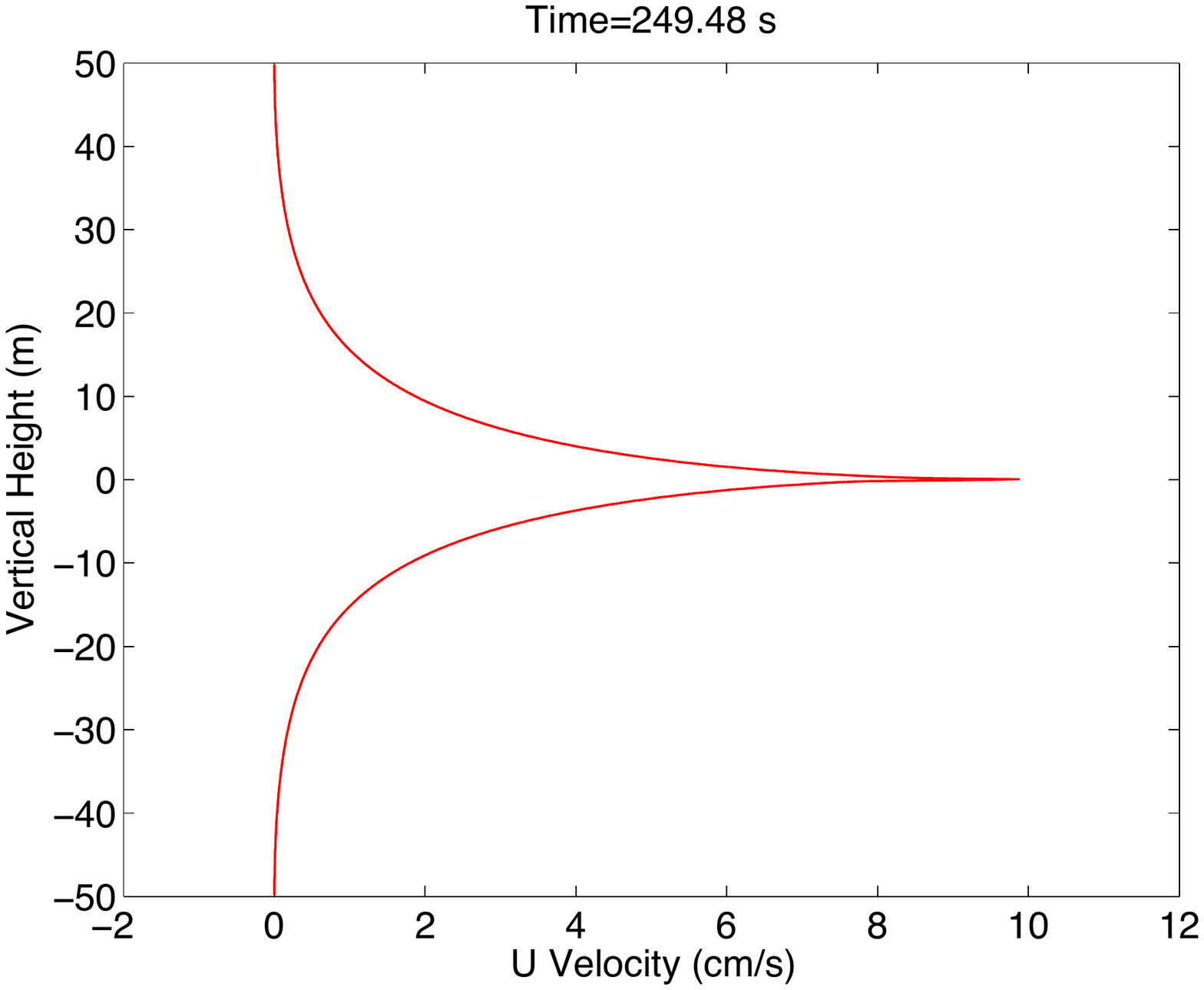}
\end{tabular}
\end{center}
\caption{\label{fig:stokesdrift} Stokes drift.    An animation of this  result is available at  \mbox{\cite{stokesdrift}} \href{http://youtu.be/lgHb5WvZLEU}{Stokes drift}.}
\end{figure}

\subsection{\label{sec:growth}Wave growth}

Following \mbox{\cite{donelan2006}} and \mbox{\cite{yang2010}}, the wave growth rate is 
\begin{eqnarray}
\label{eq:growth1}
\gamma(k) = \frac{\rho_w}{\rho_a} \frac{1}{\omega(k) S_\eta(k)} \frac{d S_\eta(k)}{dt} \;\; ,
\end{eqnarray}
where $S_\eta(k)$ is the spectral density of the free-surface elevation as a function of the wavenumber, $\omega(k)$ is the linear-wave frequency, and $k=\sqrt{k_x^2+k_y^2}$ is the magnitude of the wavenumber.  

Based on \mbox{\cite{phillips1977}} the work done by the pressure acting on the free surface is
\begin{eqnarray}
\label{eq:growth2}
\frac{d S_\eta(k)}{dt} = -S_{p u_n}(k) \;\; ,
\end{eqnarray}
where $S_{p u_n}(k)$ is the cross-spectral density of the pressure ($p$) evaluated on the free surface with the velocity normal to the free surface ($u_n=u_i n_i$).  The spectral and cross-spectral densities are calculated using height functions to the represent the free surface (see equations \ref{height1}-\ref{height3}).   The height function is single-valued, but it can discontinuous in regions where there is wave overturning.   

Due to the two-phase formulation, the pressure is continuous across the air-water interface and accounts for the flow in the air and the water in equation \ref{eq:growth2}.   By formulating the rate of  change of the spectral density in terms of the pressure and the normal velocity there is no approximation in comparison to methods that formulate it in terms of the wave drag, the wave slope, and the linear-wave phase speed in the direction of the wind.  Equation \ref{eq:growth2} is valid over all wave directions, which is important in the data assimilation of short-crested seas that are discussed in this paper.

Figures \ref{fig:growth1}a and b show the wave growth rate as a function of inverse wave age $A_*(k)=u_*/c(k)$, where $u_*$ is the friction velocity in the air and $c(k)$ is the linear-wave phase speed.   The inverse wave ages at the peak of the spectrum and at the cutoff for data assimilation are respectively $A_*(k_p)=0.065$ and $A_*(k_c)=0.22$. The results are plotted for the coarse and medium-sized data assimilations at time instances that are end of each assimilation.  The animations show the variation of the growth rates as a function of time.  Moving averages with respect to time are used to smooth the data.   The moving window is 3.19 s.   The results as a function of inverse wave age are plotted such that there is minimum of eight points per wavelength.   The height of the critical layer where the wind speed is equal to the phase speed of the wave at the peak of the spectrum is 19.6 m. 

Figures \ref{fig:growth1}a and b  and the corresponding animations show that there are regions in inverse wave age and intervals in time that have negative growth rates. The animations show that between times 17 and 27 seconds that the wave growth rate is particularly unsteady for $A_* \geq 0.2$ when there is a lot of wave breaking.    The wave growth rate is often negative between times 22 and 25 seconds.   The wave growth rate starts to recover and becomes positive between times 24 and 27 seconds.  Interestingly, there are gaps in \mbox{\possessivecite{plant1982}} data in the region where the changes in growth rate are particularly violent.  The effects of wave breaking are also evident in the PDFs of the horizontal velocities that are shown in Figures \ref{fig:sfpdf}.    We note that the most vigorous  wave breaking is occurring early in the assimilations when the waves are still coming into equilibrium.

As \mbox{\cite{oceans13}} discuss, nonlinear ocean waves are very sensitive to phase especially near the crests of long waves where short waves steepen and break \mbox{\citep{longuet1960}}.    Nonlinear waves are also susceptible to crest instabilities that can lead to wave breaking \mbox{\citep{longuet1997}}.   For sufficiently steep waves, waves whose phases are in alignment will break.   As the seaway matures, with no interactions that could lead to changes in phase, the waves that are left are those waves whose phases are not in alignment, which is a particular type of equilibrium.

Aside from the effects of wave breaking, the wave growth rate is constantly changing depending on the phase of the pressure relative to the wave.   Generally, the wave growth rate is positive but only in a time-averaged sense.    The wave growth rate fluctuates between the upper and lower limits of \mbox{\possessivecite{plant1982}}  data.   There does tend to be a dip in the growth rate for $0.2 \leq A_* \leq 0.4$ and a steep rise for $A_* > 0.4$.

Figures \ref{fig:growth2}a and b show the free-surface spectra for the coarse and medium-sized data assimilations.  The results are shown at time instances that are at the end of each assimilation.    The animations show variations of the spectra as a function of time.  The wavenumber at the cutoff is $k_c=0.7 \; {\rm rad/m}$.  The wave spectra are not resolved very well at low wavenumber for $k < 0.063 \; {\rm rad/m}$ because the length of the domain is short.  The results of the data assimilations are compared the JONSWAP spectrum that is used to assimilate the waves.   After the initial stages when the waves are coming into equilibrium ($t \leq 25 \; {\rm s}$), there is very good agreement between the data assimilations and the JONSWAP spectrum.    Considering that there is a lot of wave breaking occurring in the data assimilations, the good agreement even at the highest wave numbers is remarkable.   For $k>k_c$, the waves interact naturally with no nudging subject to nonlinear wave interactions, forcing due to wind, and dissipation due to breaking.    With no wind forcing, the tail ends of spectra would drop well below the input JONSWAP spectrum.   Nonlinear wave interactions and wind forcing counteract the dissipation due wave breaking at the tail ends of the spectra such that a $k^{-3}$ power-law behavior is maintained for $k>k_c$.

The coarse assimilation has a longer duration than the medium assimilation, and over time an energy pileup occurs at $k=k_c$ in the coarse assimilation.   The wavelength at $k=k_c$ is $\lambda_c=8.98 \; {\rm m}$.   The crosshatching that occurs in the time-averaged velocities in  Figures \ref{fig:langmuir} and \ref{fig:windstreaks1} toward the ends of the assimilations is the visual manifestation of the energy pileup.   At present, the nudging formulation using HOS as a basis  does not permit energy backscatter to occur.   We conjecture that if we had assimilated real data that energy backscatter would have been enabled and that no energy pileup would have occurred.  The fact that the energy pileup is occurring at oblique angles to the wind suggests that the Langmuir cells are contributing to the energy backscatter through their helical structure. 
\begin{figure*}
\begin{center}
\begin{tabular}{cc}
\includegraphics[trim=12mm 64mm 22mm 65mm,clip=true,angle=0,width=0.45\linewidth]{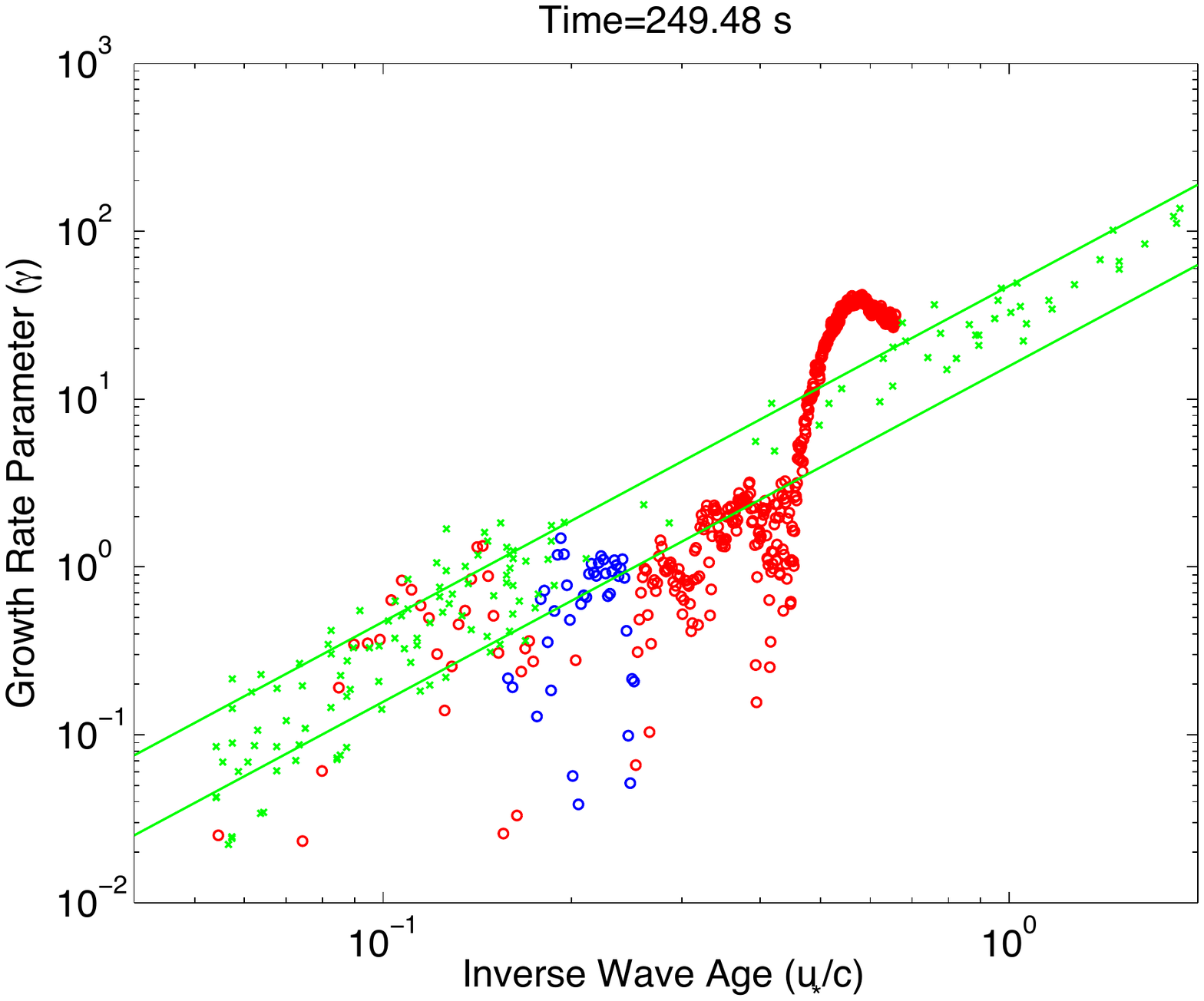} & 
\includegraphics[trim=12mm 64mm 22mm 65mm,clip=true,angle=0,width=0.45\linewidth]{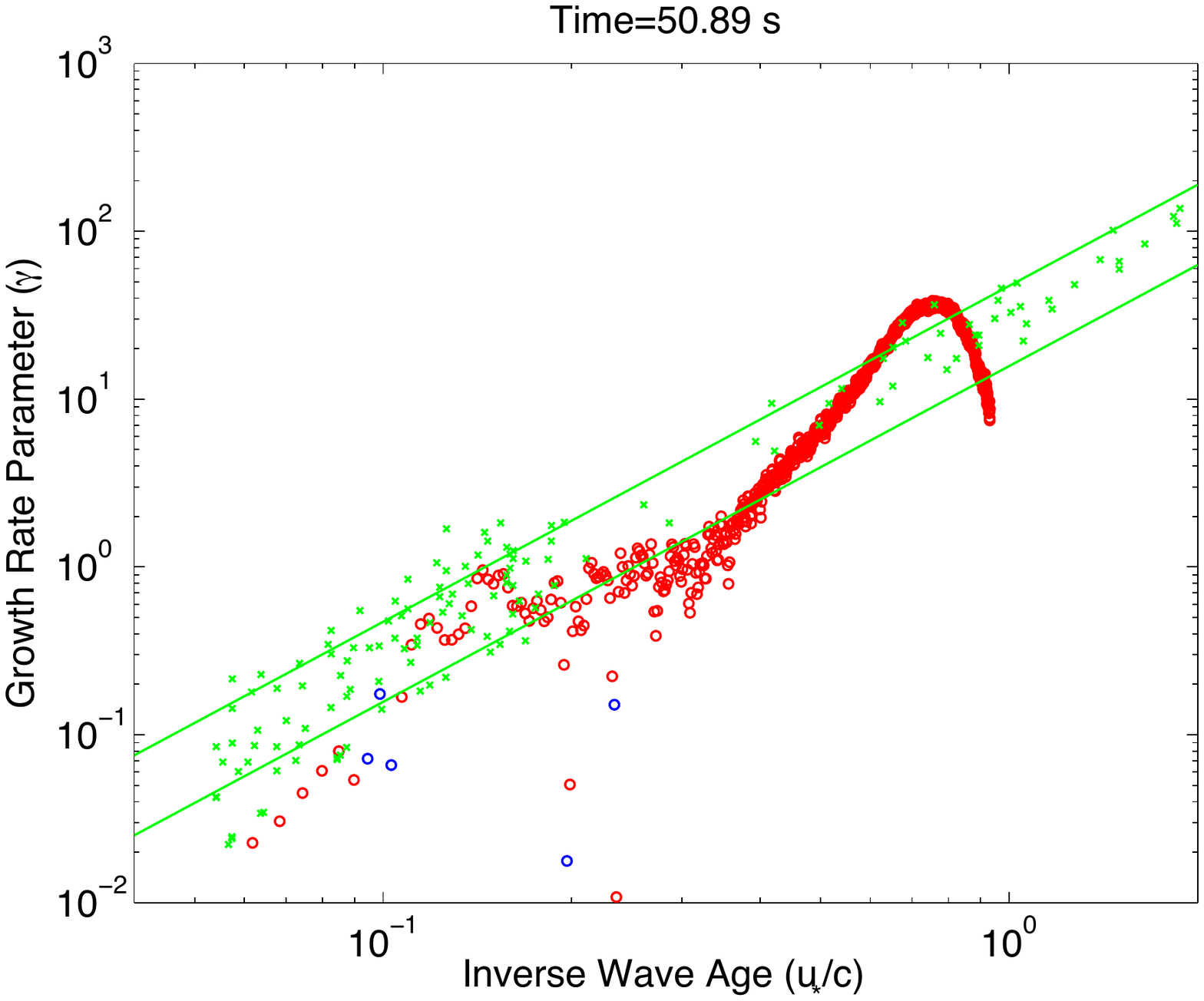}  \\
(a) & (b)
\end{tabular}
\end{center}
\caption{\label{fig:growth1}  Wave growth rates.  (a) coarse-sized assimilation. (b) medium-sized assimilation.  Positive and negative growth rates based on data assimilation  are respectively denoted by red (\textcolor{red}{$\circ$}) and blue (\textcolor{blue}{$\circ$}) circles.  \mbox{\possessivecite{plant1982}} data is labeled by (\textcolor{green}{$\times$}) crosses.   The upper and lower limits of  \mbox{\possessivecite{plant1982}} data are denoted by green lines.  Animations of the these results are available at
\mbox{\cite{growth1}} \href{http://youtu.be/wM\_Sv-sAfmg}{wave growth rate (coarse)} and 
\mbox{\cite{growth2}} \href{http://youtu.be/O41zfrLhWG8}{wave growth rate (medium)}.}
\end{figure*}

\begin{figure*}
\begin{center}
\begin{tabular}{cc}
\includegraphics[trim=12mm 62mm 22mm 65mm,clip=true,angle=0,width=0.45\linewidth]{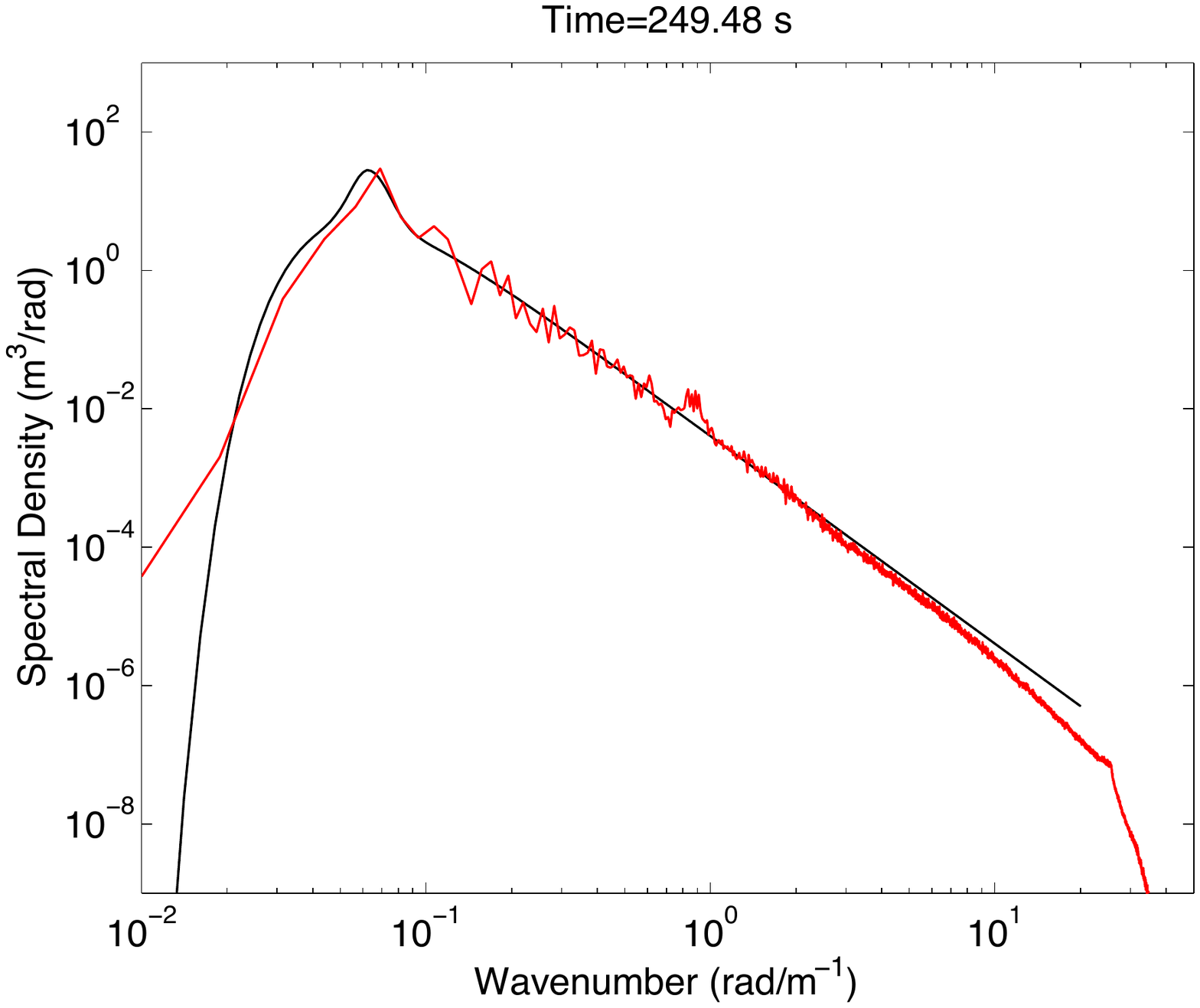} & 
\includegraphics[trim=12mm 62mm 22mm 65mm,clip=true,angle=0,width=0.45\linewidth]{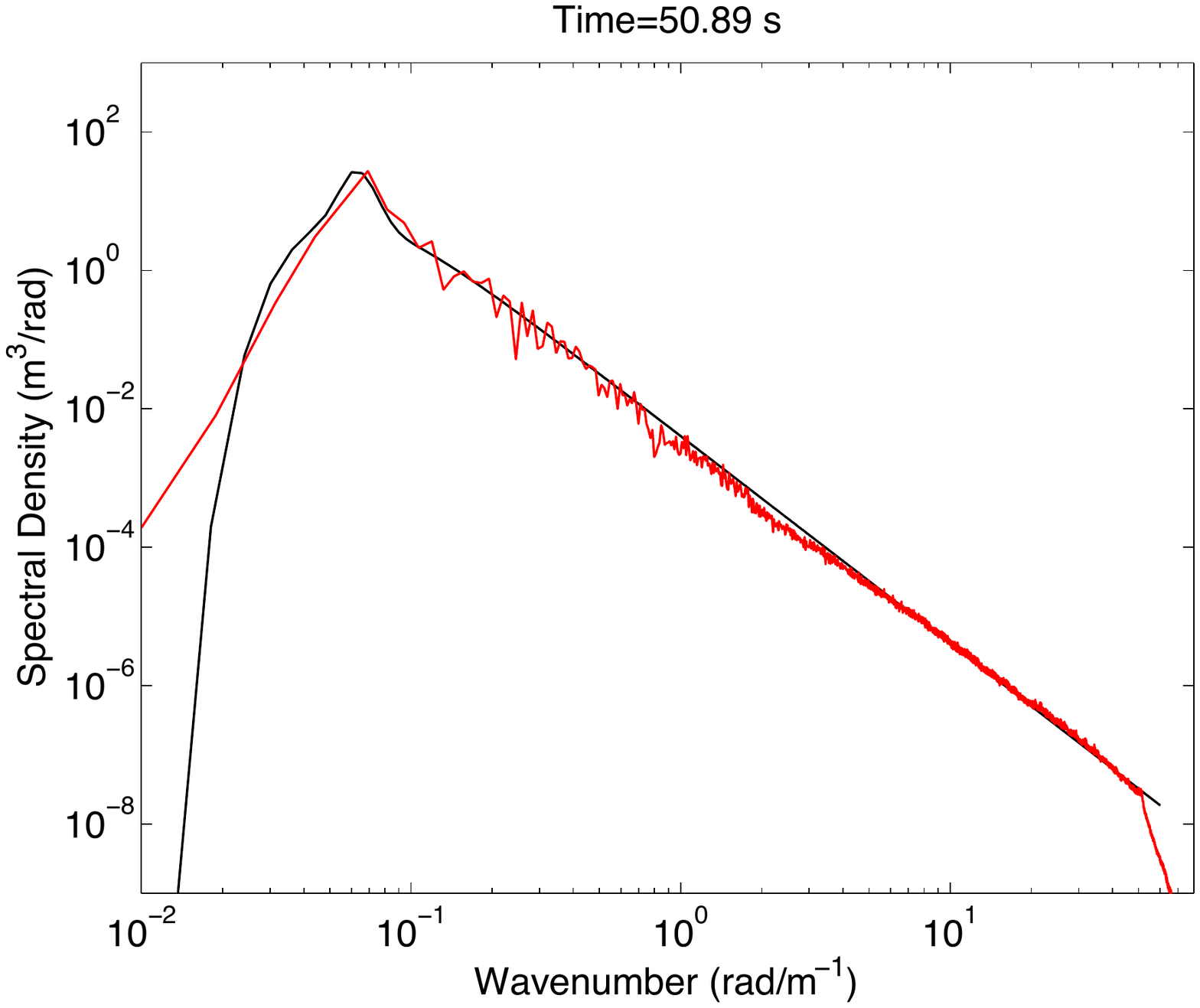}  \\
(a) & (b)
\end{tabular}
\end{center}
\caption{\label{fig:growth2}  Free-surface spectra.  (a) coarse-sized assimilation.  (b) medium-sized assimilation.   JONSWAP spectrum: \curve. NFA: \curvered.   Animations of the these results are available at
\mbox{\cite{spectrum1}} \href{http://youtu.be/2HTI5PkyKUc}{free-surface spectrum (coarse)} and
\mbox{\cite{spectrum2}} \href{http://youtu.be/YTno0xb4c8s}{free-surface spectrum (medium)}.}
\end{figure*}

\subsection{\label{sec:statistics}Statistics of free-surface quantities}

Figures \ref{fig:sfpdf}a-h show probability density functions (PDFs) for the free-surface elevation, and the u, v, and w-components of velocity evaluated on the free surface for the coarse and the medium-sized data assimilations.    The data assimilations are also compared to HOS simulations.    The results are time-averaged over 2.00 seconds.   The results are shown at time $t=20 \; {\rm s}$ when there is a strong breaking event.

The extreme statistics of all quantities in Figures  \ref{fig:sfpdf}a-h are limited by the size of the domain, which is $500 \; {\rm m} \times 125 \; {\rm m}$.   The PDFs would also be smoother in a larger domain.  \mbox{\cite{oceans13}} show how small changes in phase and amplitude affect the extreme statistics and the equilibrium of ocean waves in a larger domain based on HOS.   The animations associated with Figures \ref{fig:sfpdf}a-h show that the most extreme events occur for time $t \leq 25 \; {\rm s}$ when the waves are establishing equilibrium.  As \mbox{\cite{oceans13}} discuss,  waves whose phases are not in equilibrium break.   The most extreme events tend to occur during this breaking stage.

Based on Figures \ref{fig:sfpdf}a and b, the PDFs of the results of HOS and data assimilations agree best for the free-surface elevation.    The PDFs of the velocities in Figures \ref{fig:sfpdf}c-h differ significantly due to the effects of wave breaking.   The wave breaking is occurring when the water-particle velocity is equal to the phase velocity.   As a reference, the phase velocity of the wave at the peak of the spectrum is $c_o  =12.5 \; {\rm m/s}$.   Of the three components of velocity, the w-component of velocity shows the best agreement between the results of HOS and data assimilations due to the assimilation of the normal velocity as predicted by HOS into the data assimilations.   The extremes of the medium-sized data assimilation are larger than those of the coarse-sized data assimilation and HOS.   The extremes of the medium-sized data assimilation are smoother than those of the coarse-sized data assimilation because there are more points.

The PDFs of the free-surface elevations in Figures  \ref{fig:sfpdf}a and b are skewed slightly due to Stokes effects with steep crests and shallow troughs.   Based on the animations, the largest differences between the data assimilations and HOS occurs when $t=6.5 \; {\rm s}$ when the kurtosis is at its maximum value  (see Figures \ref{fig:skewkurt}a and c).   The highest wave crest is over 4.5 m high at that time, which is a very large wave considering that the length of the wave at the peak of the spectrum is only 100 m.

Due to the effect of the wind drift, the PDFs of the data assimilations are shifted to the left of the HOS results for u-component of the velocity in Figures  \ref{fig:sfpdf}c and d.   The PDFs for the u-component of velocity are also asymmetrical with higher velocities occurring in the crests of waves than in the troughs.   The PDFs of the data assimilations are much wider than the HOS results for the v-component of velocity in Figures \ref{fig:sfpdf}e and f due to the effects of wave breaking.    A similar effect is seen in the PDFs for the w-component of velocity in Figures \ref{fig:sfpdf}g and h.   In addition, the PDFs for the w-component of velocity are asymmetrical with higher velocities occurring on the front faces of breaking waves in the results for the data assimilations.

Figures \ref{fig:skewkurt}a-d show the skewness and the kurtosis for the coarse and medium-sized data assimilations.  The data assimilations closely track the HOS.
If real data had been assimilated in lieu of HOS, we conjecture that there would have been more extreme events in the data assimilations.

\begin{figure*}
\begin{center}
\begin{tabular}{cc}
\includegraphics[trim=14mm 65mm 22mm 65mm,clip=true,angle=0,width=0.32\linewidth]{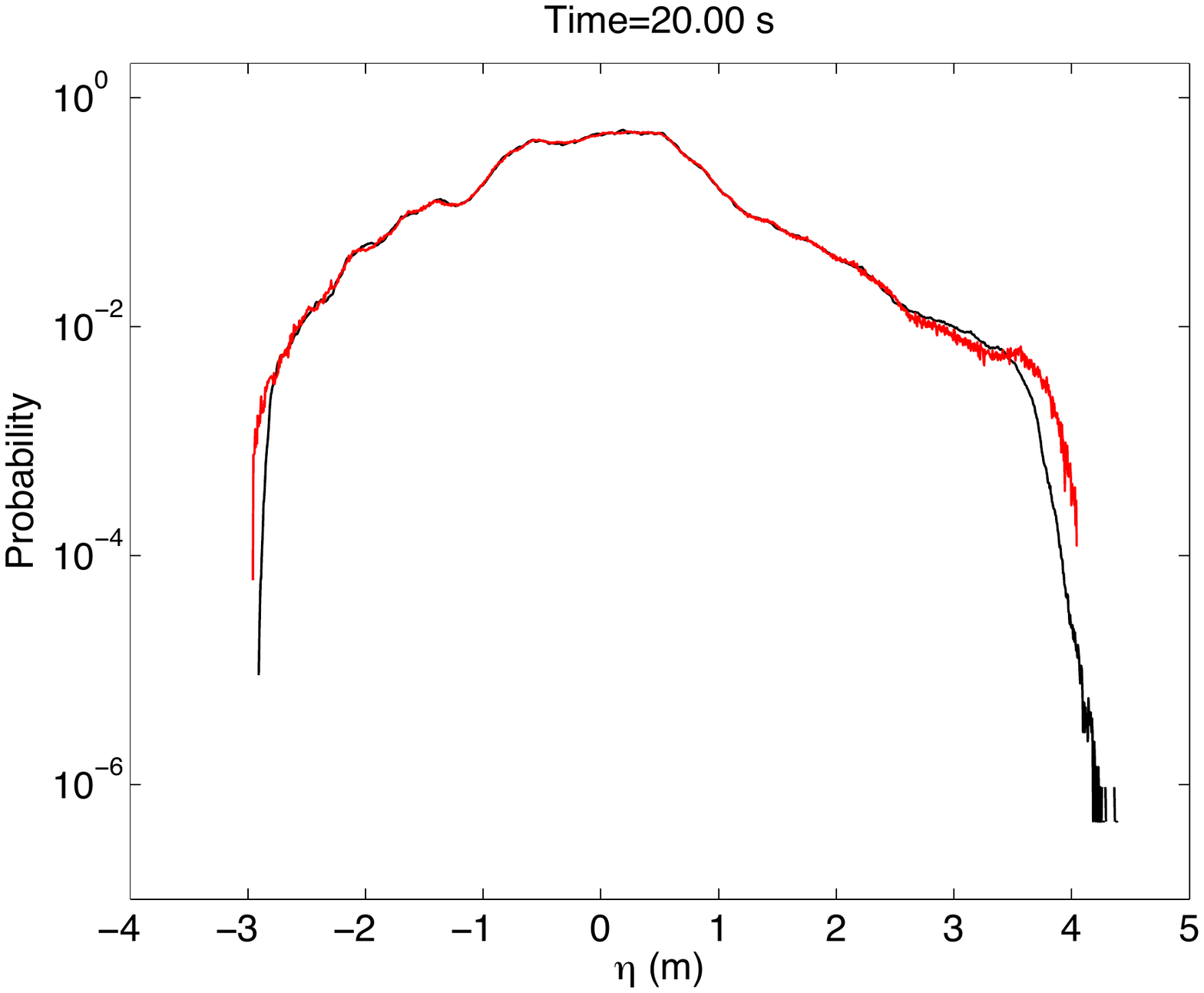} & 
\includegraphics[trim=14mm 65mm 22mm 65mm,clip=true,angle=0,width=0.32\linewidth]{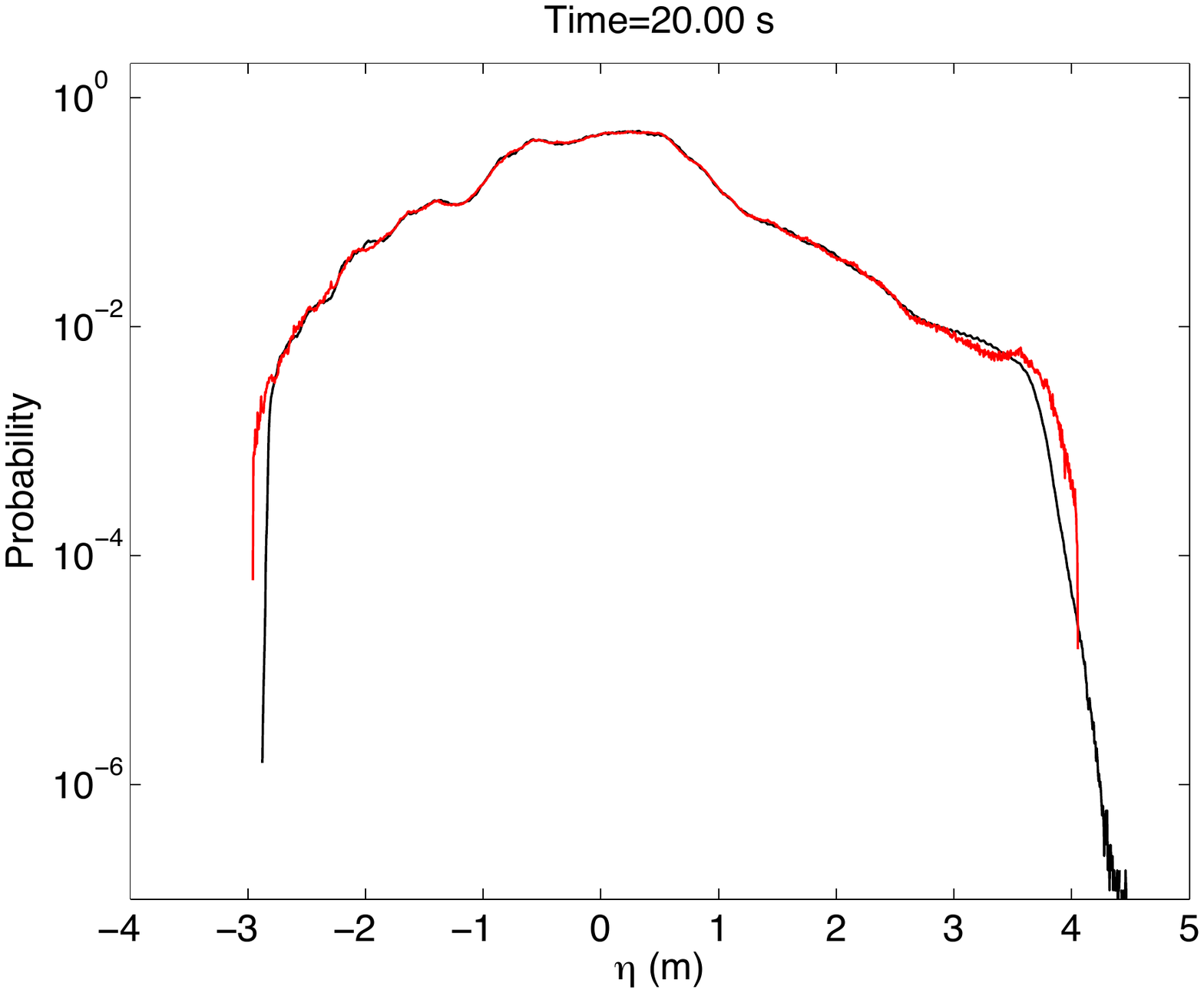}  \\
(a) & (b) \\
\includegraphics[trim=14mm 65mm 22mm 65mm,clip=true,angle=0,width=0.32\linewidth]{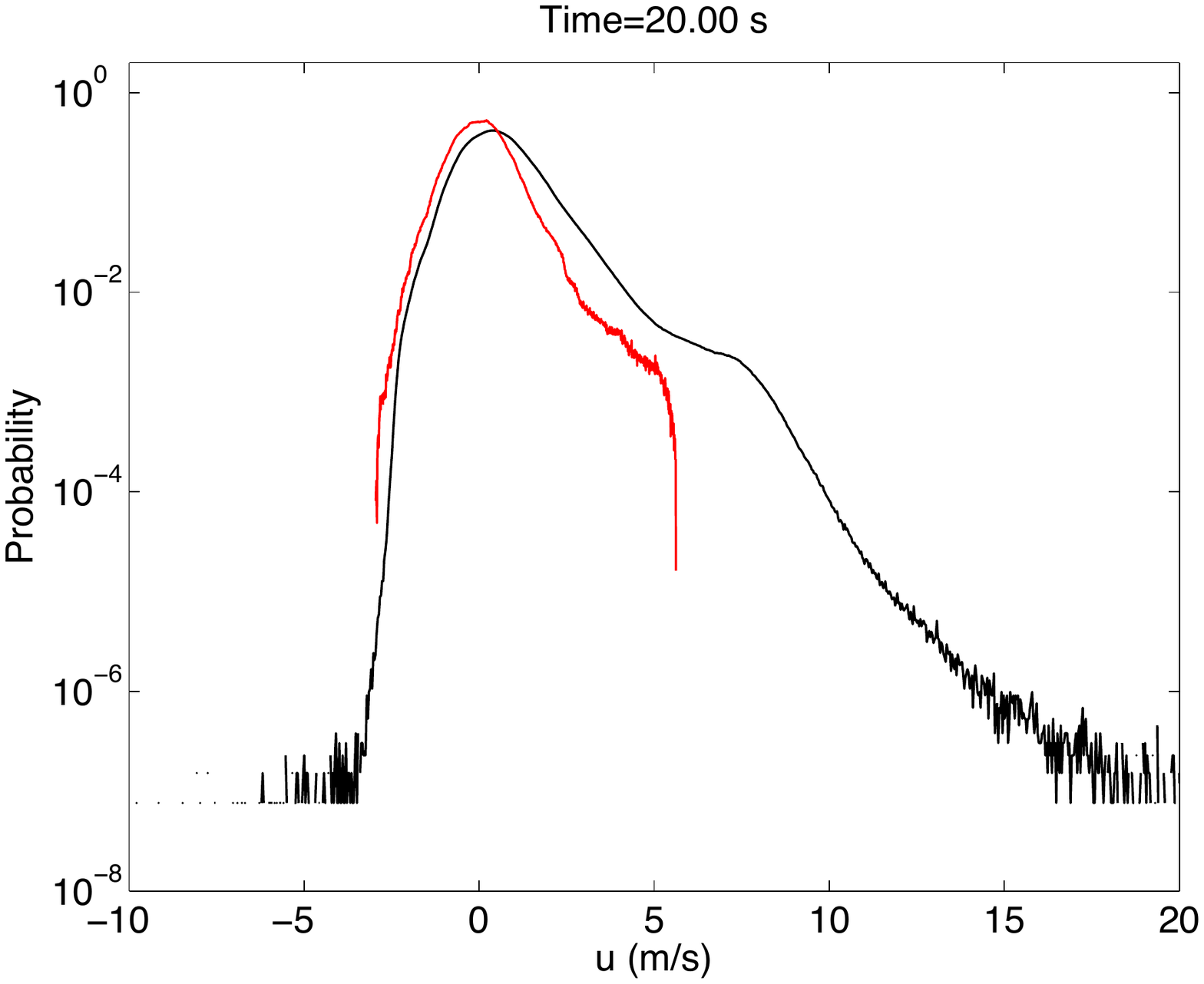} & 
\includegraphics[trim=14mm 65mm 22mm 65mm,clip=true,angle=0,width=0.32\linewidth]{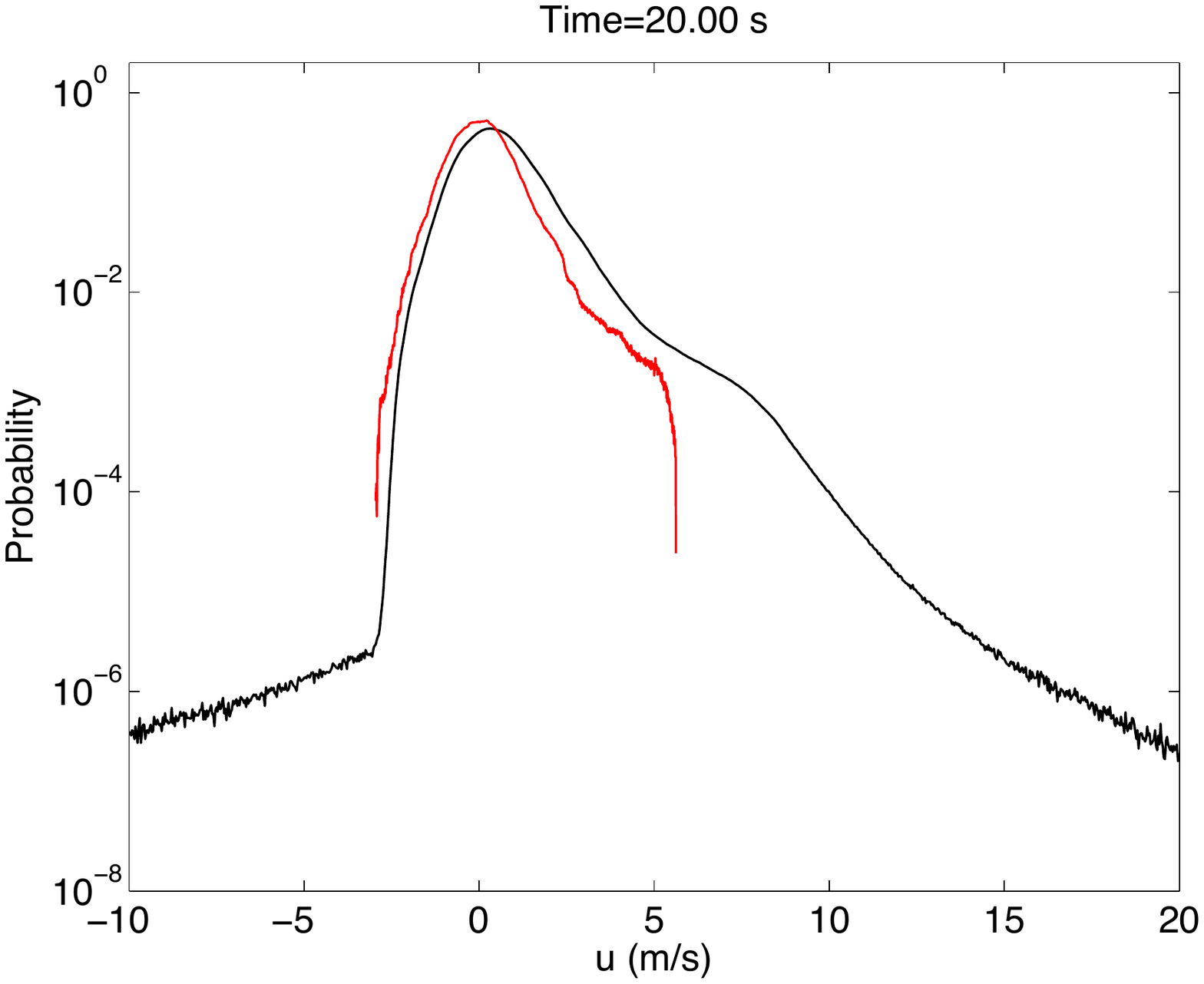}  \\
(c) & (d) \\
\includegraphics[trim=14mm 65mm 22mm 65mm,clip=true,angle=0,width=0.32\linewidth]{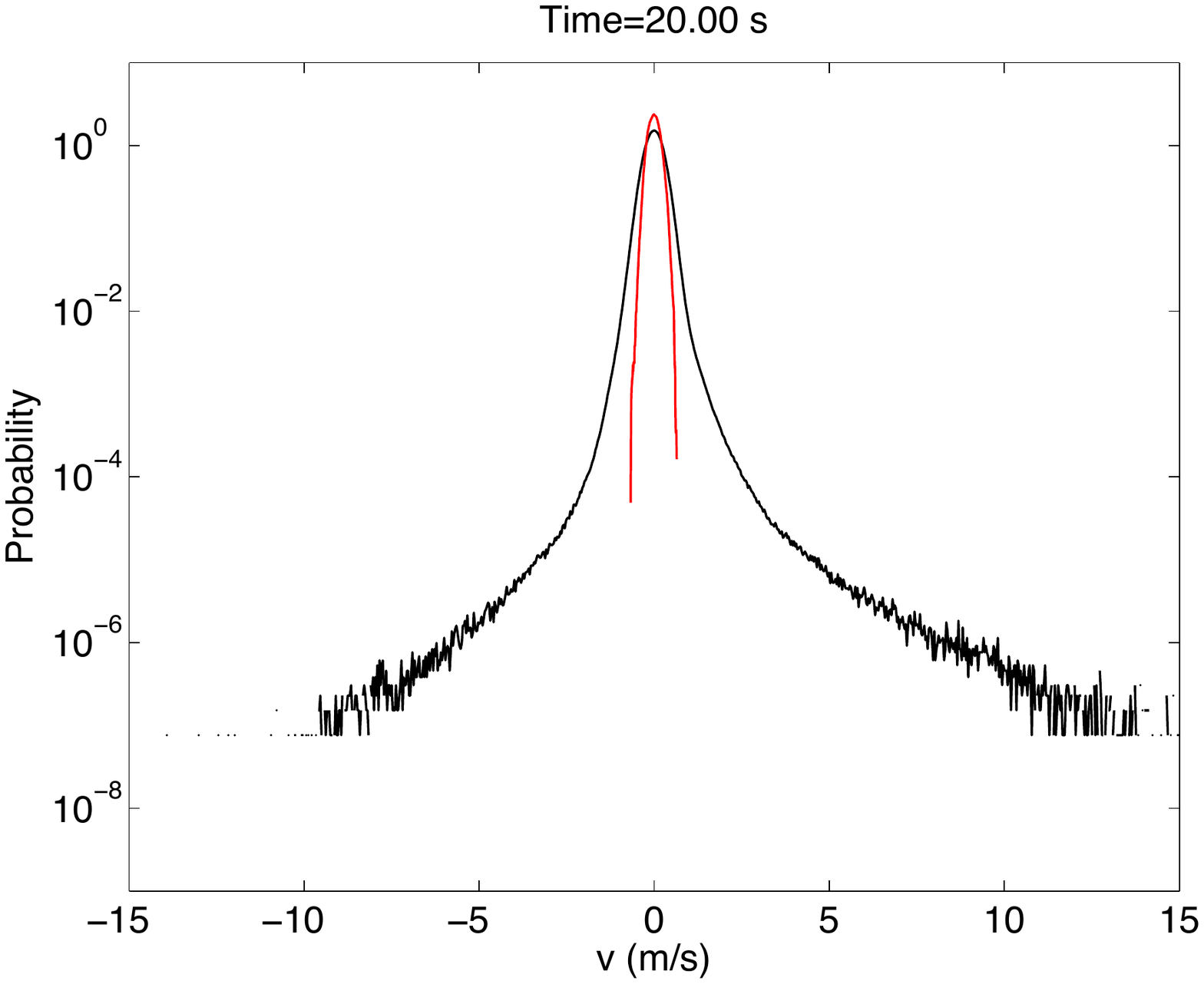} & 
\includegraphics[trim=14mm 65mm 22mm 65mm,clip=true,angle=0,width=0.32\linewidth]{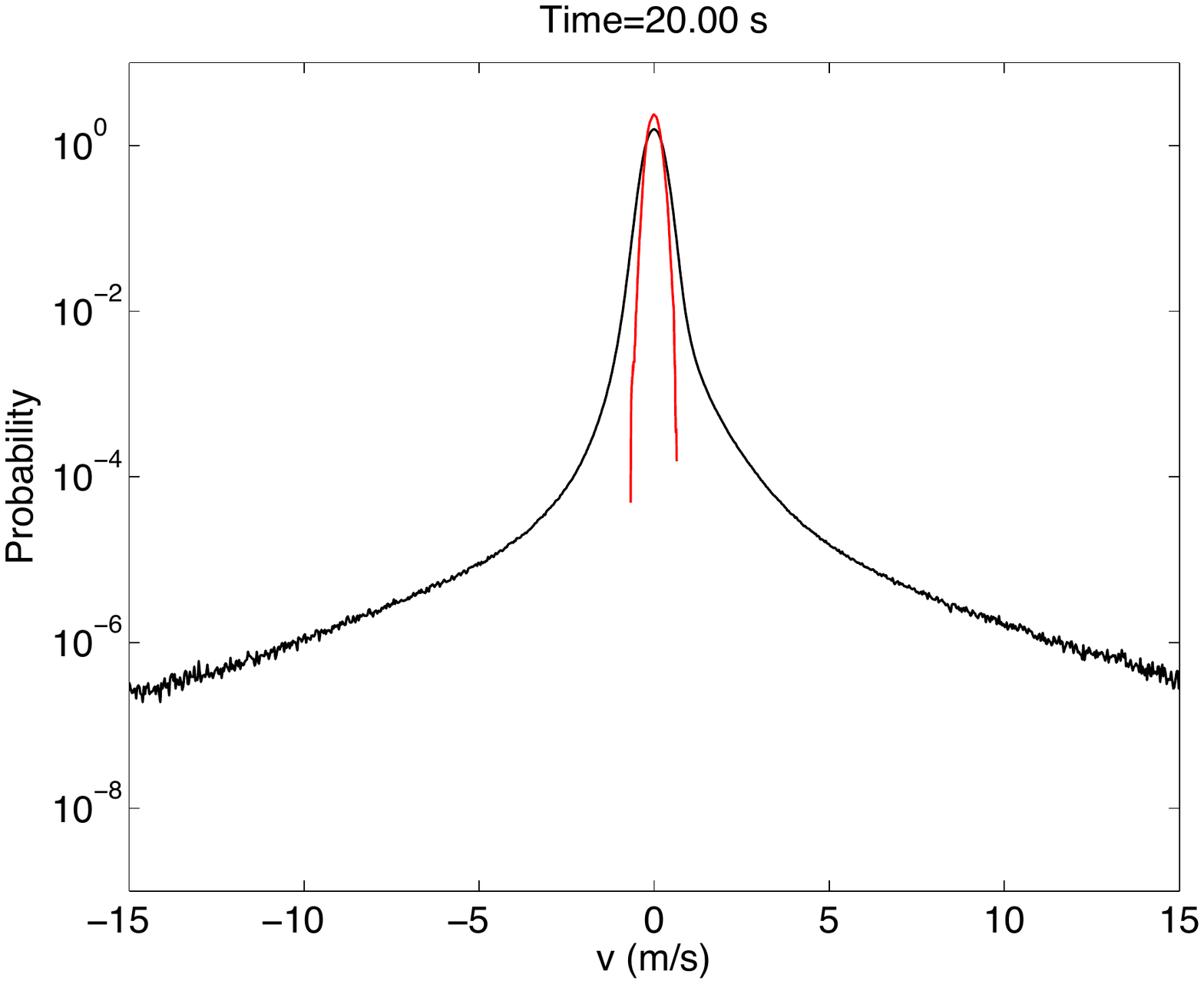}  \\
(e) & (f) \\
\includegraphics[trim=14mm 65mm 22mm 65mm,clip=true,angle=0,width=0.32\linewidth]{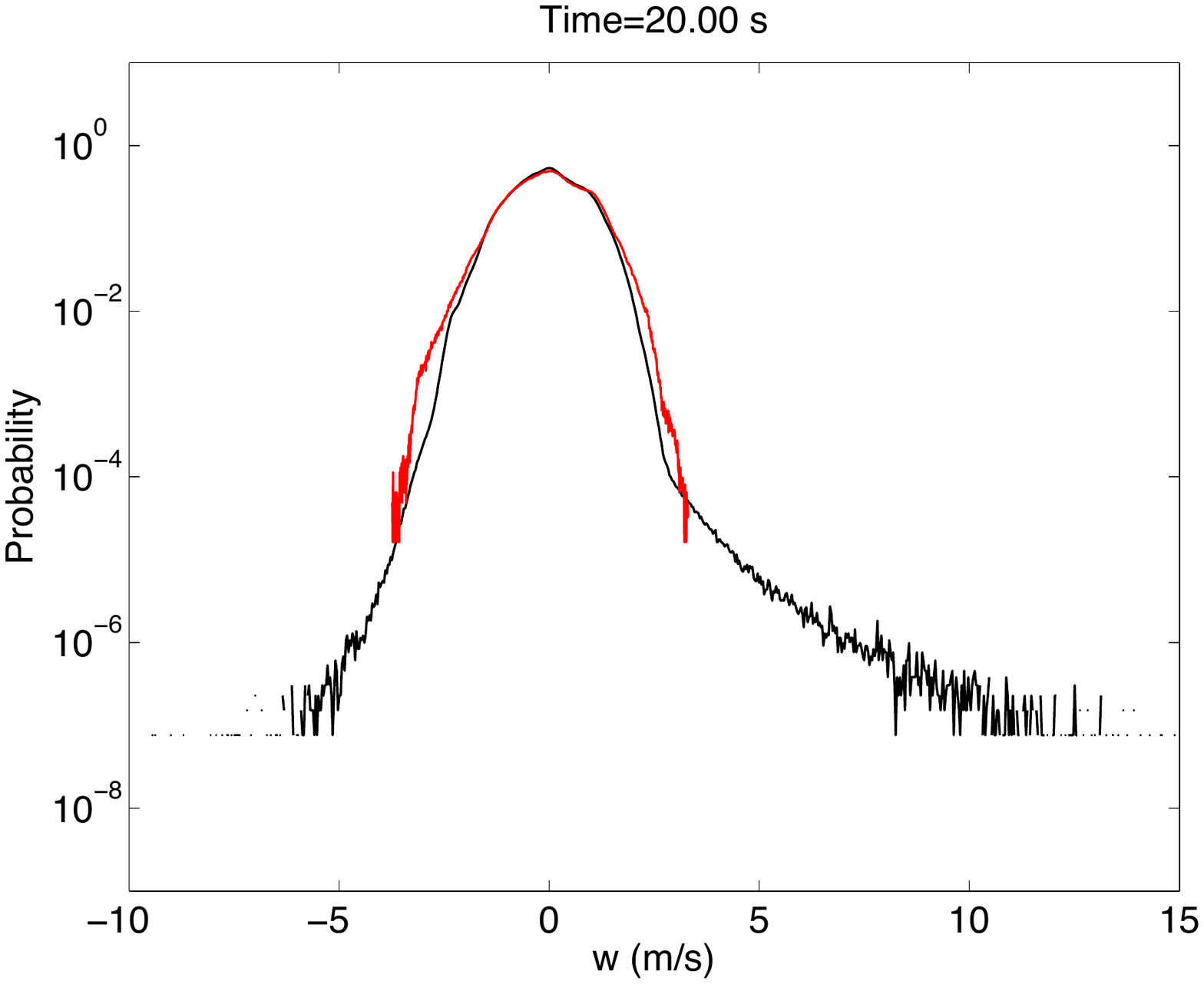} & 
\includegraphics[trim=14mm 65mm 22mm 65mm,clip=true,angle=0,width=0.32\linewidth]{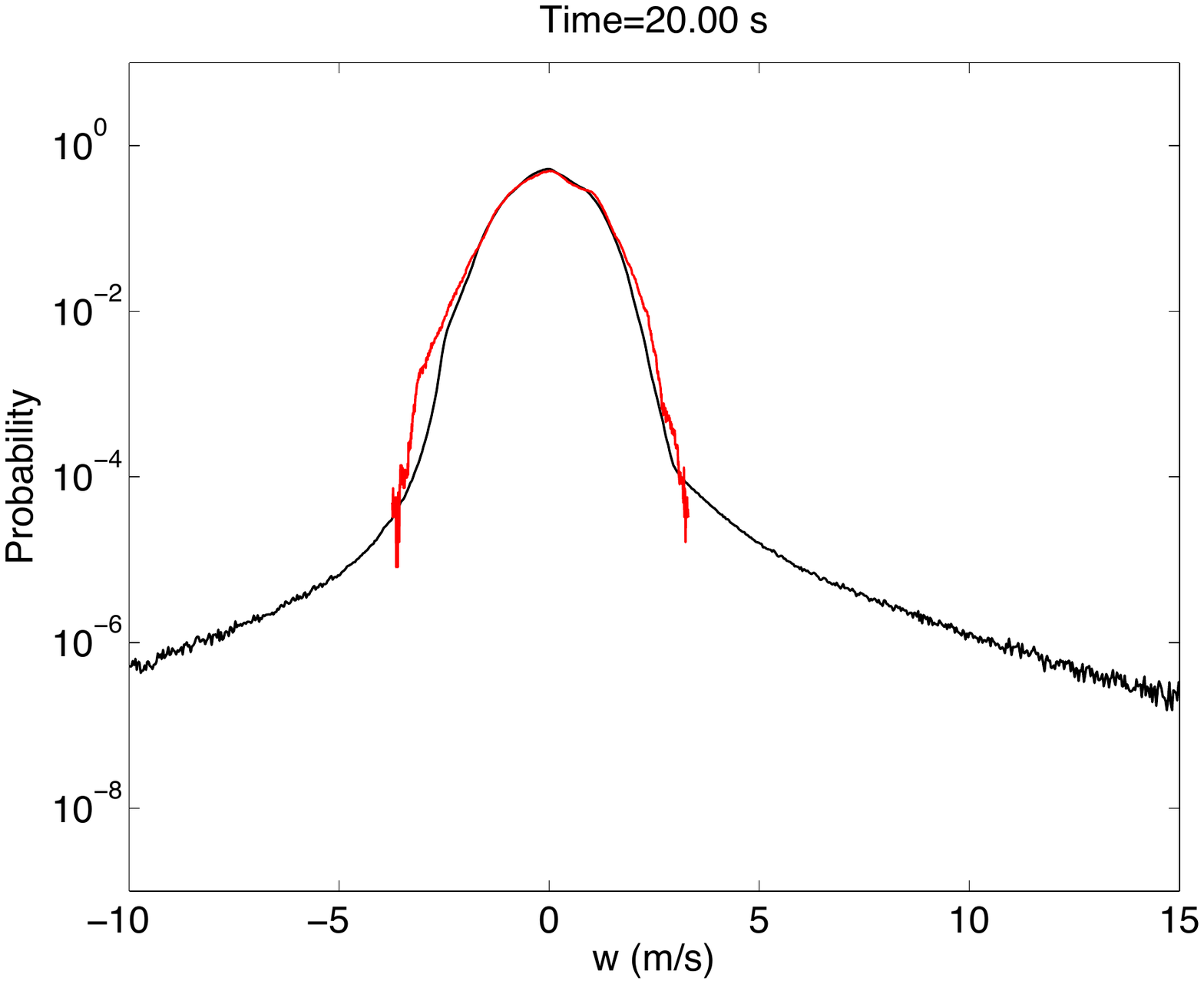}  \\
(g) & (h)
\end{tabular}
\end{center}
\caption{\label{fig:sfpdf}  Probability density functions.  Parts (a) and (b); (c) and (d); (e) and (f); and (g) and (h) are PDFs for respectively the free-surface elevation, and the u, v, and w components of velocity evaluated on the free surface.  Coarse and medium-sized assimilations are plotted respectively on the left and right.  NFA: \curve. HOS: \curvered.  Animations of the these results are available at
\mbox{\cite{run51probsurf}} \href{http://youtu.be/SnbWzJ53AZg}{$\eta$ (coarse)};
\mbox{\cite{run77probsurf}} \href{http://youtu.be/NuL9-Kmlrp8}{$\eta$ (medium)};
\mbox{\cite{run51probuvel}} \href{http://youtu.be/Sq23wY-11vU}{u (coarse)};
\mbox{\cite{run77probuvel}} \href{http://youtu.be/iZSCzE2d\_ig}{u (medium)};
\mbox{\cite{run51probvvel}} \href{http://youtu.be/6d16u6Ytaws}{v (coarse)};
\mbox{\cite{run77probvvel}} \href{http://youtu.be/914AonZBW3A}{v (medium)};
\mbox{\cite{run51probwvel}} \href{http://youtu.be/5tcwPlfPMbE}{w (coarse)}; and
\mbox{\cite{run77probwvel}} \href{http://youtu.be/wycoDnGwGOw}{w (medium)}.}
\end{figure*}

\begin{figure*}
\begin{center}
\begin{tabular}{cc}
\includegraphics[trim=12mm 66mm 22mm 70mm,clip=true,angle=0,width=0.4\linewidth]{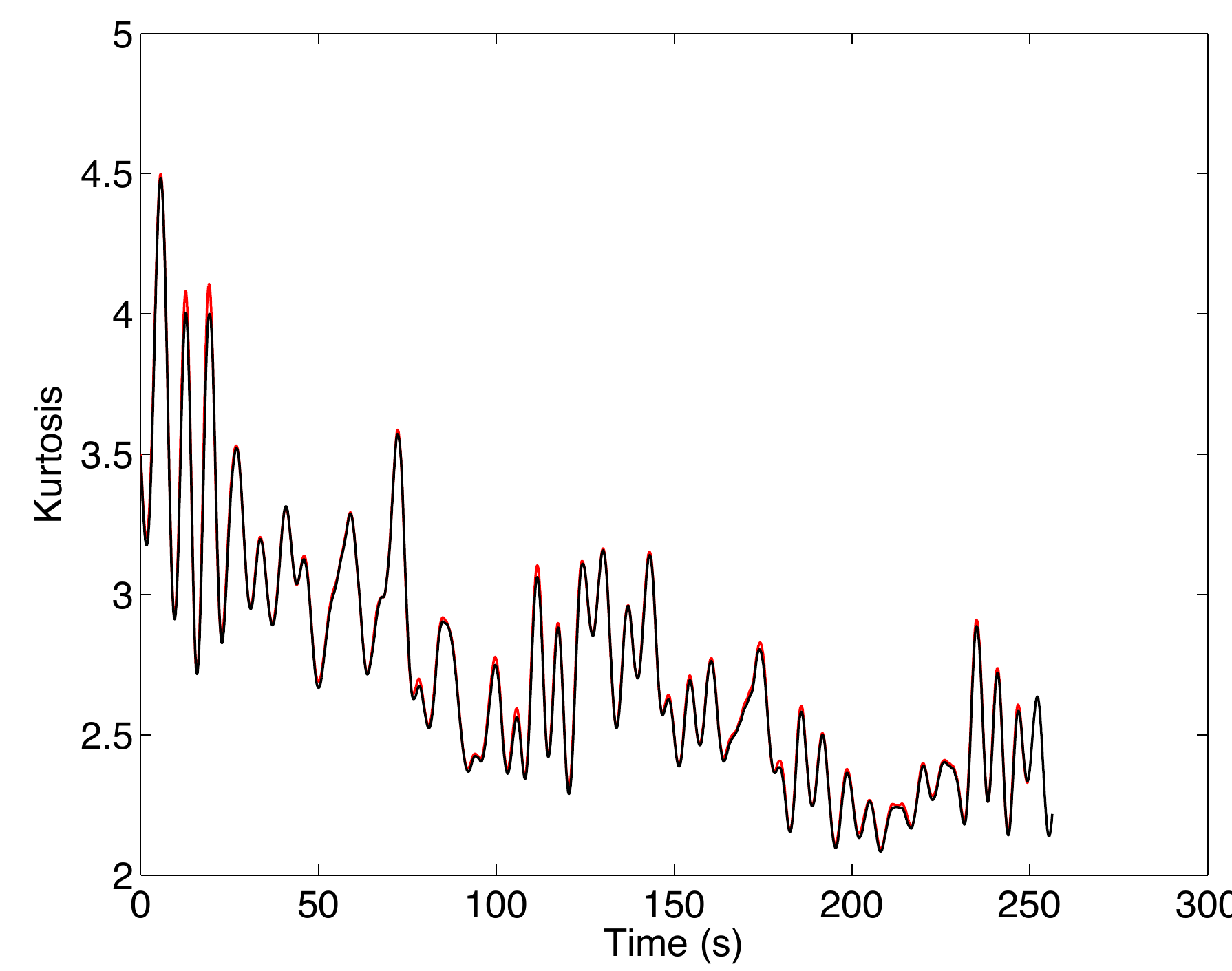} & 
\includegraphics[trim=12mm 66mm 22mm 70mm,clip=true,angle=0,width=0.4\linewidth]{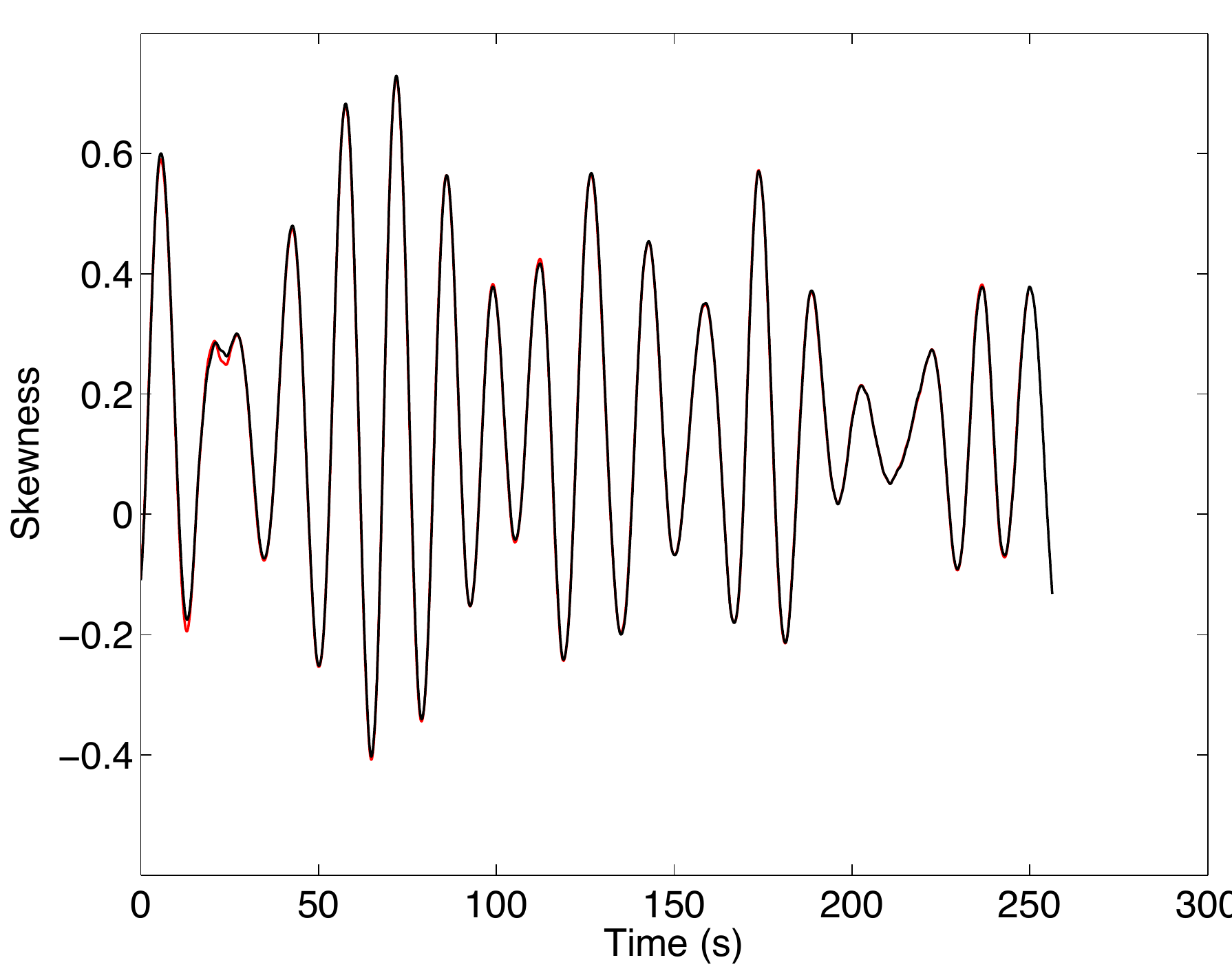}  \\
(a) & (b) \\
\includegraphics[trim=12mm 66mm 22mm 70mm,clip=true,angle=0,width=0.4\linewidth]{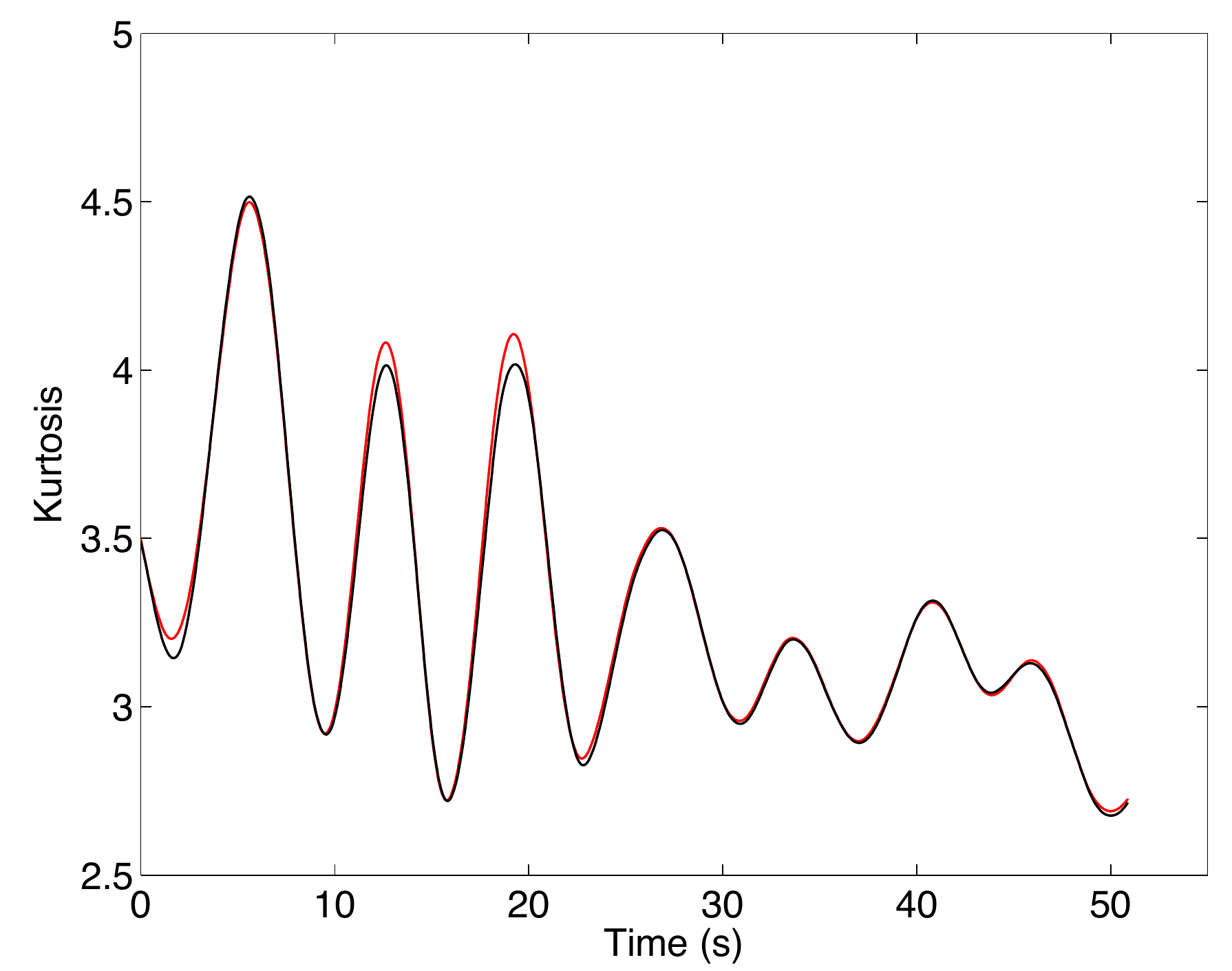} & 
\includegraphics[trim=12mm 66mm 22mm 70mm,clip=true,angle=0,width=0.4\linewidth]{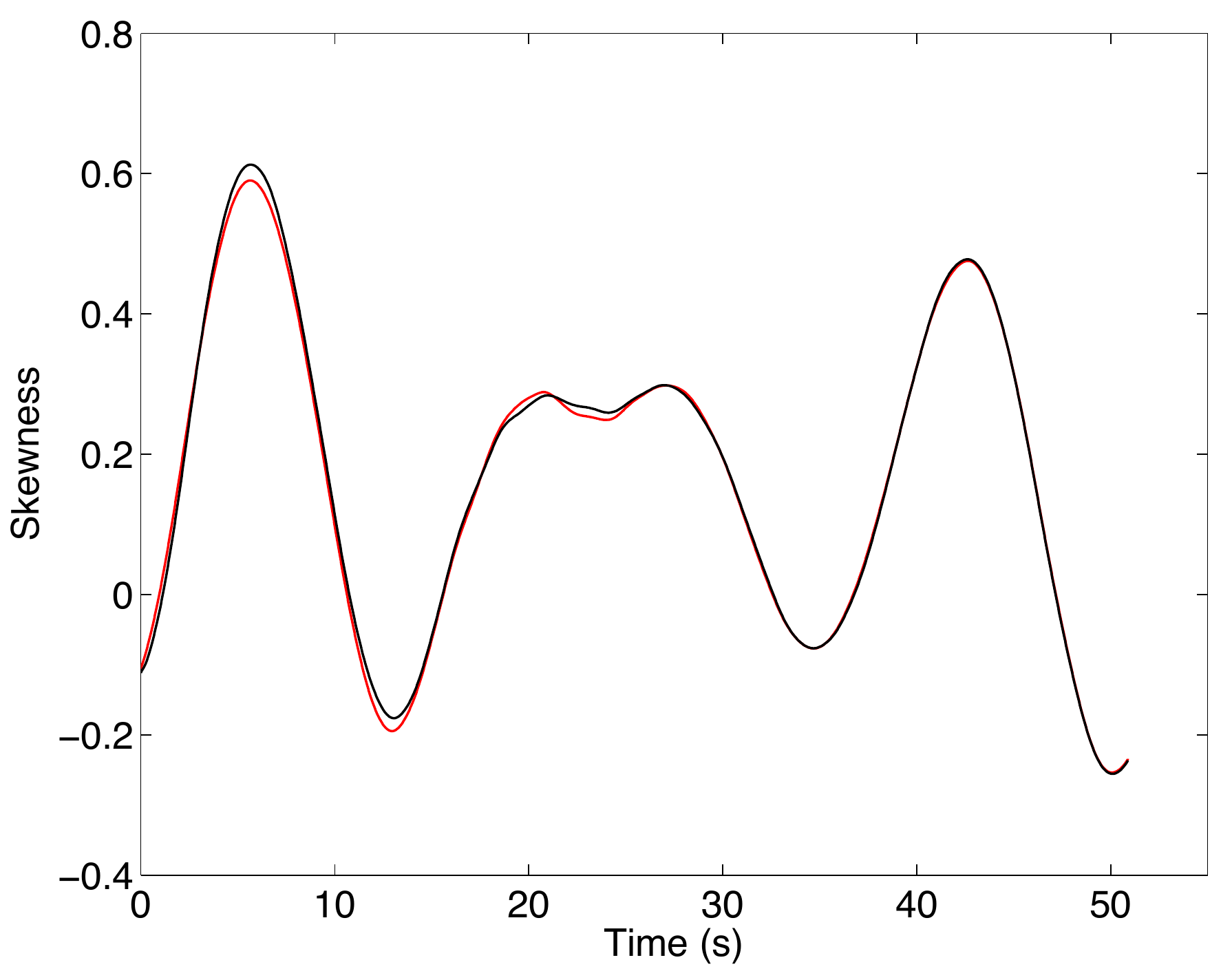} \\
(c) & (d)
\end{tabular}
\end{center}
\caption{\label{fig:skewkurt} Statistics of free-surface elevation.  Parts (a) and (b) are respectively the skewness and kurtosis for the coarse assimilation.  Parts (c) and (d) are for medium-sized assimilation. NFA: \curve. HOS: \curvered.}
\end{figure*}

\subsection{\label{sec:mixing}Mixing}

Figures \ref{fig:mixing}a-e show mixing PDFs based on tracking of Lagrangian particles.     The particles are initially seeded at various distances relative to  the free surface from $\zeta=-12.8 \; {\rm m}$ to  $\zeta=12.8 \; {\rm m}$, where $\zeta=z+\eta(x,y)$ is a free-surface-following coordinate system.  The particles are not dynamically active.   The particles can cross the free surface in regions where there is wave breaking or strong shear.  The results are based on the coarse-sized data assimilation.    The PDFs are shown at time $t=249.5 \; {\rm s}$.   The PDFs are initially delta functions at time $t=0$.   The animations show the diffusion of particles as a function of time.

Figure \ref{fig:mixing}a shows mixing PDFs in the upper ocean for depths greater than 1 m.   The animation of this result, shows that the PDF for $\zeta =-12.8 \; {\rm m}$ spreads very rapidly compared to the PDF for $\zeta =-1.6 \; {\rm m}$.    The spreading occurs for times $t \leq 4 \; {\rm s}$.     This initial spreading of the PDFs is not associated with true mixing.  Particles that are deep beneath the free surface do not follow the orbital velocities of the waves as closely as particles that are near the free surface, so the PDFs are affected differently depending on their initial depths.   

Based on the animation of Figure \ref{fig:mixing}a, the particles at the shallowest initial depth ($\zeta =-1.6 \; {\rm m}$) start to cross the free surface around time $t=90 \; {\rm s}$.   The particles at the deepest depth at $\zeta =-12.8 \; {\rm m}$  do not start mixing until around time $t=150 \; {\rm s}$, which corresponds to the first appearance of swirling jets in Figure \ref{fig:langmuir}i.   The shallowest cases have all penetrated down to $\zeta = 12 \; {\rm  m}$ after about 4 minutes.   The corresponding rate of vertical diffusion is about 5 cm/s, which is greater than the observations of \mbox{\cite{langmuir1938}}.  The vertical diffusion based on data assimilation is slightly greater than the lower bound of the experimental data in Figure 3 of  \mbox{\cite{smith2001}}.  The rate of vertical diffusion would have been even higher if the mixing had occurred when the Langmuir cells were fully formed.

Figure \ref{fig:mixing}b shows mixing PDFs in the upper ocean for depths less than 1 m.    The particles at $\zeta = 10 \; {\rm cm}$ first cross the free surface around time $t=4 \; {\rm s}$.   The mixing in the air is much more rapid than the mixing in the water.  Figure \ref{fig:mixing}c shows mixing PDFs for initial offsets that are within 20 cm of the free surface.    Particles that are slightly above the free surface diffuse down 10 m into the water within 4 minutes. 

Figures \ref{fig:mixing}d and e show mixing PDFs in the lower atmosphere for heights less than and greater than 1 m, respectively.   Particles take less than two minutes to diffuse 50 m upward into the atmosphere from the free surface.   The rate of diffusion upward into the atmosphere is about 0.4 m/s.   

\begin{figure*}
\begin{center}
\begin{tabular}{cc}
\includegraphics[trim=14mm 65mm 22mm 65mm,clip=true,angle=0,width=0.35\linewidth]{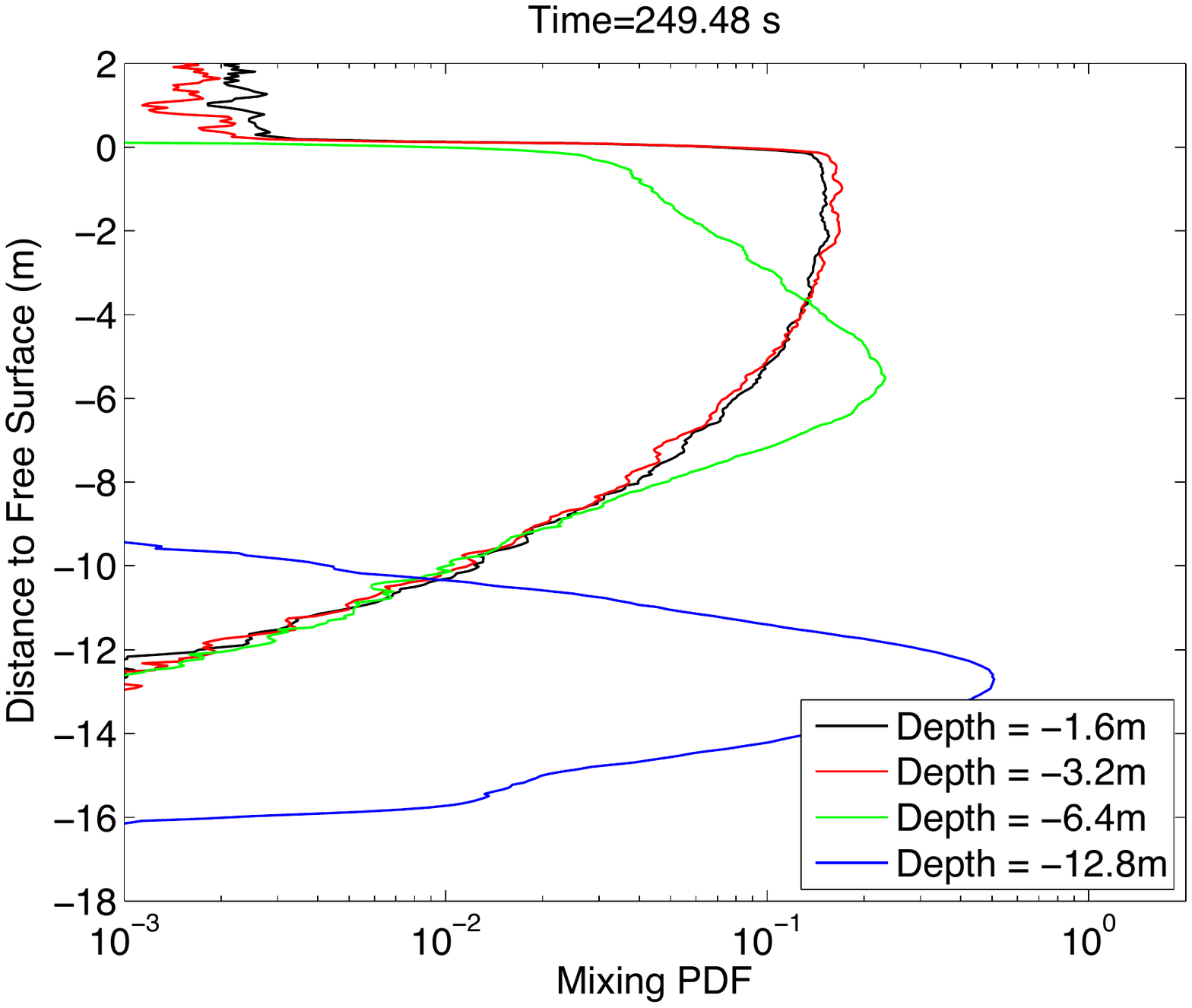} &
\includegraphics[trim=14mm 65mm 22mm 65mm,clip=true,angle=0,width=0.35\linewidth]{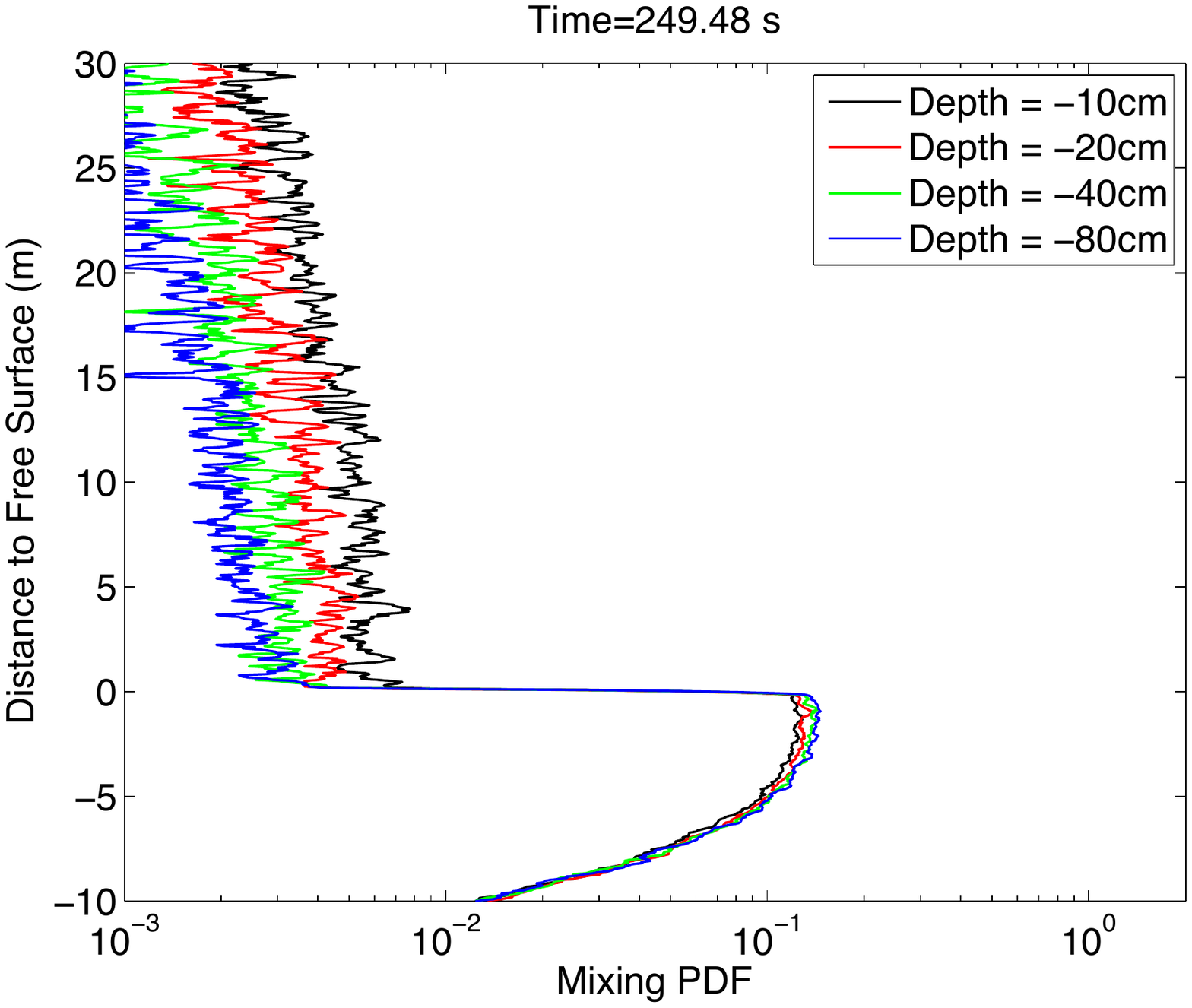} \\
(a) & (b) \\
\includegraphics[trim=14mm 65mm 22mm 65mm,clip=true,angle=0,width=0.35\linewidth]{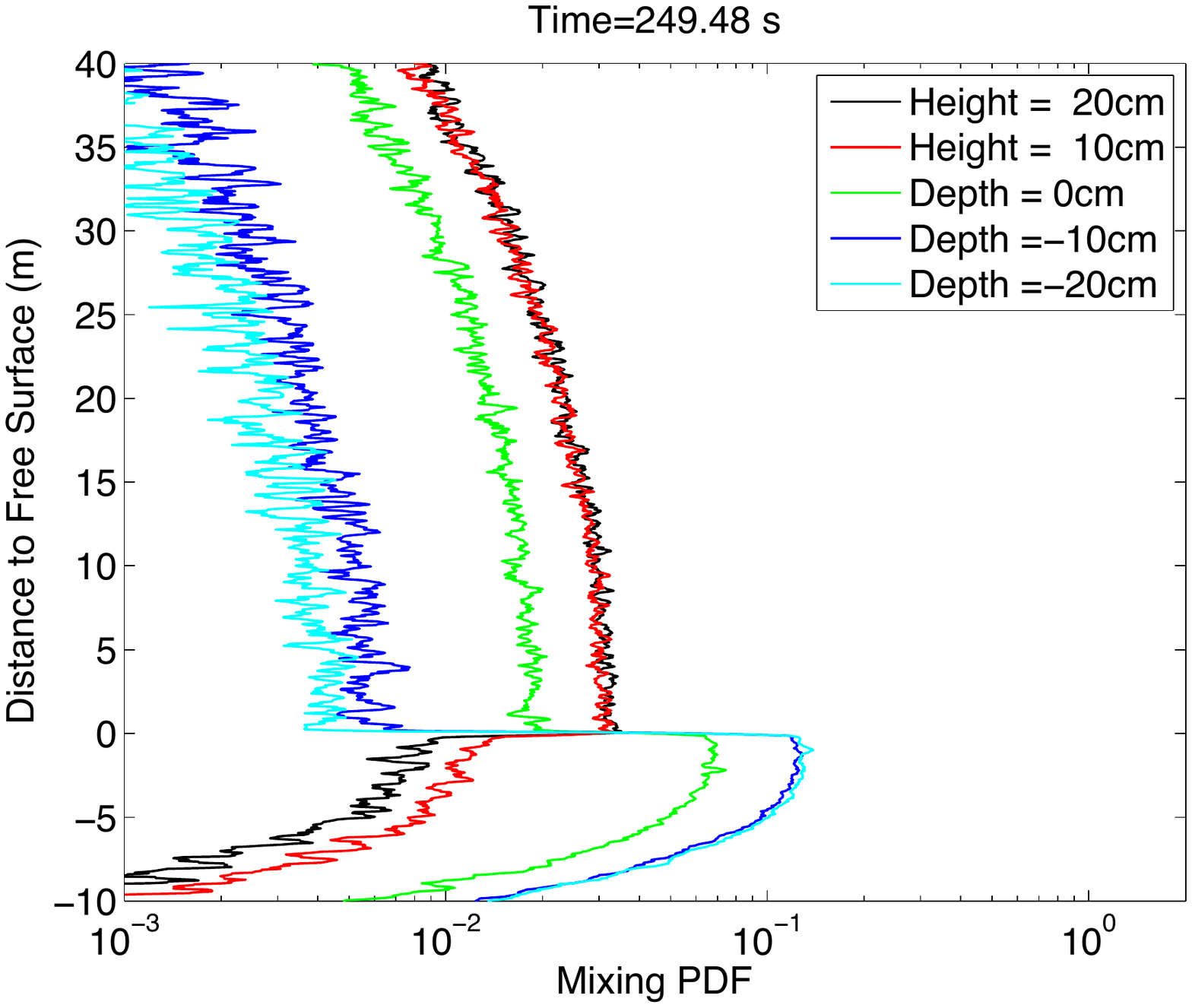} &
\includegraphics[trim=14mm 65mm 22mm 65mm,clip=true,angle=0,width=0.35\linewidth]{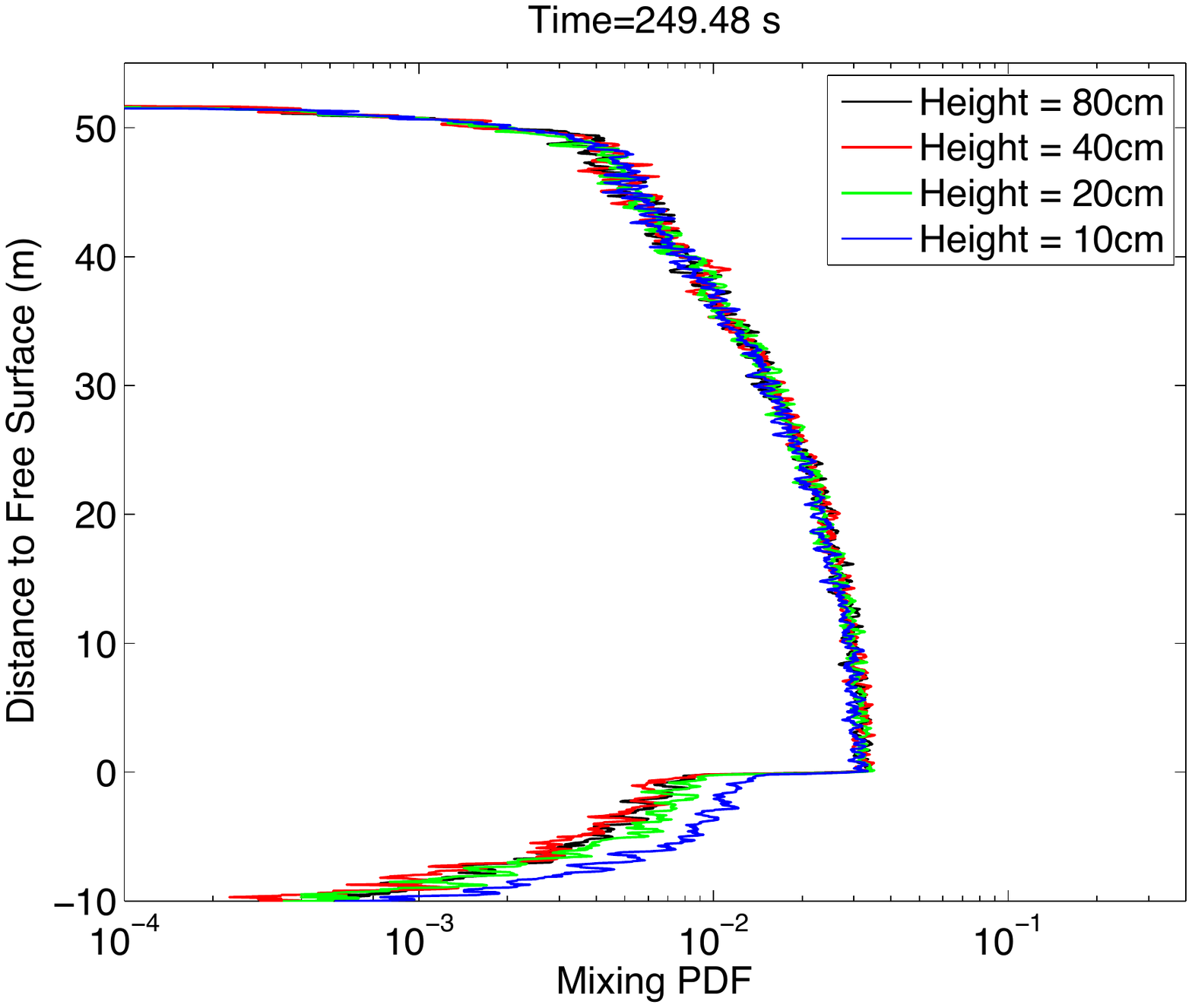} \\
(c) & (d) \\
\includegraphics[trim=14mm 65mm 22mm 65mm,clip=true,angle=0,width=0.35\linewidth]{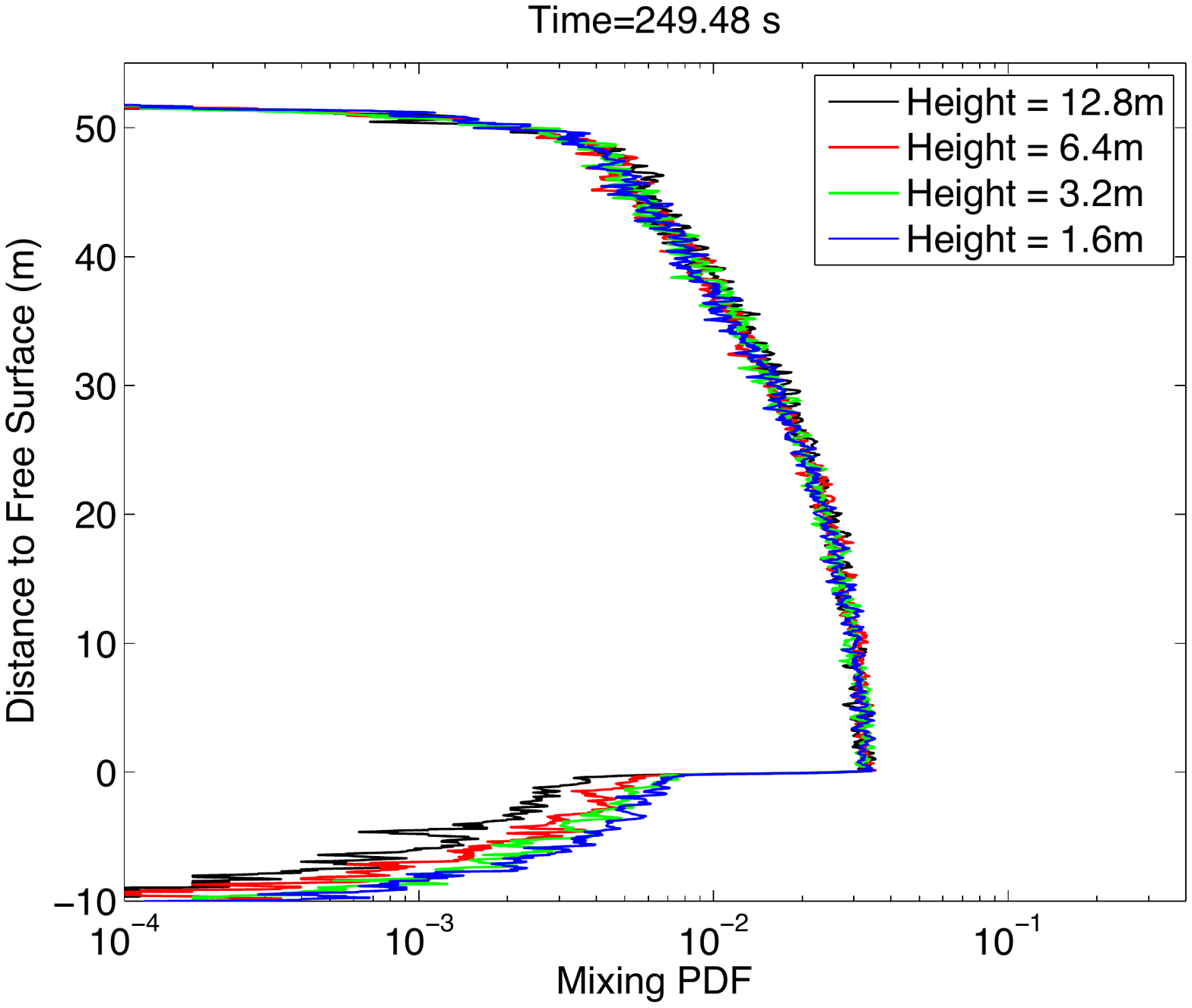} & \\
(e) &
\end{tabular}
\end{center}
\caption{\label{fig:mixing} Mixing PDFs for various initial offsets relative to the free surface for (a) $-12.8{\rm m} \leq \zeta \leq -1.6{\rm m}$; (b) $-80{\rm cm} \leq \zeta  \leq -10{\rm cm}$; (c) $-20{\rm cm} \leq \zeta \leq  20{\rm cm}$; (d) $10{\rm cm} \leq \zeta \leq 80{\rm cm}$; and (e) $1.6{\rm m} \leq \zeta \leq 12.8{\rm m}$.  Animations of these results are available at 
(b) \mbox{\cite{mixing1}} \href{http://youtu.be/Ftkl8-AiLM8}{$-12.8{\rm m} \leq \zeta \leq -1.6{\rm m}$};
(a) \mbox{\cite{mixing2}} \href{http://youtu.be/WOdB9\_NpZEA}{$-80{\rm cm} \leq \zeta \leq -10{\rm cm}$}; 
(c) \mbox{\cite{mixing3}} \href{http://youtu.be/yKH-nv3\_UOM}{$-20{\rm cm} \leq \zeta \leq  20{\rm cm}$};
(d) \mbox{\cite{mixing4}} \href{http://youtu.be/yxJY5a\_qSnk}{$10{\rm cm} \leq \zeta \leq 80{\rm cm}$};
(e) \mbox{\cite{mixing5}} \href{http://youtu.be/jJVvSXYIgLY}{$1.6{\rm m} \leq \zeta \leq 12.8{\rm m}$}.}
\end{figure*}

\section{Conclusion}

A formulation for assimilating data into NFA has been developed.    The data assimilation is illustrated by injecting HOS simulations for the wavy portion of the flow and log profiles of the wind and wind drift for the vortical portion of the flow into NFA.    The process is also applicable to the assimilation of measurements of the ocean surface, the wind, and the wind drift.   The low wavenumber portion of the wave spectrum is assimilated using nudging, and high wavenumbers are free to form naturally.  Similarly, the mean profiles of the wind and the wind drift are also assimilated using nudging, and the turbulent fluctuations form naturally.  In contrast to subgrid-scale models of turbulence, nudging permits the direct enforcement of wave and turbulence statistics. The assimilation of data into NFA permits the investigation of ocean-wave physics with higher bandwidths and more complexity than is possible using either HOS simulations or field measurements. 

Future plans include performing data assimilations of larger patches of the ocean surface with higher resolution for longer periods of time to investigate the statistics and underlying structure of breaking waves.   A larger domain with a longer duration is desirable because it will lead to wave-breaking events with more nonlinearity than those discussed in this paper.   A larger domain and a longer duration are also desirable to allow more room for windrows and wind streaks to develop.  The assimilation of real data will allow back scatter to occur that is not possible using the current formulation with HOS.   Using real data will permit the investigation of non-equilibrium effects due to the effects of wave swell, and growing and dying seas.  Alternatively, the present formulation could be modified to allow backscatter.  We believe that the nudging procedure can be generalized to enforce higher statistical moments, and to investigate non-stationary flows and turbulence control at high Reynolds numbers.   In summary, we are hopeful that future studies will be able to build upon the framework that is provided in this paper to improve our understanding of \mbox{\#{theoceansheartbeat}}.

\comment{Depending on the type, radar measures the position of the ocean surface or the velocities on the ocean surface,  which can be directly related to the wavy portion of the flow.    Optics measure the wave slope, which again can be related to wavy portion of the flow.  Options include WaMoS.  Ellen has suggested SkyBox satellite images.   Move the cutoff to lower K to check impact on backscatter.}

\comment{The separation of the flow into wavy and vortical components in combination with using nudging to enforce statistical properties of the flow is a new approach for simulating turbulence.}

\section*{Acknowledgment}
Dr.\ Thomas Drake at the Office of Naval Research sponsors this research  (contract number N00014-12-C-0568).  The first author is very grateful to the Office of Naval Research for long-term support over several decades.   The numerical simulations are supported in part by a grant from the Department of Defense High Performance Computing Modernization Program (http://www.hpcmo.hpc.mil/).  The numerical simulations have been performed on the SGI ICE X at the US Air Force Research Laboratory and the Cray XE6 at the US Army Engineer Research and Development Center.   We are grateful to Dr.\ James W. Rottman at Leidos, Inc.\  and Dr.\  Ivan Savelyev at the Naval Research Laboratory for helpful interactions.   We thank Prof.\ Stephen Thorpe at Bangor University for providing us with helpful references.  We thank Dr.\ Jochen Horstmann for providing us with SuFMoS image sequences of wind streaks.

\bibliographystyle{30onr}
\bibliography{30onr}

\end{document}